\documentclass[11pt,a4paper]{article}
\pdfoutput=1
\usepackage[subrefformat=parens,labelformat=parens]{subfig}  
\usepackage{jinstpub}
\usepackage[english]{babel}
\usepackage[utf8x]{inputenc}
\usepackage[T1]{fontenc}

\usepackage{amsmath}
\usepackage{graphicx}
\usepackage[colorinlistoftodos]{todonotes}
\usepackage[titletoc,title]{appendix}
\usepackage{dirtytalk}       
\usepackage{array,tabularx}   
\usepackage{mathrsfs}         
\usepackage{enumitem} 
\usepackage{booktabs} 

\newenvironment{conditions*} 
  {\par\vspace{\abovedisplayskip}\noindent
   \tabularx{\columnwidth}{>{$}l<{$} @{${}={}$} >{\raggedright\arraybackslash}X}}
  {\endtabularx\par\vspace{\belowdisplayskip}}

\makeatletter
\newcommand*{\rom}[1]{\expandafter\@slowromancap\romannumeral #1@}
\makeatother

\title{Calibration of the charge and energy loss per unit length of the MicroBooNE liquid argon time projection chamber using muons and protons}

\collaboration{MicroBooNE Collaboration}

\author[j]{C.~Adams}
\author[l]{M.~Alrashed}
\author[k]{R.~An}
\author[c]{J.~Anthony}
\author[cc]{J.~Asaadi}
\author[p]{A.~Ashkenazi}
\author[hh]{S.~Balasubramanian}
\author[i]{B.~Baller}
\author[q]{C.~Barnes}
\author[t]{G.~Barr}
% ...

% ...
\affiliation[c]{University of Cambridge, Cambridge CB3 0HE, United Kingdom}
% ...
\affiliation[i]{Fermi National Accelerator Laboratory (FNAL), Batavia, IL 60510, USA}
\affiliation[j]{Harvard University, Cambridge, MA 02138, USA}
\affiliation[k]{Illinois Institute of Technology (IIT), Chicago, IL 60616, USA}
\affiliation[l]{Kansas State University (KSU), Manhattan, KS, 66506, USA}
% ...
\affiliation[p]{Massachusetts Institute of Technology (MIT), Cambridge, MA, 02139, USA}
\affiliation[q]{University of Michigan, Ann Arbor, MI, 48109, USA}
% ...
\affiliation[t]{University of Oxford, Oxford OX1 3RH, United Kingdom}
% ...
\affiliation[cc]{University of Texas, Arlington, TX, 76019, USA}
% ...
\affiliation[hh]{Wright Laboratory, Department of Physics, Yale University, New Haven, CT, 06520, USA}

  \emailAdd{microboone\_info@fnal.gov}

\abstract{ We describe a method used to calibrate the position- and time-dependent response of the MicroBooNE liquid argon time projection chamber anode wires to ionization particle energy loss. The method makes use of crossing cosmic-ray muons to partially correct anode wire signals for multiple effects as a function of time and position, including cross-connected TPC wires, space charge effects, electron attachment to impurities, diffusion, and recombination. The overall energy scale is then determined using fully-contained beam-induced muons originating and stopping in the active region of the detector. Using this method, we obtain an absolute energy scale uncertainty of 2\% in data. We use stopping protons to further refine the relation between the measured charge and the energy loss for highly-ionizing particles. This data-driven detector calibration improves both the measurement of total deposited energy and particle identification based on energy loss per unit length as a function of residual range. As an example, the proton selection efficiency is increased by 2\% after detector calibration. } 

\date{\today}

\begin{document}

\maketitle
\flushbottom

\section{Introduction}
\label{sec:Intro}

Liquid argon time projection chamber (LArTPC) technology provides both particle tracking and energy loss per unit length reconstruction with high resolution. The basic working principle of a LArTPC neutrino detector is that neutrinos first interact with an argon nucleus and produce charged and neutral secondary particles.  The charged secondary particles travel through liquid argon and mainly lose their energy by ionizing and exciting argon atoms.  The ionization electrons travel under an applied electric field to a set of anode wire planes. The charge is measured on the anode wire planes in order to reconstruct particle trajectories and energies. Excited argon atoms also produce scintillation light, which is detected by the photon detectors. In this paper, we address the calibration of the charge collected at the anode wires for the MicroBooNE LArTPC~\cite{uBDetJINST}.

MicroBooNE is a 170-ton LArTPC with dimensions $2.56\times2.33\times10.36$ m$^3$ (horizontal drift dimension$\times$height$\times$length). The nominal electric field inside the TPC is 0.273 kV/cm, which leads to a nominal electron drift velocity of 0.11 cm/$\mu$s. The drifted charge from particle interactions is read out in three planes with a plane spacing and a wire pitch of 3 mm. The 3456 collection plane wires are vertical and the angle between the induction and collection plane wires is 60 degrees. The readout window size is 4.8 ms and the ADC sampling rate is 2 MHz. Scintillation photons are observed by 32 photo-multipliers (PMTs)~\cite{Conrad:2015xta}. MicroBooNE started collecting neutrino events in Booster Neutrino Beam (BNB) at Fermilab in October 2015.

The first step of energy loss per unit length ($dE/dx$) reconstruction in LArTPC detectors involves the extraction of charge information from the signals (waveforms) on the anode plane wires. The MicroBooNE experiment uses several signal processing techniques~\cite{uBsignalJINST, uBsignalJINST_2} including noise filtering and signal deconvolution for charge extraction. However, the total charge extracted in this way normally does not equal the total charge produced from ionization for a number of reasons. The reasons are distortions in detector response due to cross-connected TPC channels~\cite{uBnoiseJINSTpre}, space charge effects (SCE)~\cite{sc2}, electron attachment to impurities~\cite{ICARUS1O2,ICARUS2O2,ICARUSpm,longbo,argoneut}, diffusion~\cite{diffusion2015Li} and recombination~\cite{recomb, recombICARUS}. To trace back  the exact amount of charge released from the original interaction, we have to correct for each of these effects starting from the ionization charge per unit length in the direction of the particle's travel, $dQ/dx$, as reconstructed from the signal
collected on the TPC wires. Throughout this paper $dx$ refers to the length of a short segment of the particle track, which should not be confused with an infinitesimal distance in x (drift) direction of the TPC.

\label{sec:micro_calibration}

The data-driven detector calibration consists of several steps. In the first step we correct the position- and time-dependence of the detector response to ionization charge using data from cosmic ray muons (CR) which enter the detector from either the anode or cathode side and exit through the opposite face (\say{crossing CR}). This correction process is known as the $dQ/dx$ calibration or equalization calibration of the detector. Once the detector response is corrected to be uniform throughout the TPC and in time, we determine a calibration for converting $dQ/dx$ to $dE/dx$, by using stopping muons from neutrino interactions. Here stopping muons from neutrino interactions refers to the muons produced in the charged current interactions of neutrinos which ultimately decay inside the detector with a Bragg peak in their $dE/dx$ profile. This process is known as the overall $dE/dx$ calibration of the detector. In the last step, stopping protons from neutrino interactions are used to further refine the relation between the measured charge and the energy loss for highly ionizing particles.

The MicroBooNE coordinate system is shown in figure 1 (left). MicroBooNE uses a right-handed coordinate system in which the $y$ axis is vertical, the $x$ axis is horizontal, perpendicular to the anode and cathode planes, and the $z=x\times y$ is also horizontal and along the beam direction. The TPC signal formation in MicroBooNE is illustrated in figure 1 (right), where the induction wire planes are referred to as the “U” and “V” planes and the collection wire plane is referred to as the “Y” plane~\cite{uBDetJINST}. 
In this analysis we focus on calibrating only the collection wire plane of the detector, as this is the wire plane predominantly used for calorimetry in a LArTPC. The calibration of induction planes is generally more difficult because the response of the induction wires is highly dependent on the angle of the tracks relative to the wires. This is caused by the cancellation of overlapping induction signals for large-angle tracks due to the bipolar signal shape. New techniques are being developed to improve the reconstruction of the induction plane signals~\cite{uBsignalJINST,uBsignalJINST_2}, which will allow reliable $dE/dx$ measurements on induction plane wires in the future. 

The two-step calibration procedure presented in this paper is similar to calibration techniques developed for other calorimeters such as the MINOS detectors~\cite{Hartnell:2005uq}. 

\begin{figure}[!ht]
\centering
\includegraphics[width=0.39\textwidth, height=5.0cm]{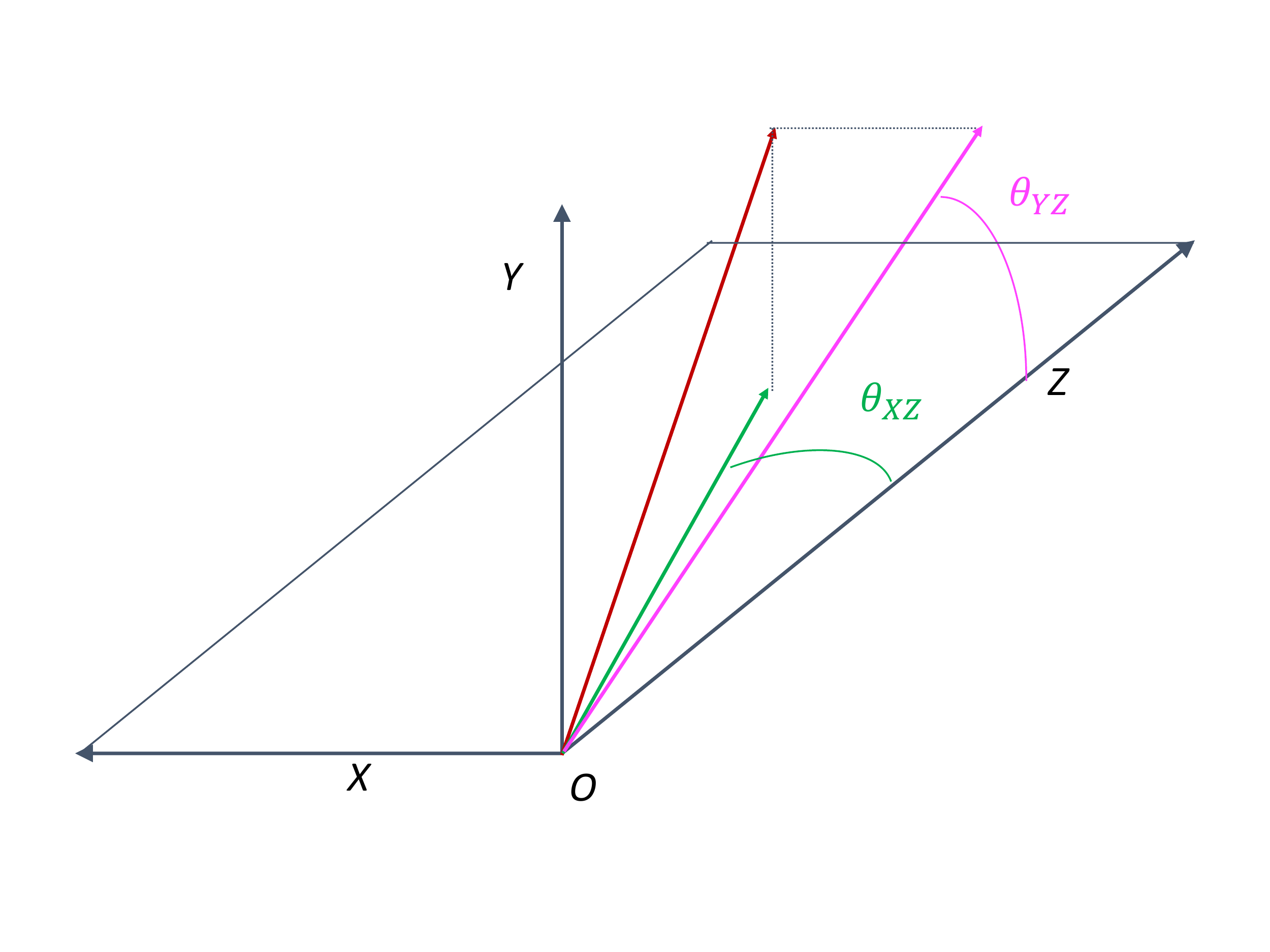}\hspace{0em}
\includegraphics[width=0.60\textwidth, height=6.5cm]{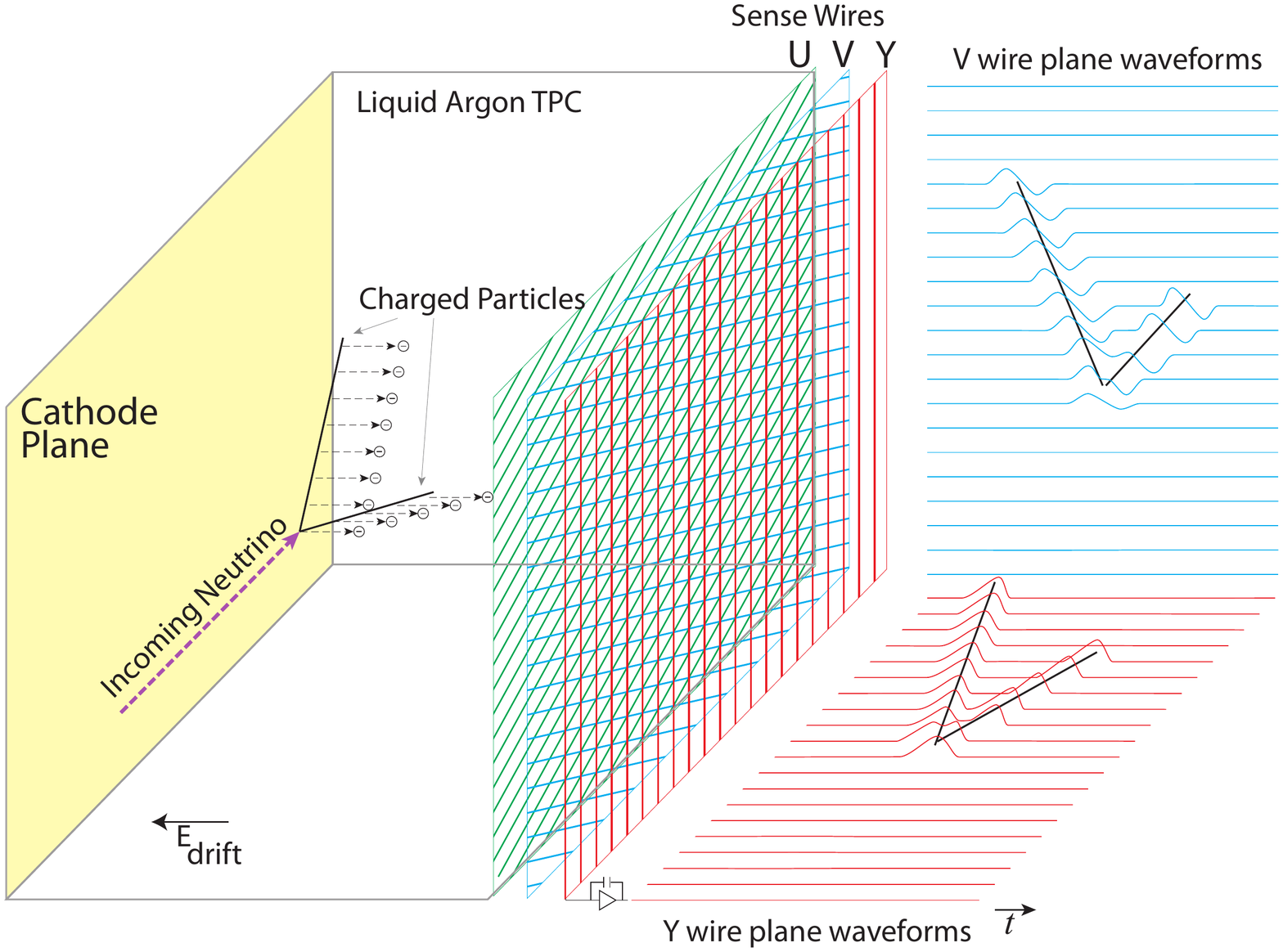}
\caption{(Left) Definition of coordinates $x$, $y$, and $z$, and the angles $\theta_{XZ}$ and $\theta_{YZ}$ of MicroBooNE Detector. $x$ is along the drift direction with the anode at $x=0$ cm and the cathode at $256$ cm; $y$ is in the vertical direction with $y=-116$ cm at the bottom of TPC and $y=116$ cm at the top of TPC; $z$ is along the beam direction with $z=0$ cm at the upstream edge of TPC and $z=1036$ cm at the far end. (Right) Diagram illustrating the signal formation in a LArTPC with three wire planes \cite{uBDetJINST}. U wire plane waveforms are not shown in the diagram.}
\label{fig:theta_xz_yz_def} 
\end{figure}

\section{dQ/dx Calibration}\label{sec:dqdx}

\subsection{Introduction}

\label{sec:rel_cal}
The goal of the $dQ/dx$ calibration procedure is to make the detector response uniform in space and over time. There are many effects that can produce a nonuniform detector response. The dominant effects are described in the following sections.

\subsubsection{Cross-connected TPC channels}
\label{sec:short_wire}
\begin{figure}[!ht]
\centering
\includegraphics[width=0.6\textwidth]{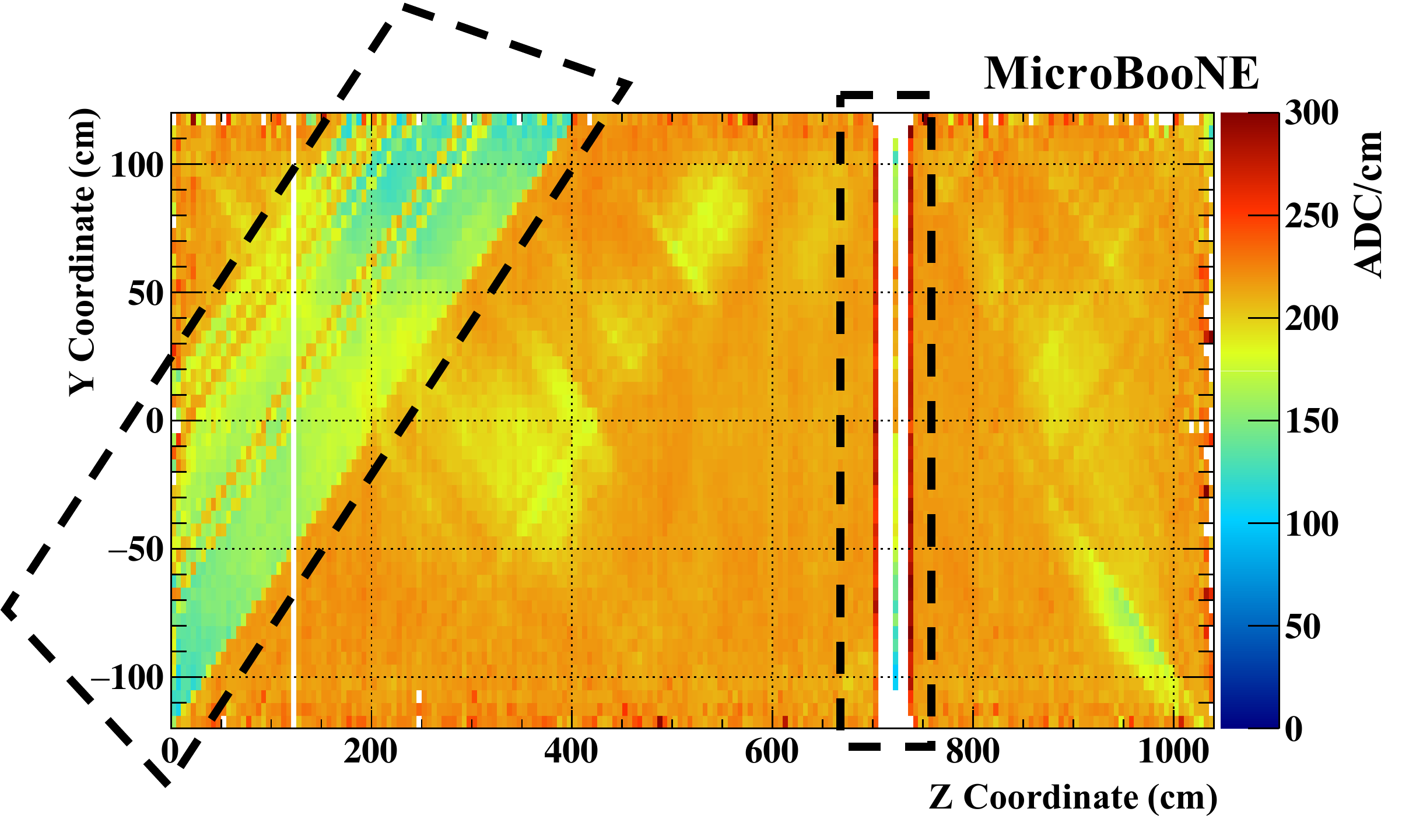}
\caption{Distribution of $dQ/dx$ in the collection plane as a function of Y and Z coordinates in data. The region inside the highlighted boundaries shows the effect of cross-connected TPC channels. The colors in the plot represent the median $dQ/dx$ value for a given 5 cm $\times$ 5 cm cell in the yz plane. 
}
\label{fig:x_yz_var} 
\end{figure}

Cross-connected wires, which can distort the electric field between wire planes~\cite{uBsignalJINST_2}, affect the $dQ/dx$ response of roughly 20\% of the detector volume.
Figure~\ref{fig:x_yz_var} shows the distribution of $dQ/dx$ in the collection plane as a function of Y and Z coordinates in data. The highlighted diagonal region represents a ``shorted-U'' region where multiple U plane channels are shorted to one or more V plane channels. The highlighted vertical region represents a ``shorted-Y'' region where multiple Y plane channels are shorted to one or more V plane channels. 
By calibrating out these effects we recover the affected areas for useful physics.

\subsubsection{Space charge effects (SCE)}
\label{sec:spe}
\begin{figure}[!ht]
\centering
\includegraphics[width=0.6\textwidth]{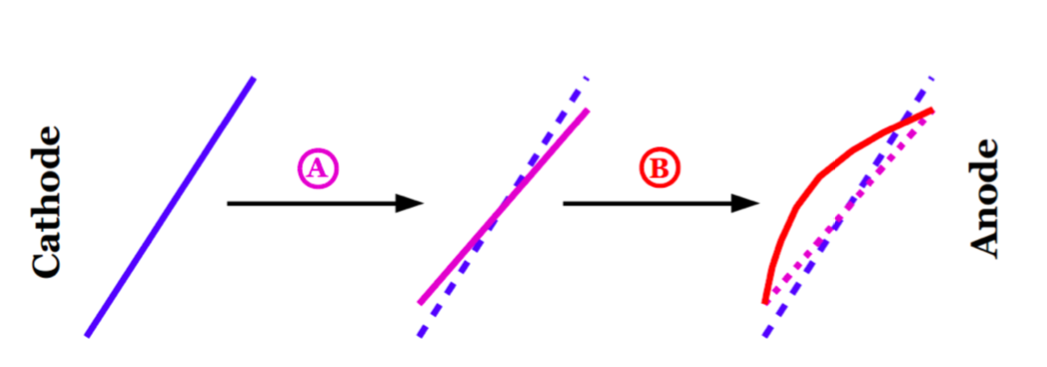}
\caption{Diagram showing the effects  of space charge on track reconstruction. The impact is two-fold: the reconstructed track could be squeezed by two extremes in the transverse directions of the TPC, as indicated in the rotation A, and bent towards cathode, as indicated by transformation B.
(Image from Ref.~\cite{sc2} used by permission of its creator.)}
\label{fig:spe_cartoon} 
\end{figure}

Since MicroBooNE is a surface-based detector, there is a significant flux of CR tracks in the detector volume. Because of this, there is a significant accumulation of slow-moving positive argon ions inside the detector, which is enough to distort the uniformity of the drift electric field. These distortions in the electric field inside the TPC have two significant effects:
\begin{itemize}
\item distortions in the magnitude of the drift electric field
compared to a uniform electric field, and
\item distortions in the electric field direction compared to the nominal
direction parallel to the x-axis.
\end{itemize}
When the magnitude of the electric field is nonuniform, the recombination of electrons and ions is affected. This recombination effect~\cite{recomb} is sensitive to changes in the electric field. When the drift electric field is relatively low, recombination becomes dominant compared to the recombination at higher electric fields. Moreover, space charge effects can lead to spatial distortions in the trajectories of reconstructed particle tracks and electromagnetic showers. The positive argon ions built up in the detector tend to drag ionization electrons closer to the middle of the detector.
The space charge effects are expected to be stronger at the detector edges transverse to the drift direction. The cumulative effect of the space charge in track reconstruction leads to squeezing of the reconstructed track in transverse directions and bending towards the cathode. See Figure~\ref{fig:spe_cartoon}.

\noindent

\label{sec:ana_met}

\begin{figure}[!ht]
\centering
\includegraphics[width=0.49\textwidth]{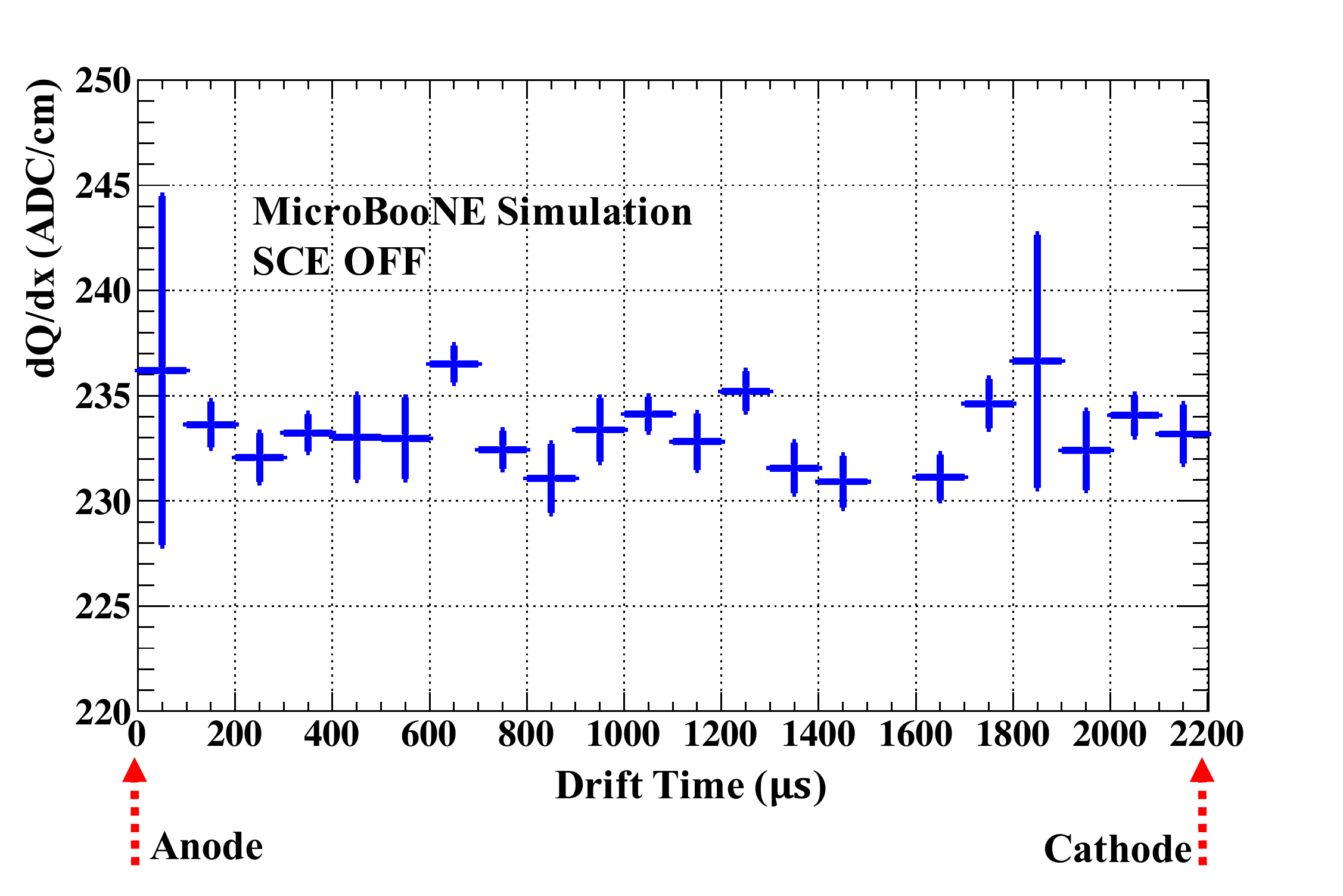}\hspace{0em} 
\includegraphics[width=0.49\textwidth]{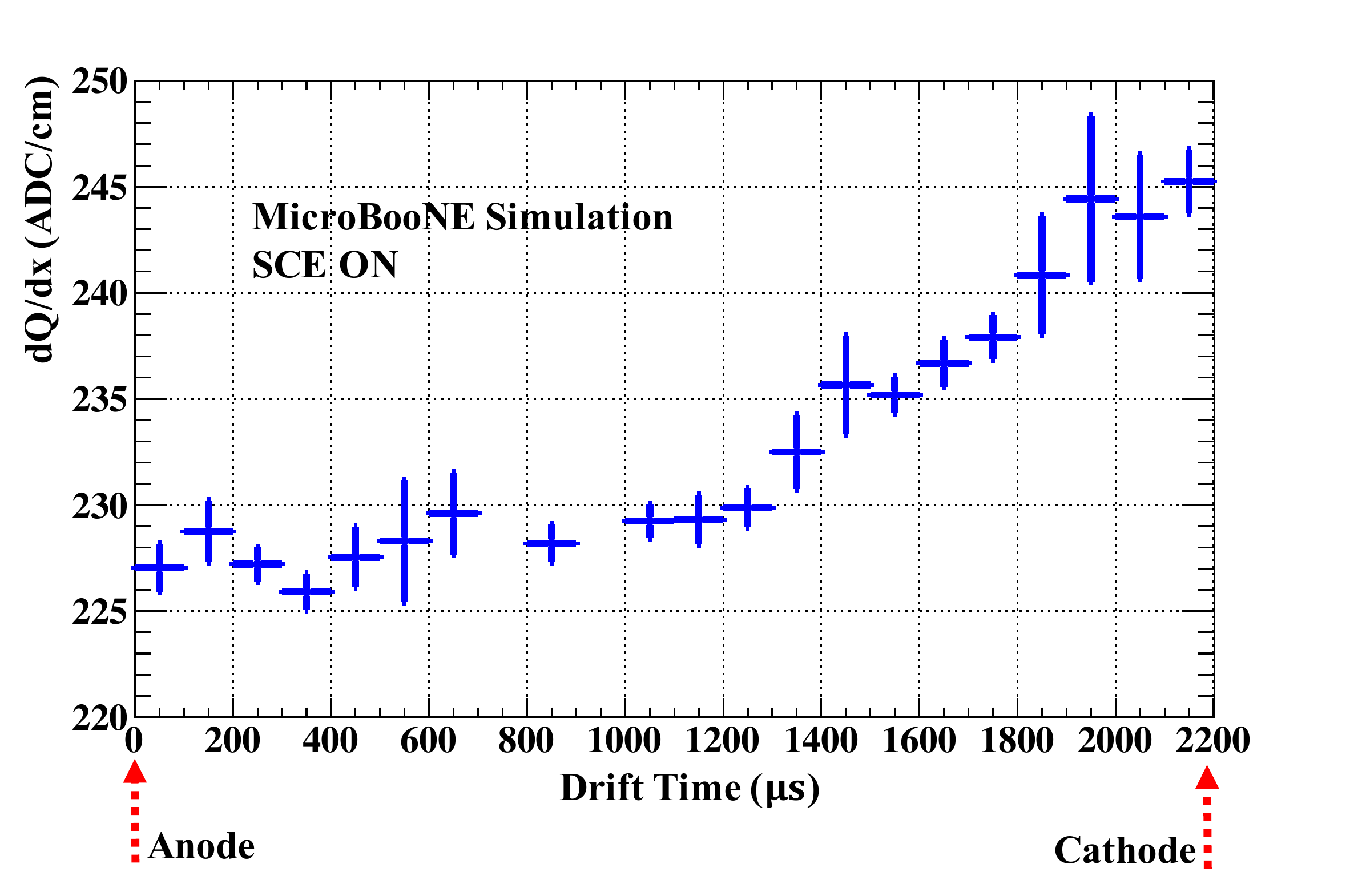} 
\caption{Plots of $dQ/dx$ {\it vs.} drift time generated using samples of simulated isotropic single muons. (Left) Space charge effects are turned off. (Right) Space charge effects are turned on. In both of the samples diffusion is completely turned off and the electron lifetime is set very high. Both plots are created using collection plane information and the uncertainties shown are statistical.}
\label{fig:sp_plots} 
\end{figure}

Figure~\ref{fig:sp_plots} shows how space charge effects~\cite{sc2} implemented in the MicroBooNE simulation change $dQ/dx$ values over the entire drift distance of the MicroBooNE LArTPC. As seen in the right plot of figure~\ref{fig:sp_plots}, the $dQ/dx$ values closer to the cathode are higher compared to that at the anode. The accumulation of positive ions causes the electric field magnitude closer to the cathode to be approximately 10$\%$ higher than at the anode. The higher field suppresses electron-ion recombination near the cathode. In addition, due to spatial distortions, tracks reconstructed closer to the cathode are bent and squeezed making reconstructed $dx$ values smaller. Thus we observe a higher collected charge per unit track length closer to the cathode.

\noindent

\subsubsection{Electron attachment to impurities}
\label{sec:e_life}
When a cloud of ionization electrons drifts to the anode, electronegative contaminants such as H\textsubscript{2}O and O\textsubscript{2} can capture some of the drifting electrons, reducing the $dQ/dx$ measured at the wire planes.
The capture rate is inversely dependent on the strength of the applied electric field, as at higher electric field magnitude, ionization electrons have a greater chance of making it to the anode plane before attaching to electronegative impurities.

Equation \ref{eqn:e_life} describes the depletion of a cloud of ionization
electrons due to capture by electronegative contaminants as the electrons drift towards the anode: 
\begin{equation}
\label{eqn:e_life}
\frac {n_e(t_{\mathrm{collected}})}{n_e(t_0)} = \exp\left(\frac{-(t_{\mathrm{collected}}-t_{0})}{\tau}\right),
\end{equation}
where $t_{0}$ is the start time, $n_{e}(t_{0})$ is the initial number of electrons at time $t_{0}$, $n_{e}(t_{\mathrm{collected}})$ is the number of electrons collected by anode plane wires after a time $t_{\mathrm{collected}}$ and $\tau$ is the electron lifetime.

\begin{figure}[!ht]
\centering
\includegraphics[width=0.49\textwidth]{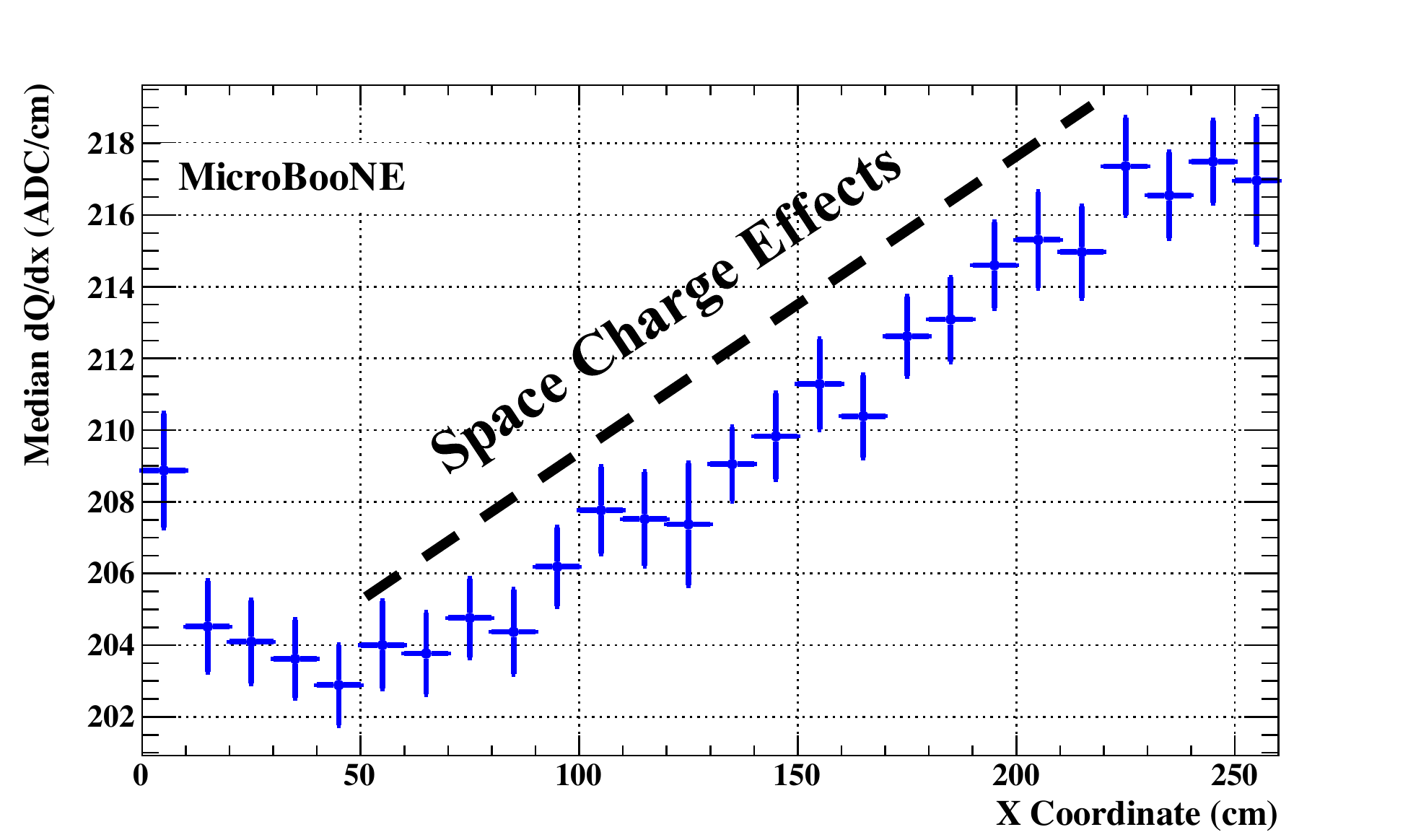}
\includegraphics[width=0.49\textwidth]{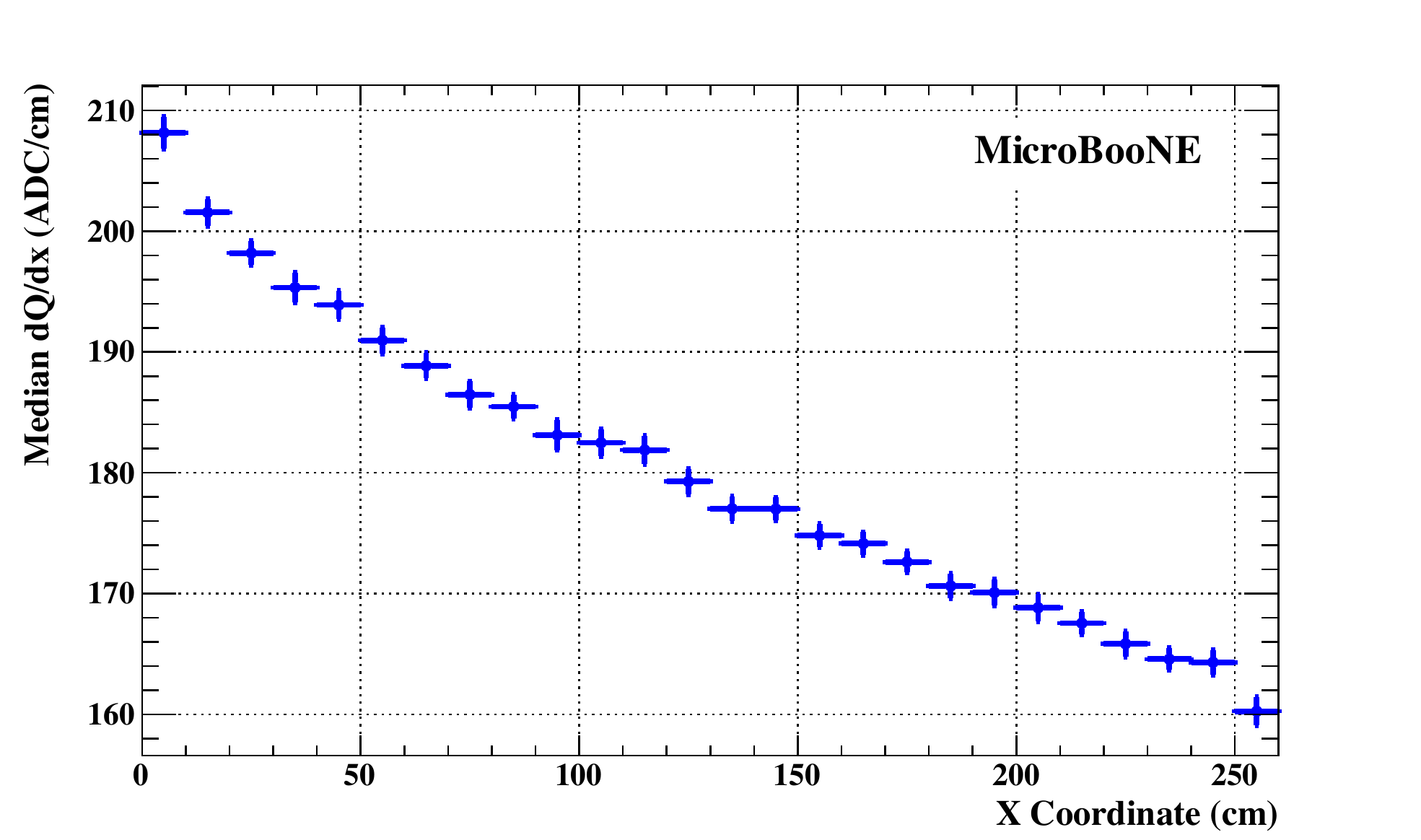} 
\caption{$dQ/dx$ as a function of drift distance under different purity conditions in MicroBooNE data. Here the median $dQ/dx$ value is plotted for 10 cm wide bins in drift direction on the collection plane. (Left) $dQ/dx$ as a function of drift distance in high argon purity conditions, with free electron lifetime exceeding 20~ms, on the date Feb-25-2016. (Right) $dQ/dx$ as a function of drift distance in low argon purity conditions, corresponding to the date March-31-2016, when purity was unusually low (see Figure \ref{fig:time_corre_gbl_mdn}).}
\label{fig:low_e_life} 
\end{figure}

The electron lifetime depends on the amount of electronegative contaminants present in the medium, where higher electron lifetime corresponds to lower argon contamination levels. Figure~\ref{fig:low_e_life} shows the effect of electron attachment to impurities on $dQ/dx$ values under different purity conditions. In high argon purity conditions space charge effects become dominant, causing $dQ/dx$ to increase from anode to cathode. In low argon purity conditions $dQ/dx$ value drops from the anode to the cathode due to electron attachment to impurities.

\subsubsection{Diffusion}
\label{sec:diff}

Both longitudinal and transverse diffusion can be studied.
The cloud of ionization electrons tends to get smeared out in the direction of the drift because of longitudinal diffusion. This widens and lowers the pulse height of the signal at longer drift distance, which can lead to loss of signal if the pulse-height is below  reconstruction threshold. The effect of longitudinal diffusion is small in practice because the electronics shaping time is selected to be comparable to the diffusion smearing. In addition, transverse diffusion charge can allow charge to be detected on
multiple wires, which can smear the detected signal and reduce the resolution of charge reconstruction. This smearing is not corrected in the signal processing procedure used for this analysis.

\subsubsection{Temporal variations}
The detector response can change over time because of effects such as drift of the electronics gains, changes in temperature, different running conditions, etc. Note the time referred to here is the calendar time, not to be confused with drift time in the TPC. The most significant time-dependent change affecting MicroBooNE calibrations are changes in argon purity.

\subsubsection{dQ/dx equalization strategy}

The general strategy of the $dQ/dx$ calibration is to separate detector nonuniformities into yz plane, x (drift) direction, and calendar time variations, and calibrate them in sequence using CR. More details will be discussed in section~\ref{sec:ana_met_dqdx}. 

CR are the standard candle for uniform energy deposition throughout the detector. These CR have typical momenta in the range of 4 - 5~GeV, which results in a peak $dE/dx$ of $\sim$~1.7 MeV/cm . 

It should be noted that the calibration scheme described here is an approximation because contributions such as space charge effects cannot be completely factorized in separate yz plane and drift directions: for a given $x$ value, the variations in the yz plane are different. Ideally we should carry out this calibration by voxelizing the detector into small three-dimensional cubes and derive a calibration constant for each cube. But limited statistics make the voxelation approach impractical. Moreover, spatial distortions introduced by space charge effects that impact track reconstruction would not be completely addressed by such a voxel-based calibration scheme.

\subsection{Data sample}
\label{sec:data_sample_dqdx}

The $dQ/dx$ calibration of the detector is carried out using both data and Monte Carlo (MC) simulation. For MC simulation we use samples of simulated CORSIKA CR events~\cite{Heck:1998vt} overlaid with neutrino interactions simulated with the {\sc genie} generator~\cite{Andreopoulos:2009rq}. The simulation of particle propagation in the MicroBooNE detector is based on Geant4~\cite{Agostinelli:2002hh}. The drift of ionization electrons to the wire planes and propagation of scintillation light to the PMTs are modeled in LArSoft~\cite{larsoft}, which induces effects such as recombination, diffusion, space charge effects and electronics readout including cross-connected TPC channels. The electron lifetime is set to infinity in the simulation so there is no electron attachment to impurities. For data, we use data collected in the MicroBooNE detector with a trigger coincident with the beam and with a random trigger anti-coincident with the beam
from February to October of 2016, and a second period from September 2017 to March 2018.
In both MC and data we use the Pandora~\cite{pandora} pattern recognition program, combined with a Kalman filter implemented in the LArSoft framework to fit the CR tracks. 

\subsection{Event selection}

\label{sec:sel_cut_rel}

\begin{figure}[!ht]
\centering
\includegraphics[width=0.8\textwidth]{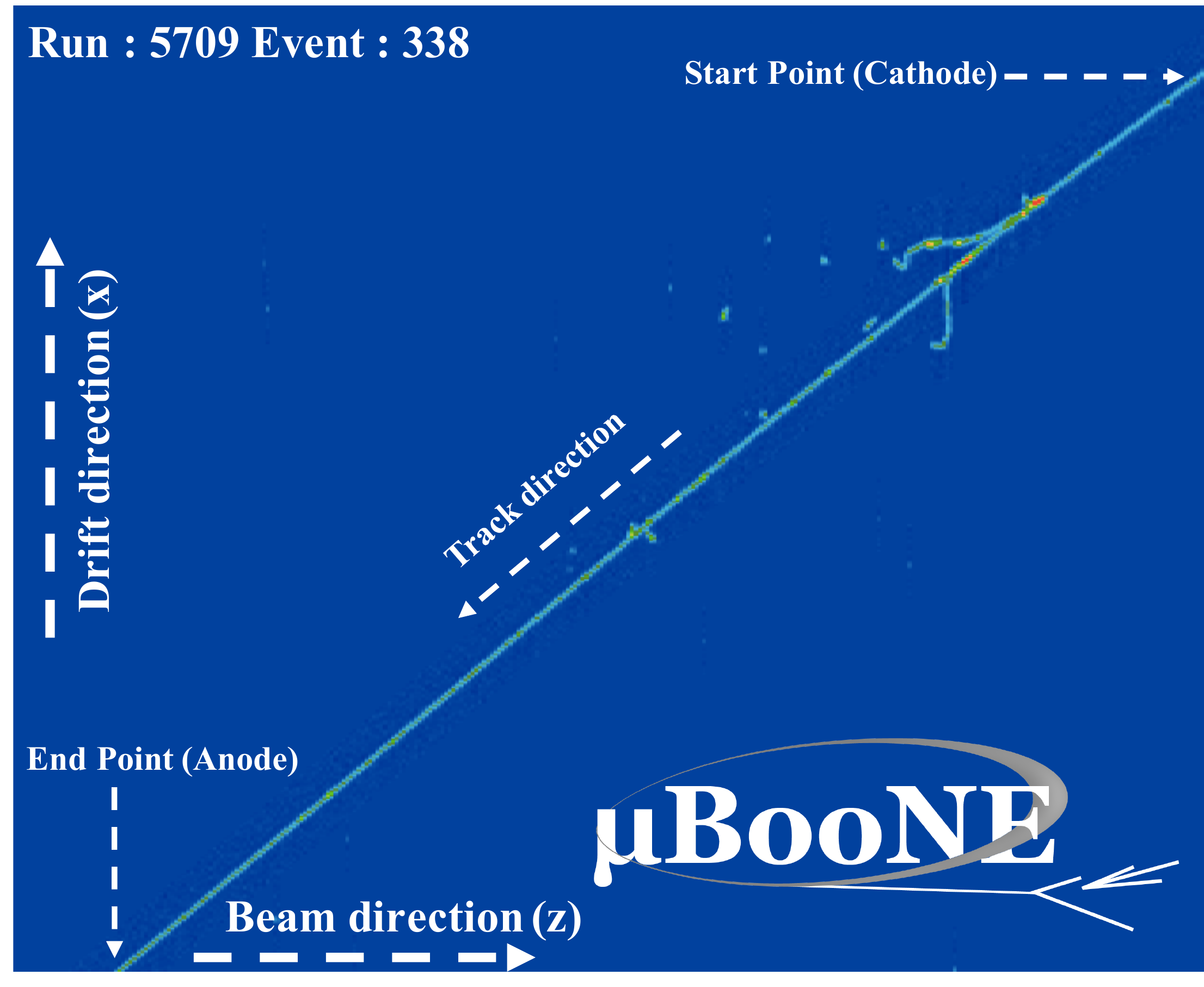} 
\caption{An event display showing examples of anode-cathode crossing CR.}
\label{fig:cross_event_display} 
\end{figure}
\begin{figure}[!ht]
\centering
\includegraphics[width=0.6\textwidth]{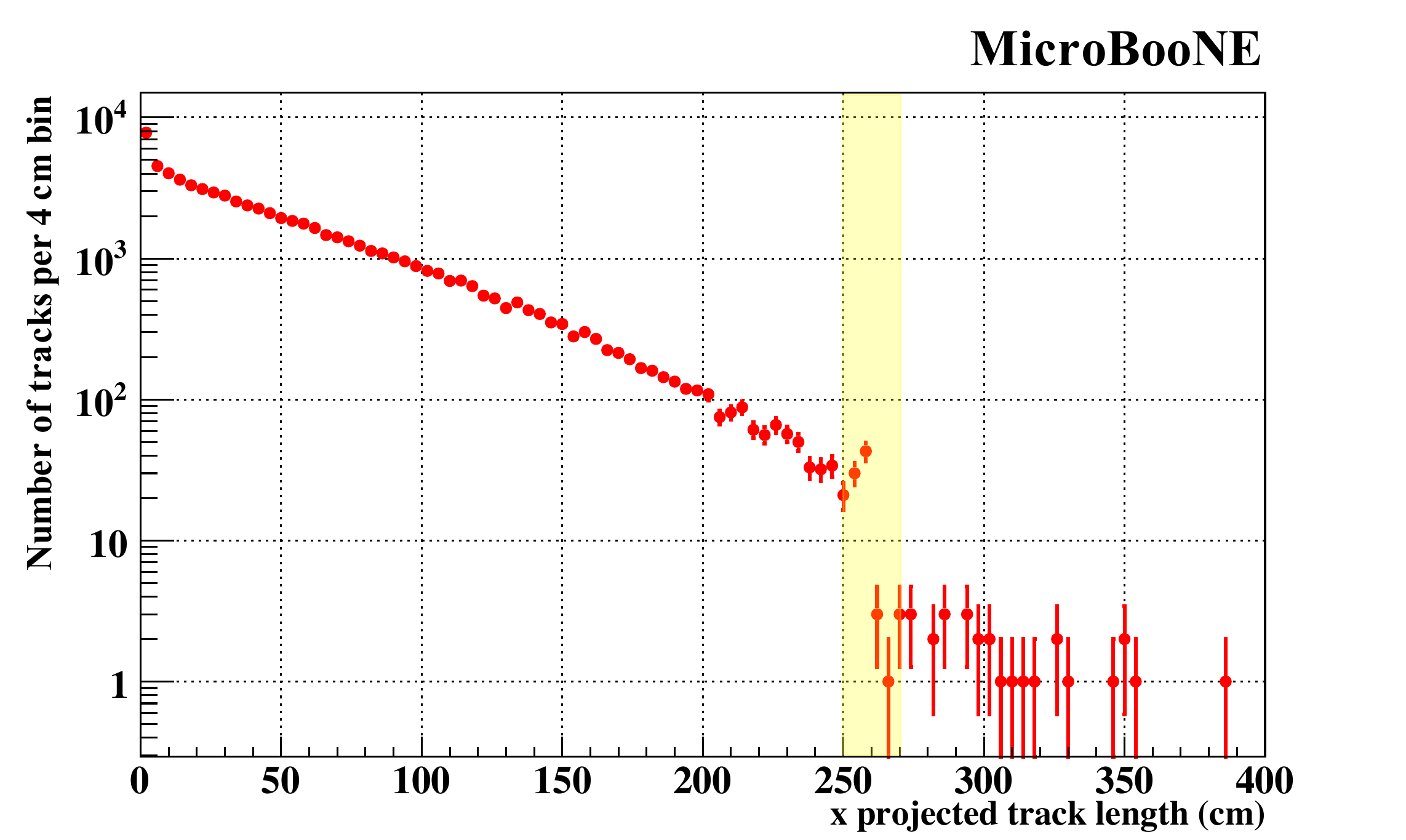} 
\caption{Distribution of x projected track length of all CR tracks in a single run 5709 taken on April 1, 2016. The colored band shows anode-cathode crossing CR candidates in MicroBooNE data. }
\label{fig:cosmics_in_mb} 
\end{figure}

For the $dQ/dx$ calibration of the detector, we choose to use the anode-cathode crossing CR. The main reason is that the anode-cathode crossing CR span the entire drift distance, which make them a very valuable sample to study any effects that depend on the drift distance. 
Each CR in the selection is ``tagged'' with an initial time $t_0$, corresponding to the time that the electrons produced nearest the anode are detected. Crossing CR have a wide spatial but limited angular coverage. Figure~\ref{fig:cross_event_display} shows an example of anode-cathode crossing CR in MicroBooNE data. Figure~\ref{fig:cosmics_in_mb} shows the x projected track length distribution of all CR. It has been estimated that $\sim$ 0.13$\%$ of the tracks are anode-cathode crossing CR. Out of a total of 69429 tracks from a single run 5709 taken on April 1, 2016, 87 are crossing CR and 52 tracks survived after angular cuts. 

\begin{figure}[!ht]
\centering \includegraphics[width=0.49\textwidth]{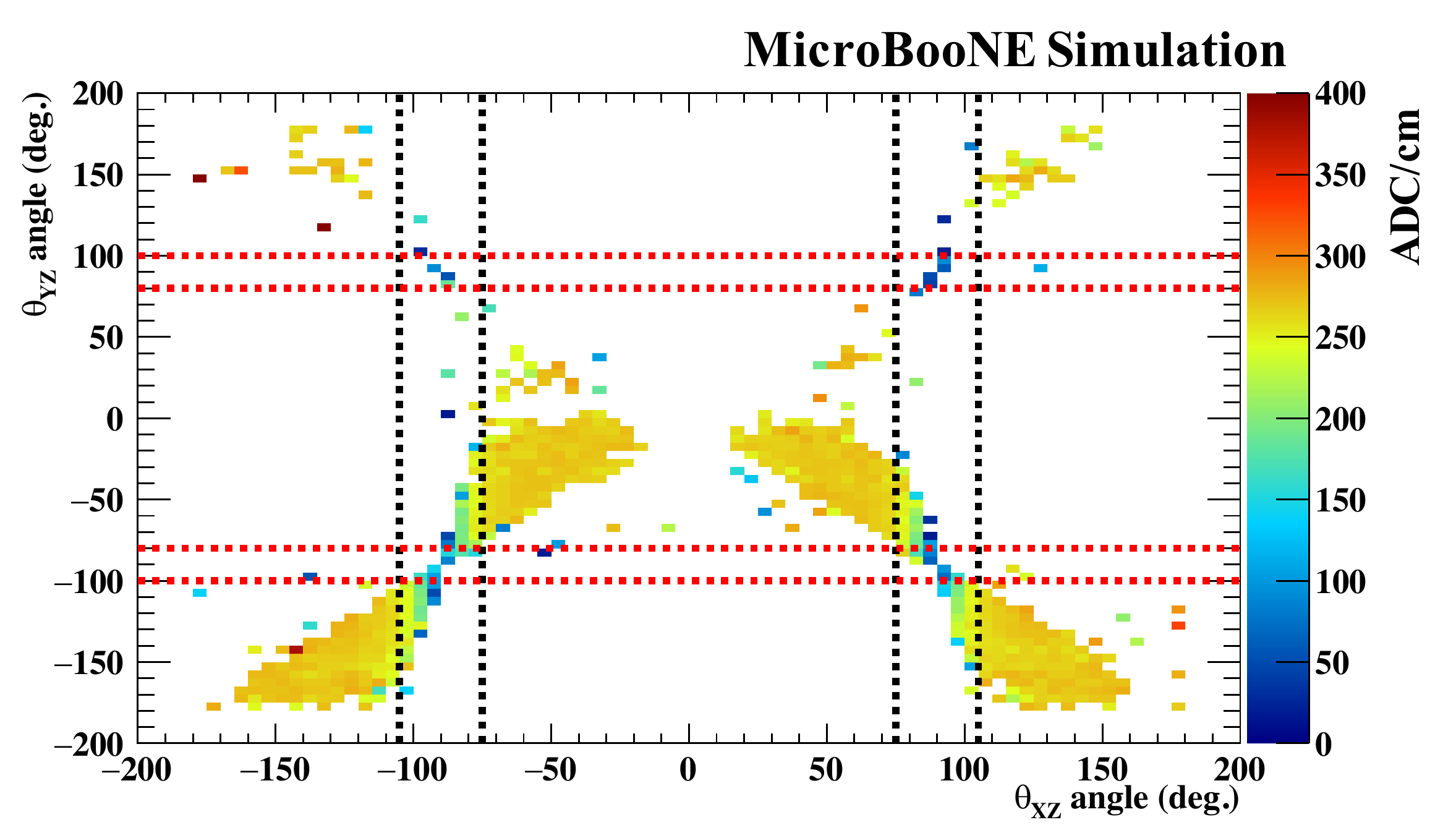} 
\includegraphics[width=0.49\textwidth]{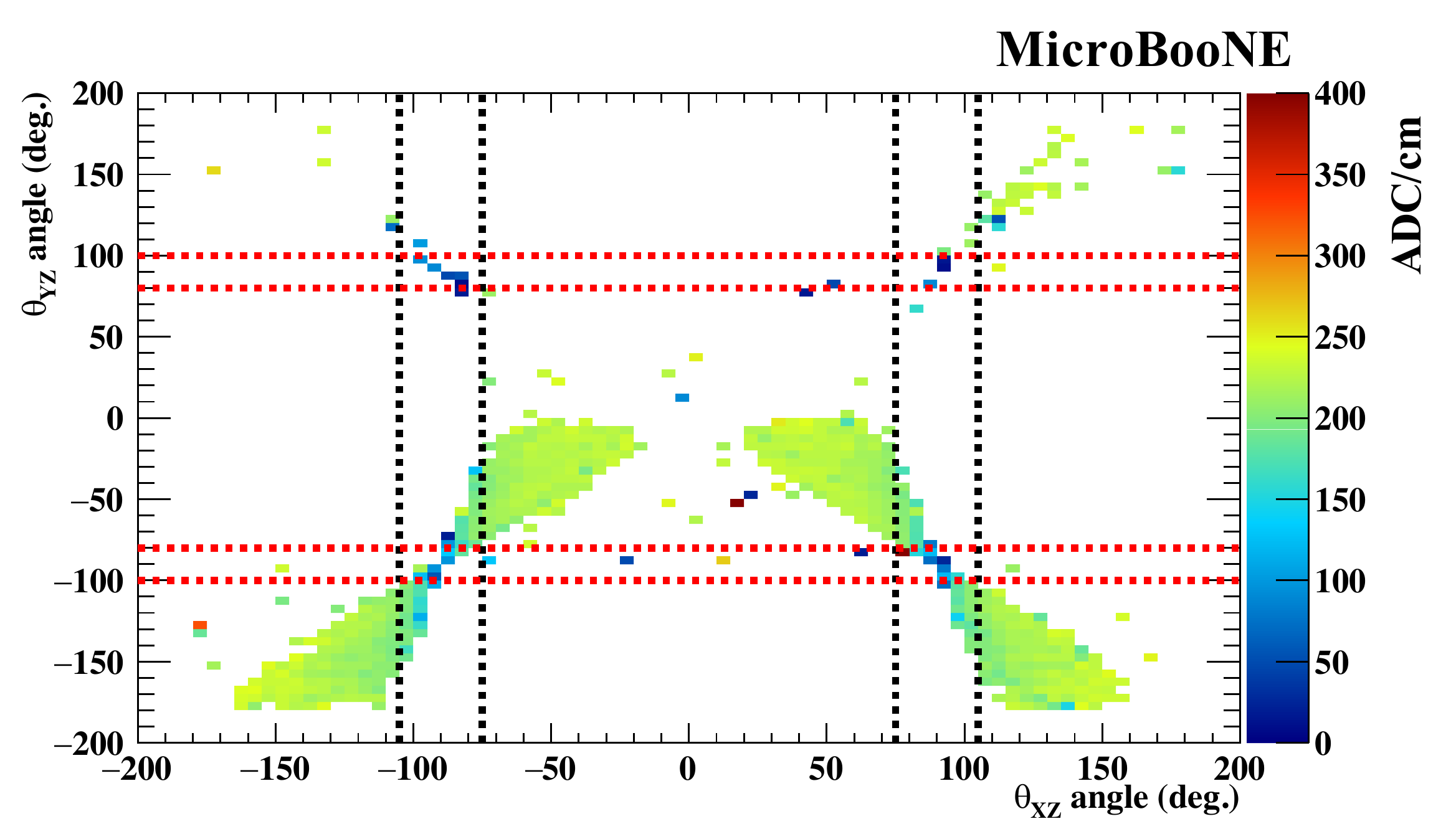}
\caption{Average $dQ/dx$ values in the phase space of $\theta_{XZ}$ and $\theta_{YZ}$ in the collection plane for crossing CR tracks. The color scale represents average $dQ/dx$ for a track which has a given $\theta_{XZ}$ and $\theta_{YZ}$ angular orientation. The units of the Z axis (color scale) are in ADC/cm. Bin size used is 5\textsuperscript{0}$\times$5\textsuperscript{0}. The regions inside the dashed lines show the angles excluded for crossing CR tracks as discussed in section~\ref{sec:sel_cut_rel} for the collection plane. (Left) MC. (Right) Data.}
\label{fig:ang_cut_pl2} 
\end{figure}

The tracks used in the $dQ/dx$ calibration satisfy the following selection criteria:
\begin{itemize}
\item 250 cm $<$ Track projected x length $<$ 270 cm: Any track that satisfies this selection cut is considered to be an anode-cathode crossing CR. A 20 cm wide window is selected in order to account for imperfect reconstruction of track start and end positions.

\item The absolute value of track angle $\theta_{XZ}$ (see figure~\ref{fig:theta_xz_yz_def}) should not be in the range of 75$^{\circ}$ to 105$^{\circ}$: This selection ensures that we are excluding tracks which are nearly orthogonal to the wire planes, which can be mis-reconstructed \cite{uBsignalJINST}. See figure~\ref{fig:ang_cut_pl2}.

\item The absolute value of track angle $\theta_{YZ}$ (see figure~\ref{fig:theta_xz_yz_def}) should not be in the range of 80$^{\circ}$ to 100$^{\circ}$: This selection helps to remove tracks that are nearly parallel to the collection plane wires. See figure~\ref{fig:ang_cut_pl2}. The difference between MC and data in these figures is due to the fact that the MC simulation of detector effects such as space charge effects and the wire field response is not in a perfect agreement with data.
\end{itemize}

\subsection{Analysis method}
\label{sec:ana_met_dqdx}
The $dQ/dx$ calibration of the collection plane response is carried out in three separate steps.
\begin{enumerate}
\item Detector calibration in the yz plane. This step aims to remove the effects of space charge, cross-connected TPC channels, and transverse diffusion.
This step is performed using all crossing tracks occurring over a period of several months.
\item Detector calibration in the drift (x) direction, which may be time-dependent. This step removes effects of electron attachment to impurities, space charge, and longitudinal diffusion. A separate correction is derived for each day of data.
\item Detector calibration in time. This step removes any temporal variations in the overall detector response, and is only performed for data. There is currently no time dependence in the MC simulation.
\end{enumerate}

\subsubsection{Method for detector calibration in the yz plane}
\label{sec:yz_cal}
\begin{figure}[!ht]
\centering
\includegraphics[width=0.49\textwidth]{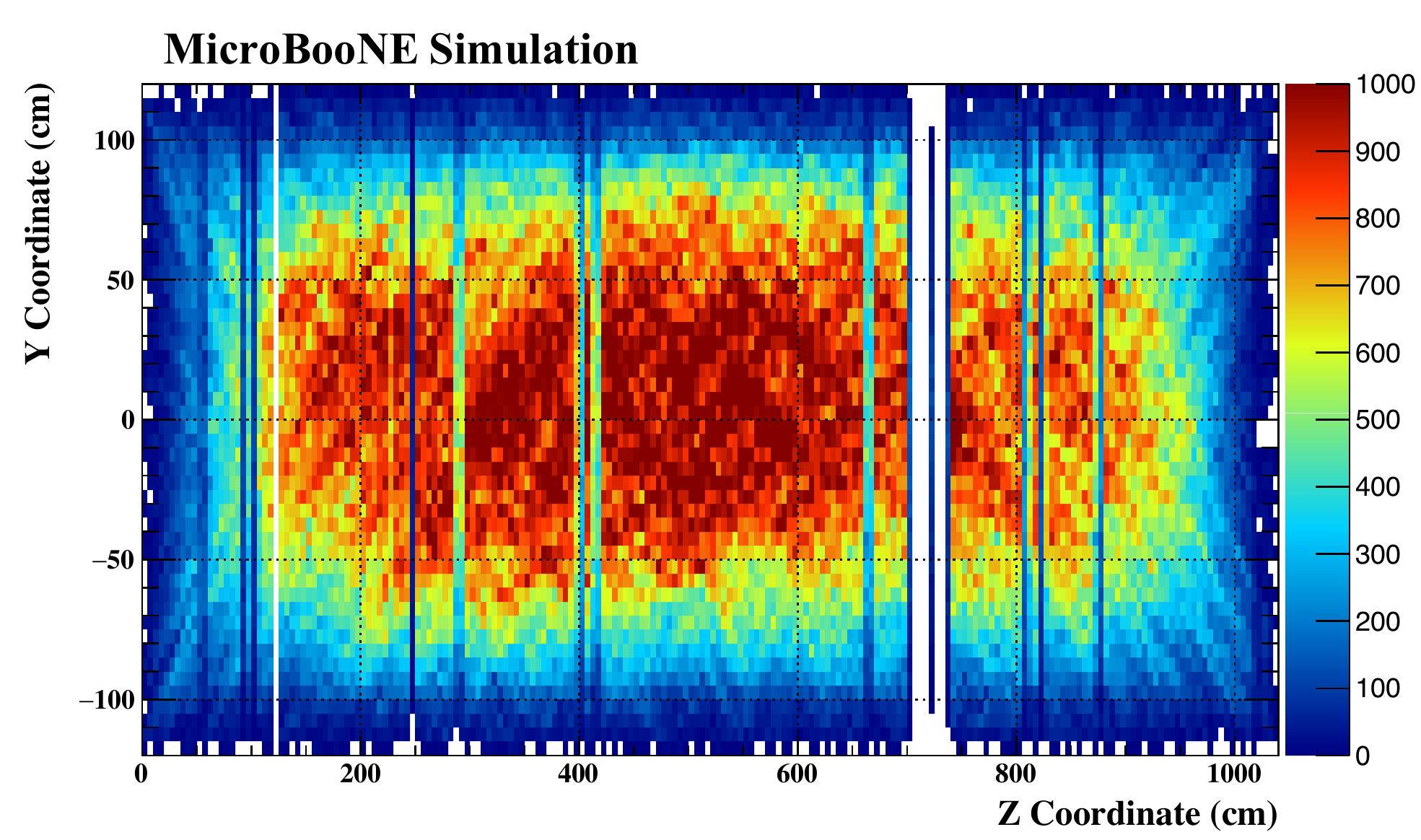}
\includegraphics[width=0.49\textwidth]{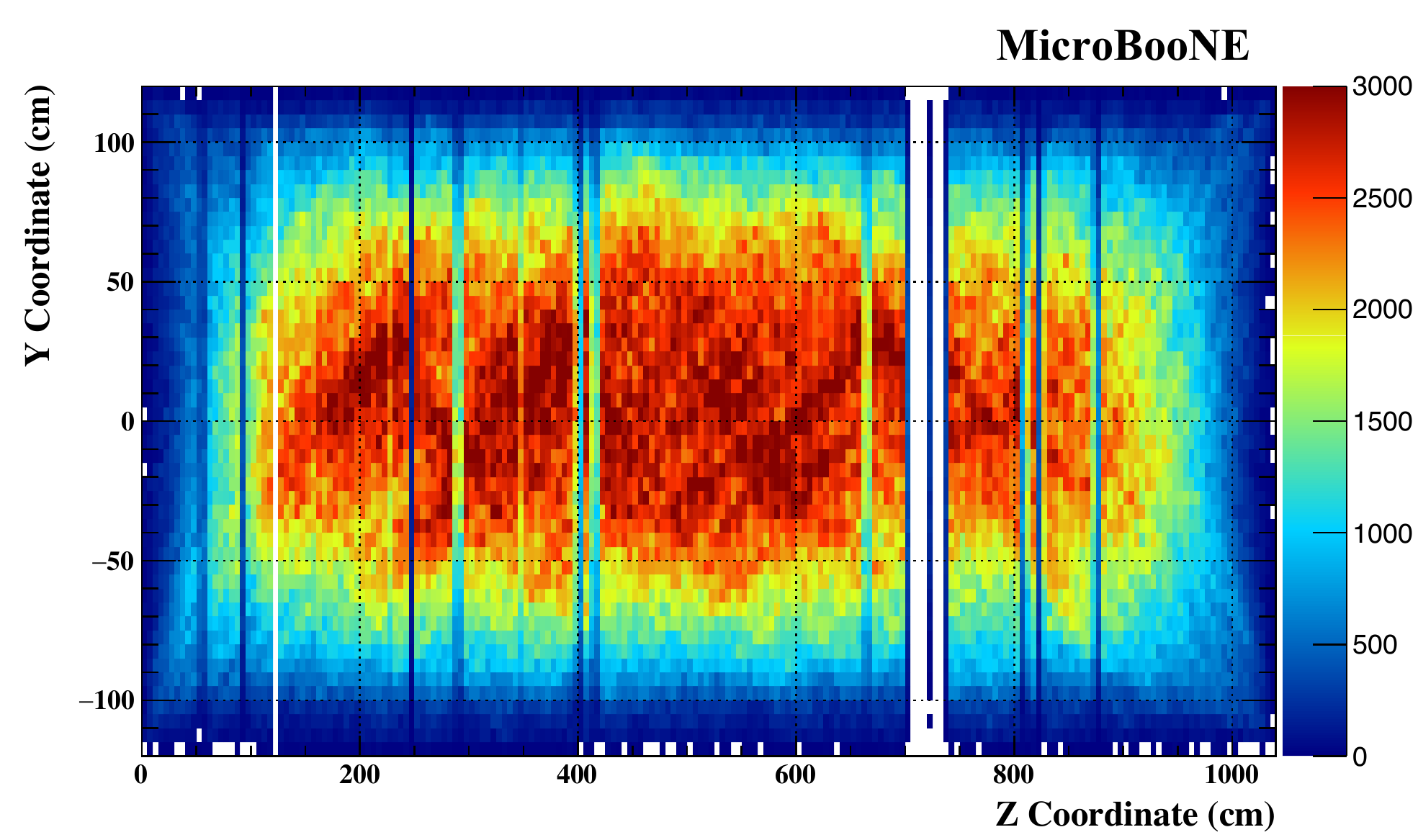} 
\caption{Distributions of number of hits in 5cm by 5cm yz plane cells in the collection plane. Left is for MC and right is for data.}
\label{fig:yz_plane_hits} 
\end{figure}

We segment the yz plane into 5\,cm by 5\,cm cells.
Each three-dimensional (3D) hit of each crossing CR track 
is assigned to a cell based on the $y$ and $z$ coordinates of the hit as
reconstructed by Pandora, without correction for SCE-induced distortion.
For each cell containing more than five hits, we calculate a median 
charge deposit per unit length, $[(dQ/dx)(y_i,z_i)]_\textrm{Local}$, where $y_i$ and $z_i$ are
the central coordinates of each cell. Figure~\ref{fig:yz_plane_hits} shows the distribution of number of hits for each yz plane cell in the collection plane for both MC and data respectively. The statistics is low in the peripheral cells because of the spatial distortion caused by space charge. However, this is not a problem since 1) yz correction factors are still quite uniform near the TPC edges as shown in figures~\ref{fig:mc_yz_x_pl2_co_factors} and~\ref{fig:yz_data_1}; 2) all the analyses will define a fiducial volume to exclude activities in the peripheral region. The use of the median $dQ/dx$ mitigates effects of delta ray contamination, misreconstructed calorimetric information, and long tails of the $dQ/dx$ distribution.
We also calculate a global median charge deposit per unit length, $[dQ/dx]_\textrm{Global}$ using
all 3D hits of all crossing CR tracks. The correction factor for each cell is
defined as
\begin{equation}
\label{eqn:cor_fact_yz}
C(y_{i},z_{i}) = \frac{[dQ/dx]_\textrm{Global}}{[(dQ/dx)(y_i,z_i)]_\textrm{Local}}\,.
\end{equation}
If a cell has five or fewer space points, the correction factor for that cell is calculated from the $C(y_i,z_i)$ values of nearby cells.

\subsubsection{Method for detector calibration in the drift direction}
\label{sec:x_cal}
We segment the drift direction into bins of 10\,cm in data and bins of 5\,cm in MC.
These bin sizes are chosen to minimize statistical fluctuations in the correction factors.
Each 3D hit of each crossing CR track on a given day
is assigned to a bin based on the reconstructed $x$ coordinate of the hit,
with $x$ position corrected only for the $t_0$ of the CR as described in section
\ref{sec:sel_cut_rel}, and charge deposit corrected only by the $C(y_i,z_i)$ factor.
For each bin containing more than five hits, we calculate a median 
charge deposit per unit length, $[(dQ/dx)'(x_i)]_\textrm{Local}$, where $x_i$ is the
central position of the bin, and the prime on $(dQ/dx)'$ denotes that the yz correction has been applied.
The correction factor for each bin in the drift direction is defined as
\begin{equation}
\label{eqn:cor_fact_x}
C(x_{i}) = \frac{[(dQ/dx)']_\textrm{Global}}{[(dQ/dx)'(x_i)]_\textrm{Local}}\,,
\end{equation}
where $[(dQ/dx)']_\textrm{Global}$ is the global median $dQ/dx$ value after applying the yz correction to each 3D hit.

In data, we derive separate drift direction correction factors for each day as space charge effects and electron lifetime can change over time. To get a reliable set of correction factors, we need a data sample which contains enough statistics. For this purpose, we derive correction factors in the drift direction only for the days where there are more than 40 crossing tracks after all the angular cuts. In the case of lack of statistics for a given day, we use the drift direction correction factors derived for the immediate neighboring day. For MC we derive only a single set of drift direction correction factors for the collection plane as there is no time variation of space charge effects and the electron lifetime is set to infinity. 

\subsubsection{Time dependent calibration of the detector}
\label{sec:t_cal}
\begin{figure}[!ht]
\centering
\includegraphics[width=0.49\textwidth]{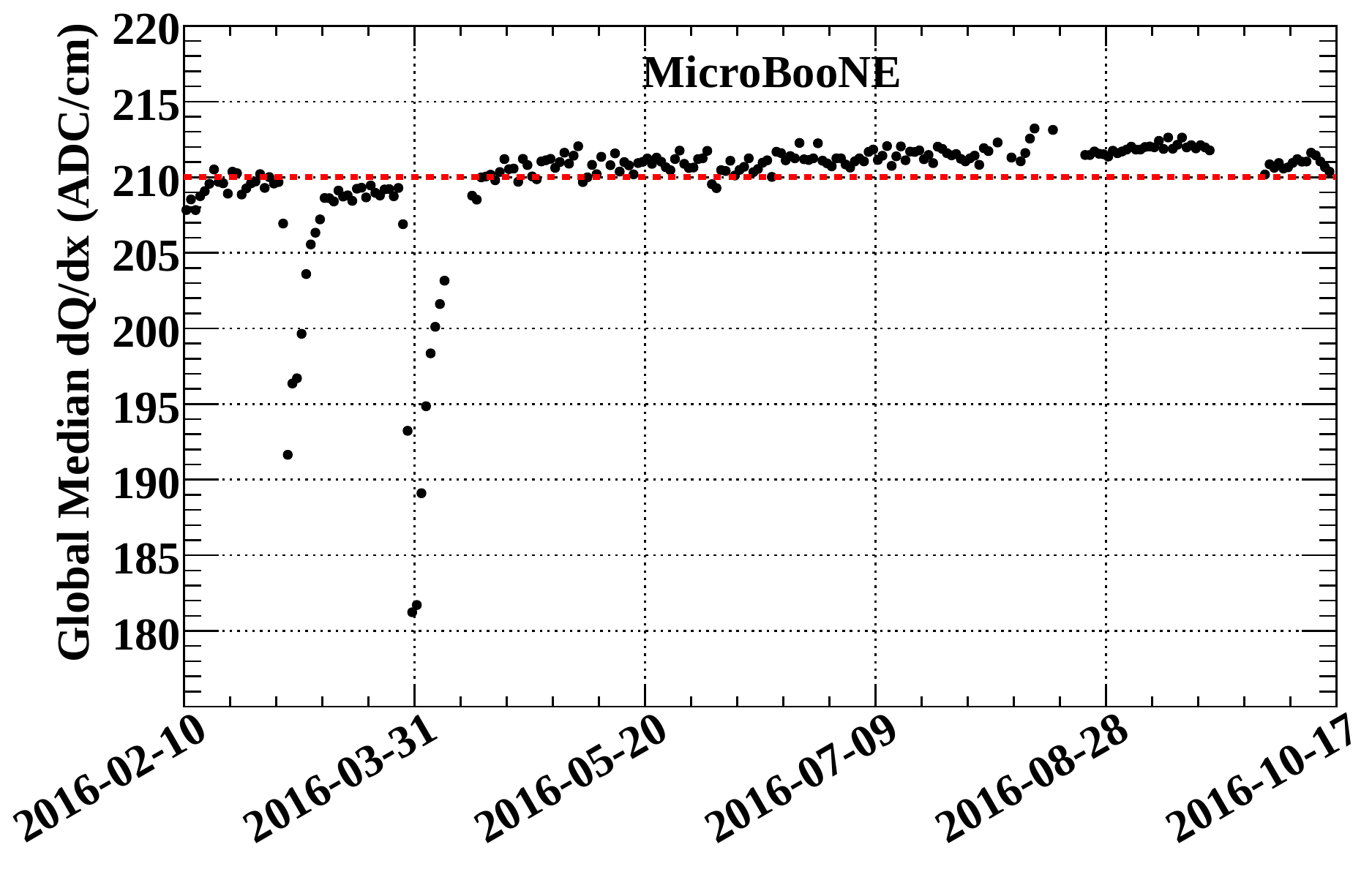}
\includegraphics[width=0.49\textwidth]{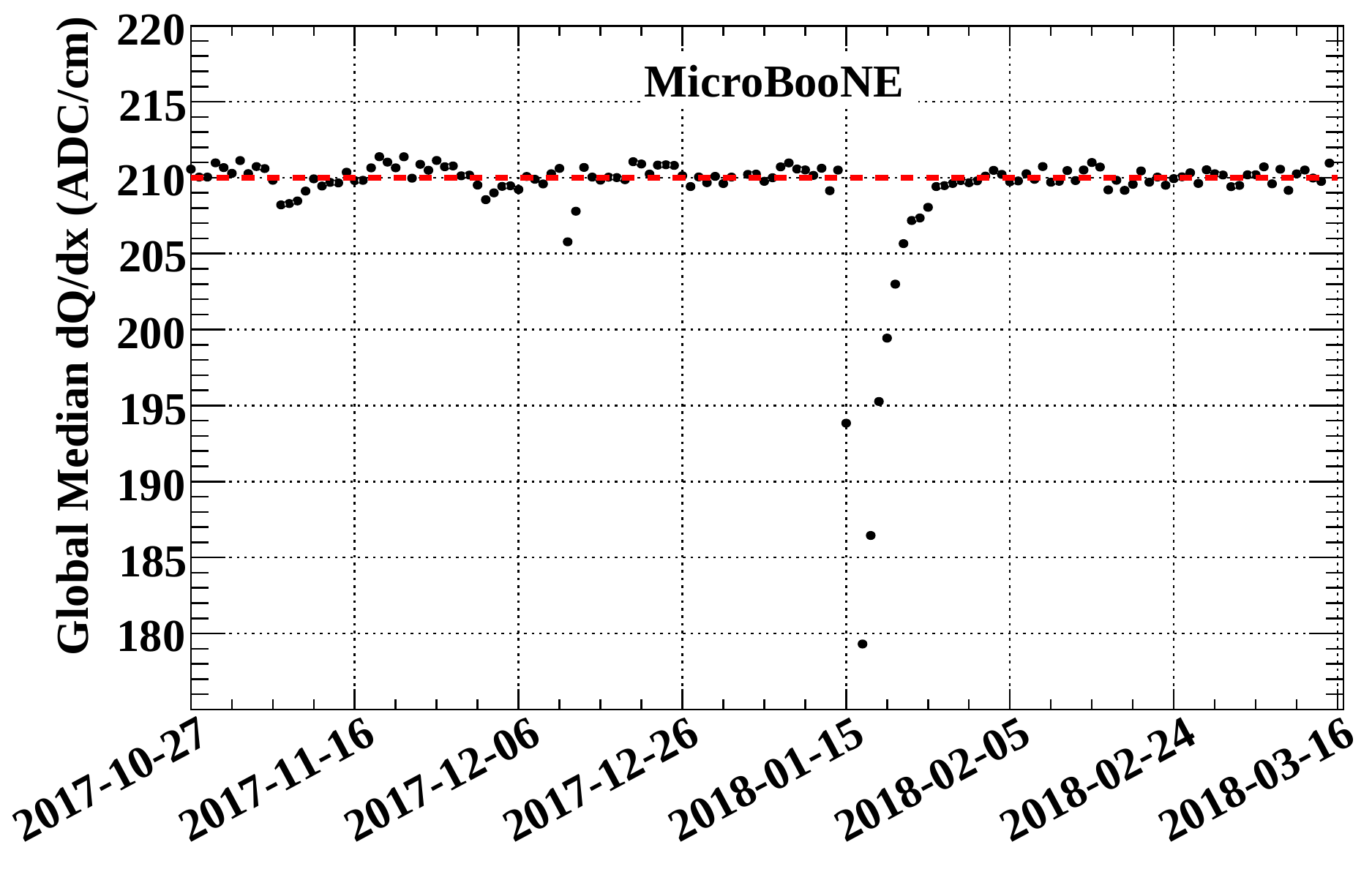}
\caption{The global median $dQ/dx$ per day over time in the collection plane
after correcting for yz plane and drift direction irregularities for each day.
The red dashed lines show the reference value. Left: the global median $dQ/dx$ from February 10, 2016 to October 17, 2016. 
Right: the global median $dQ/dx$ from October 27, 2017 to March 16, 2018. 
 Dips from the reference values across several days are due to time periods in which argon purity was low (e.g.\ during recirculation pump maintenance). 
}
\label{fig:time_corre_gbl_mdn}
\end{figure}

After applying the yz and drift direction factors to each 3D hit, the 
daily global median value of $dQ/dx$ of all hits on crossing CR tracks varies over
time, as shown in figure~\ref{fig:time_corre_gbl_mdn}. 
Dips lasting several days reflect times when the argon purity was low, e.g., during recirculation pump maintenance.
We choose a reference value 
$[dQ/dx]_\textrm{Ref} \equiv 210\textrm{\,ADC/cm}$ for the global median $dQ/dx$ in the collection plane, since all global median $dQ/dx$ values calculated for each day fluctuate around this value after yz- and drift-dependent correction. We then define the overall time dependent correction as
\begin{equation}
\label{eqn:time_cor}
C(t) = \frac{[dQ/dx]_\textrm{Ref}}{[(dQ/dx)''(t)]_\textrm{Global}}~,
\end{equation}
where $[dQ/dx]_\textrm{Ref}$ is the reference $dQ/dx$ value in the collection plane, and $[(dQ/dx)''(t)]_\textrm{Global}$ is the global median $dQ/dx$ value in the collection plane after correcting for yz plane and drift direction irregularities for each day. 

Values of $dQ/dx$ fully calibrated to the reference value are obtained by multiplying the reconstructed values by the factor $C(y_i,z_i) C(x_i) C(t)$. For MC simulated data without temporal variation we define $C(t)=1$.

\subsection{Results}
\subsubsection{dQ/dx calibration (Simulation)}
\label{sec:res_rel_cal_mc}
\begin{figure}[!ht]
\centering
\includegraphics[width=0.49\textwidth]{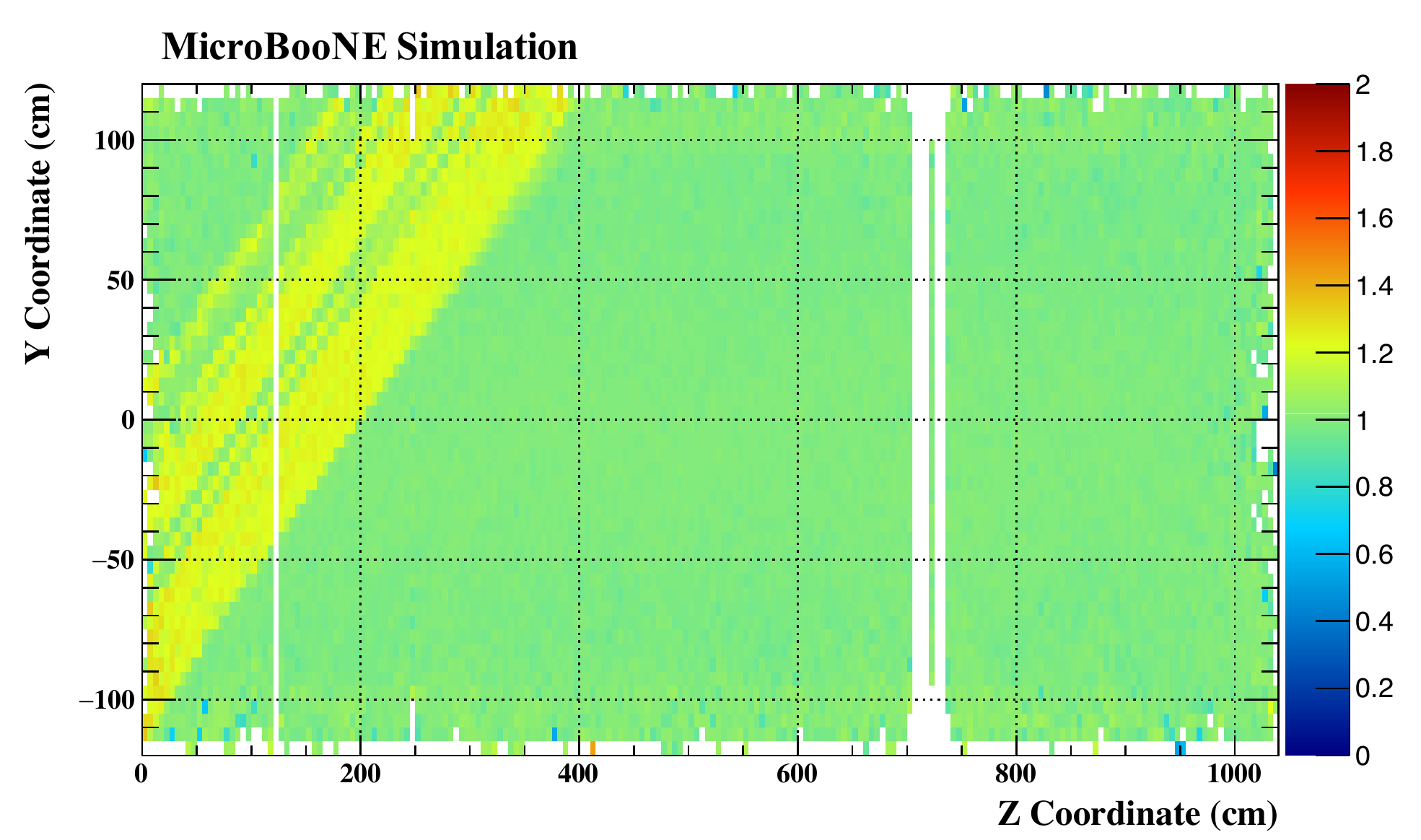} 
\includegraphics[width=0.49\textwidth]{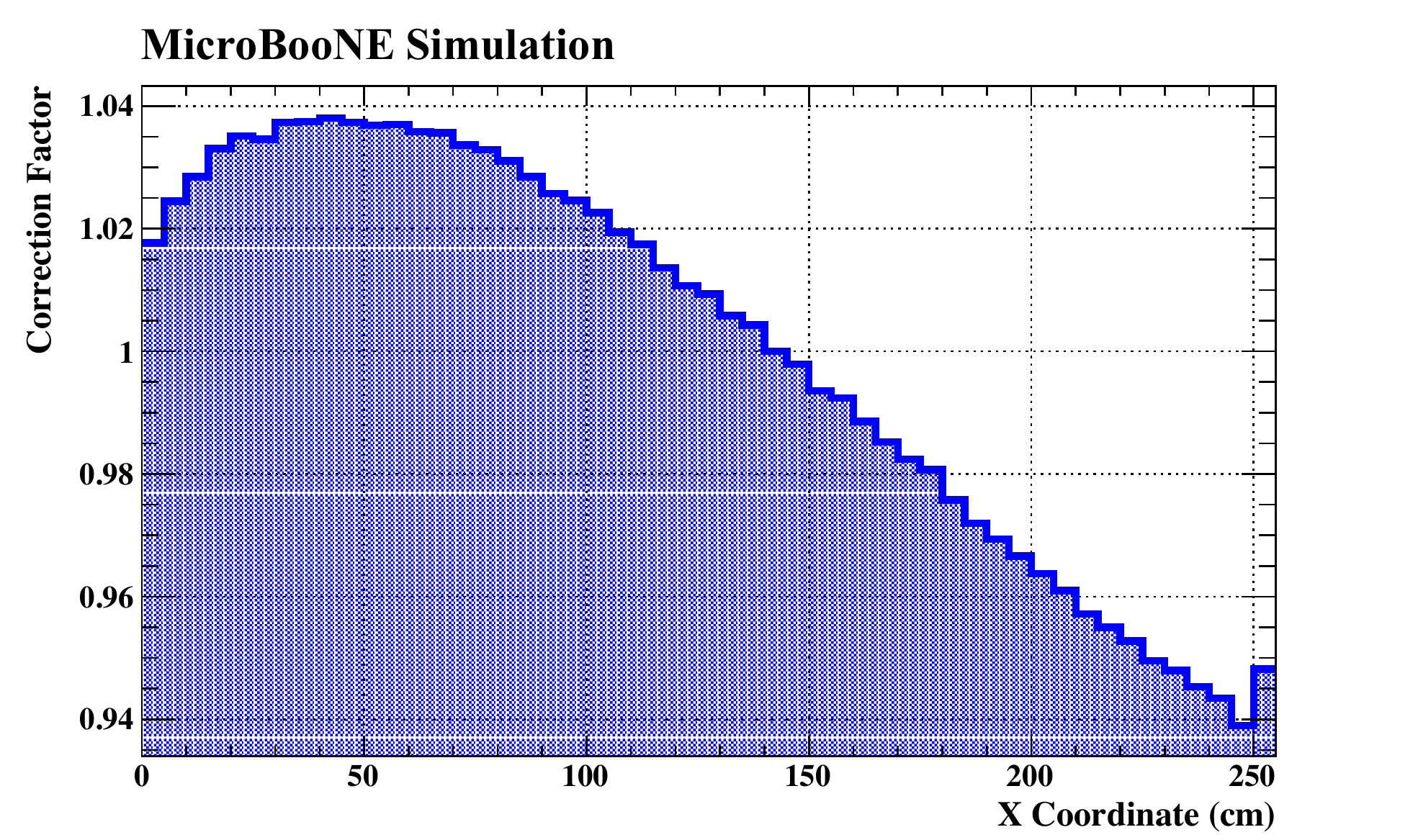} 
\caption{Left: yz correction factors in the collection plane in MC. Here the Z axis color represents correction factors for a given 5 cm $\times$ 5 cm cell in the yz plane. 
Right: Drift direction correction factors in MC simulation. }
\label{fig:mc_yz_x_pl2_co_factors} 
\end{figure}

The $dQ/dx$ calibration of the detector in MC is performed using a single large dataset described in section~\ref{sec:data_sample_dqdx}. Figure~\ref{fig:mc_yz_x_pl2_co_factors} shows the extracted yz correction factors $C(y,z)$ and drift direction x correction factors $C(x)$. The variation of the correction factors with respect to the drift distance could be attributed to electron attachment to impurities as well as space charge effects as shown in Figure~\ref{fig:low_e_life}. The bump in the last bin of the drift direction correction distribution is due to low statistics in that bin.

\subsubsection{dQ/dx calibration (data)}
\label{sec:res_rel_cal_dat}
\begin{figure}[!ht]

\center

\includegraphics[width=0.47\textwidth]{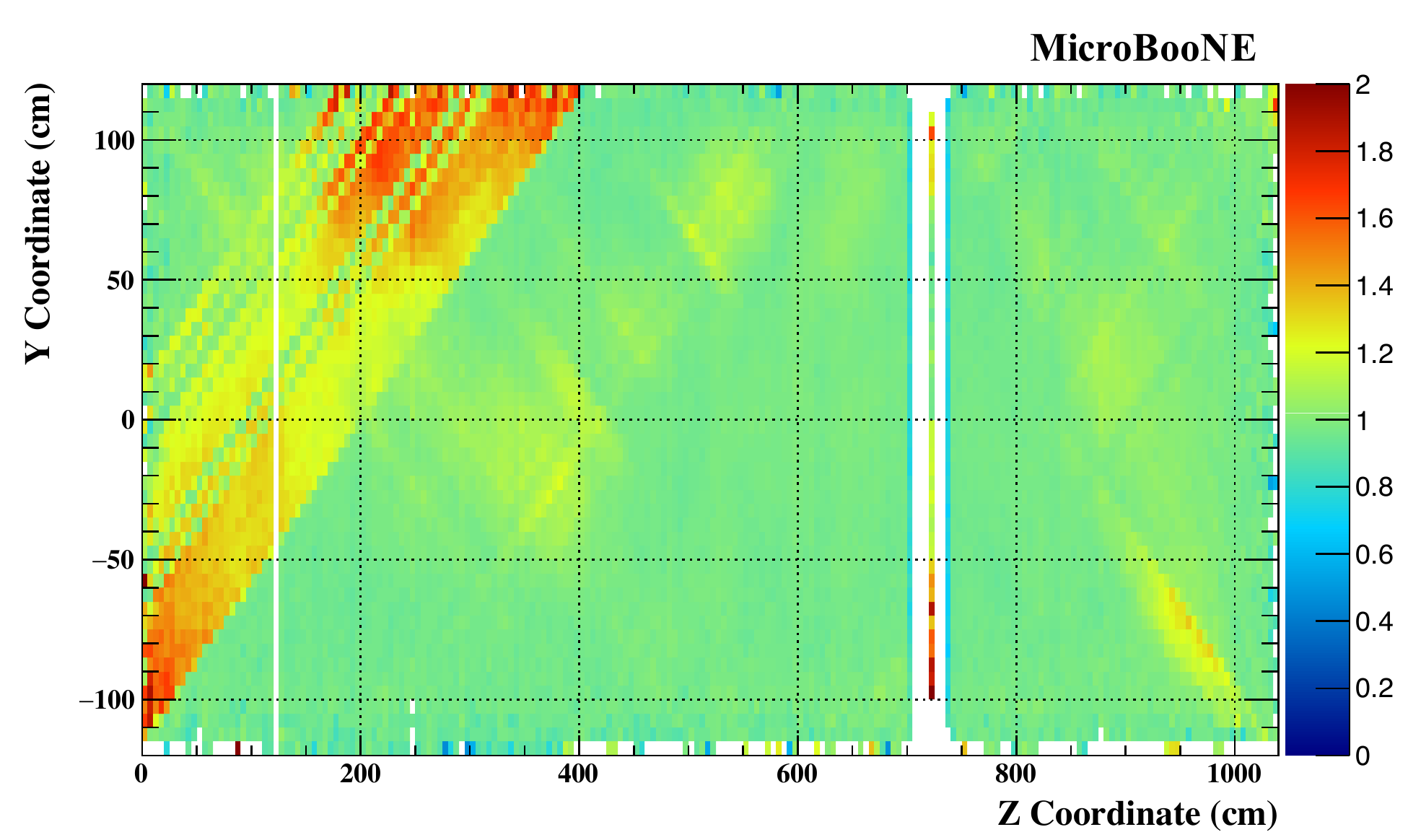} 
\includegraphics[width=0.52\textwidth]{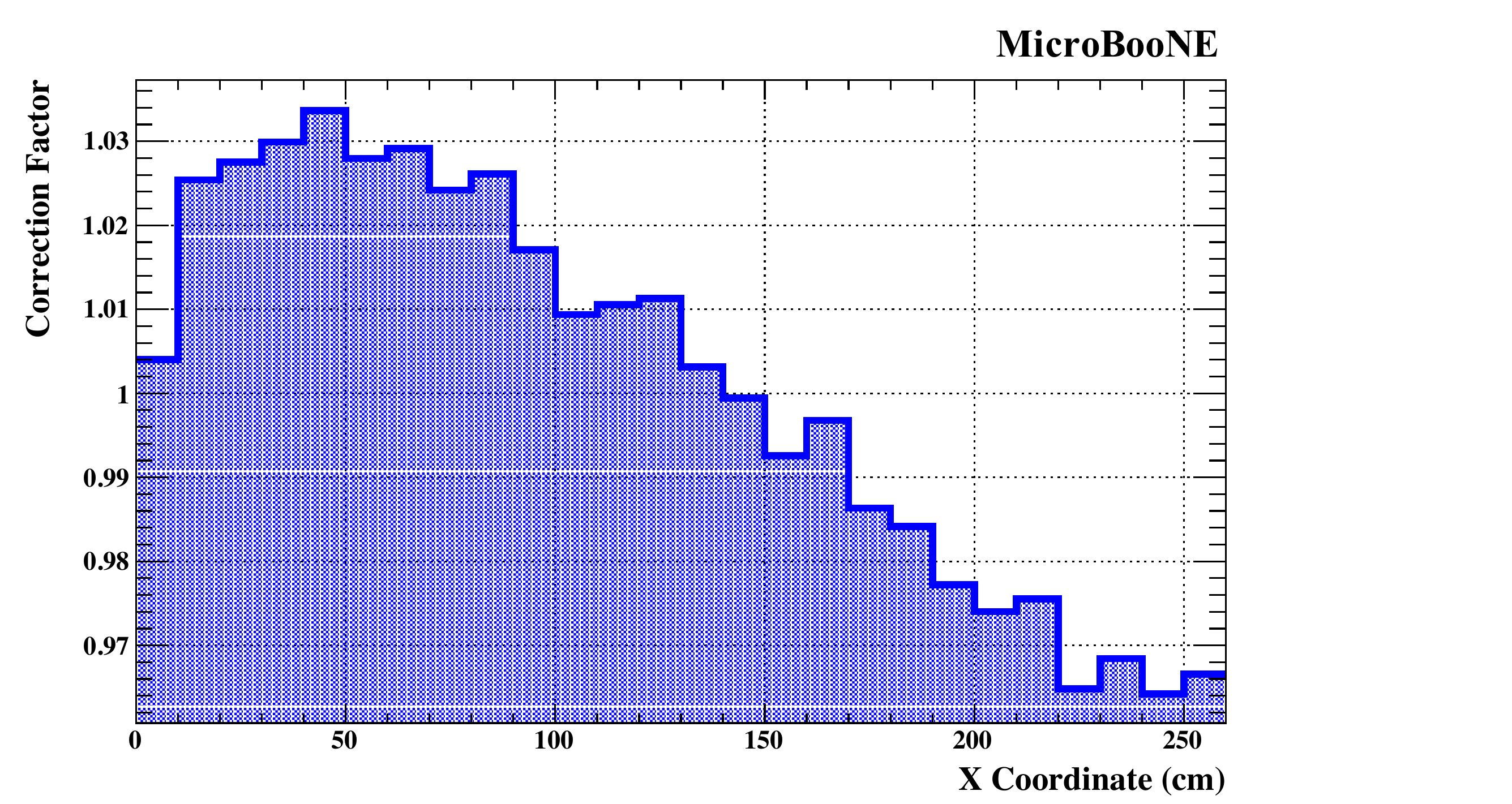} 
\caption{Left: yz correction factors for data taken from February to October in 2016. Here the Z axis color represents correction factors for a given 5 cm $\times$ 5 cm cell in the yz plane. Right: A set of drift direction x correction factors for data taken on February 25, 2016.
}
\label{fig:yz_data_1} 
\end{figure}

The yz and drift direction x correction factors are derived for two separate datasets as described in section~\ref{sec:data_sample_dqdx}.
In the first dataset, we have data combined from February 2016 to October 2016; and in the second dataset, we have data combined from September 2017 to March 2018. See figure~\ref{fig:yz_data_1} 
 for the variation of the yz correction factors for data taken from February to October in 2016 and drift direction x correction factors for data taken on February 25, 2016 in the collection plane. The distributions for data taken from September 2017 to March 2018 are similar.

%
%

\begin{figure}[!ht]
\center
\includegraphics[width=0.49\textwidth]{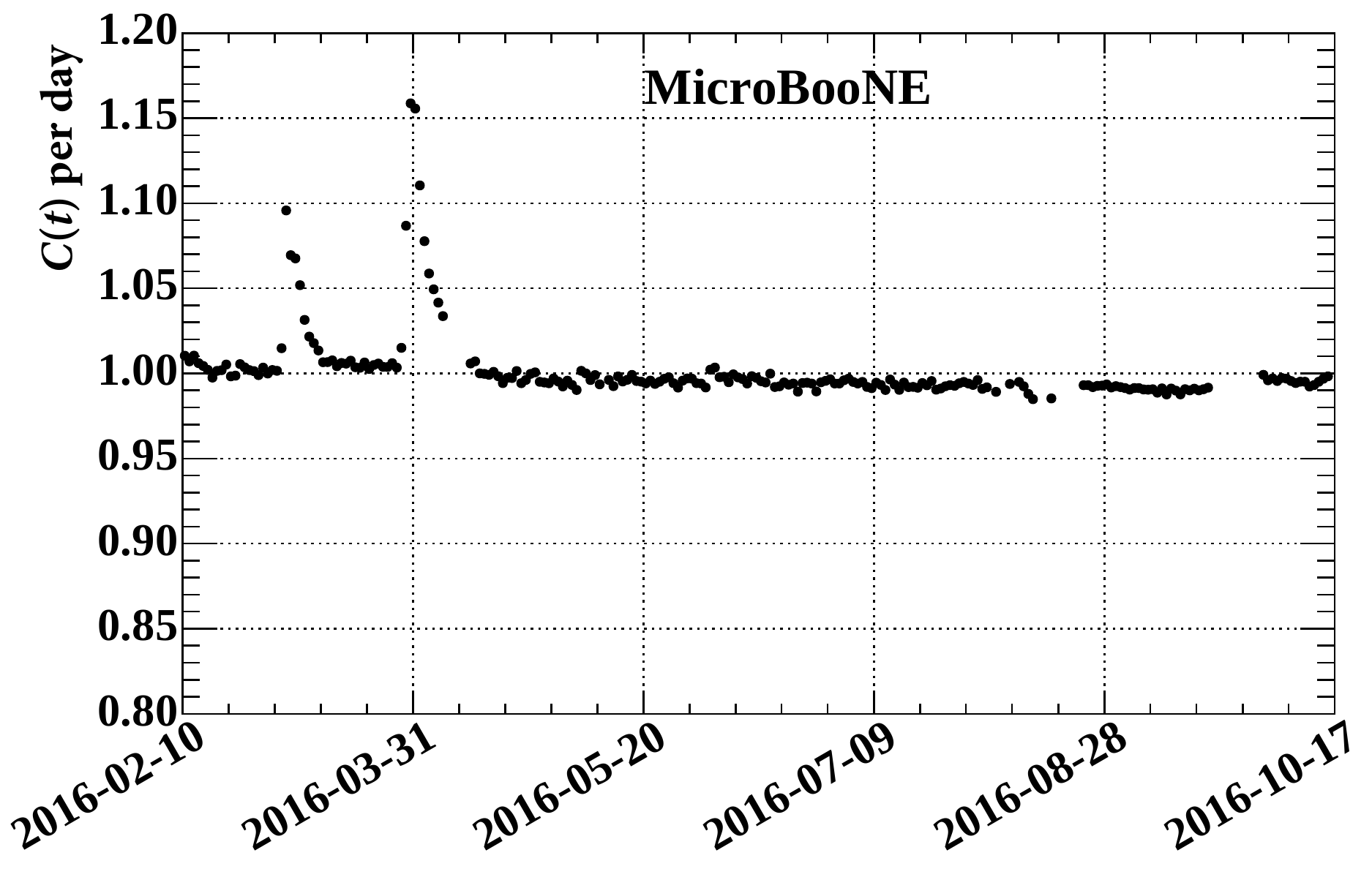}
\includegraphics[width=0.49\textwidth]{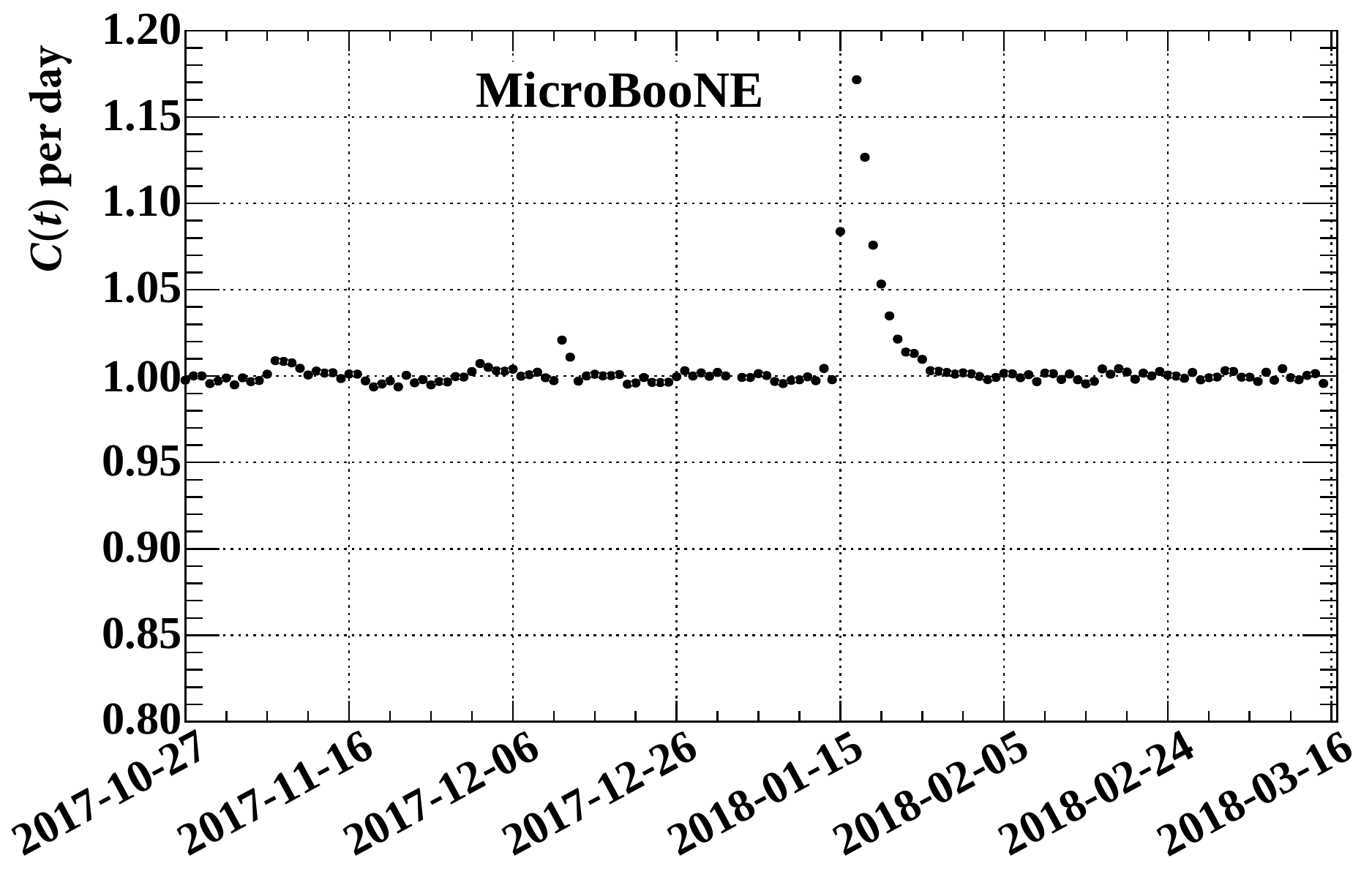}
\caption{Left: The time correction $C(t)$ shown for data taken from February 2016 to October 2016. Right: The time correction $C(t)$ shown for data taken from September 2017 to March 2018.}
\label{fig:time_cor_data} 
\end{figure}

\begin{figure}[!ht]
\center
\includegraphics[width=0.49\textwidth]{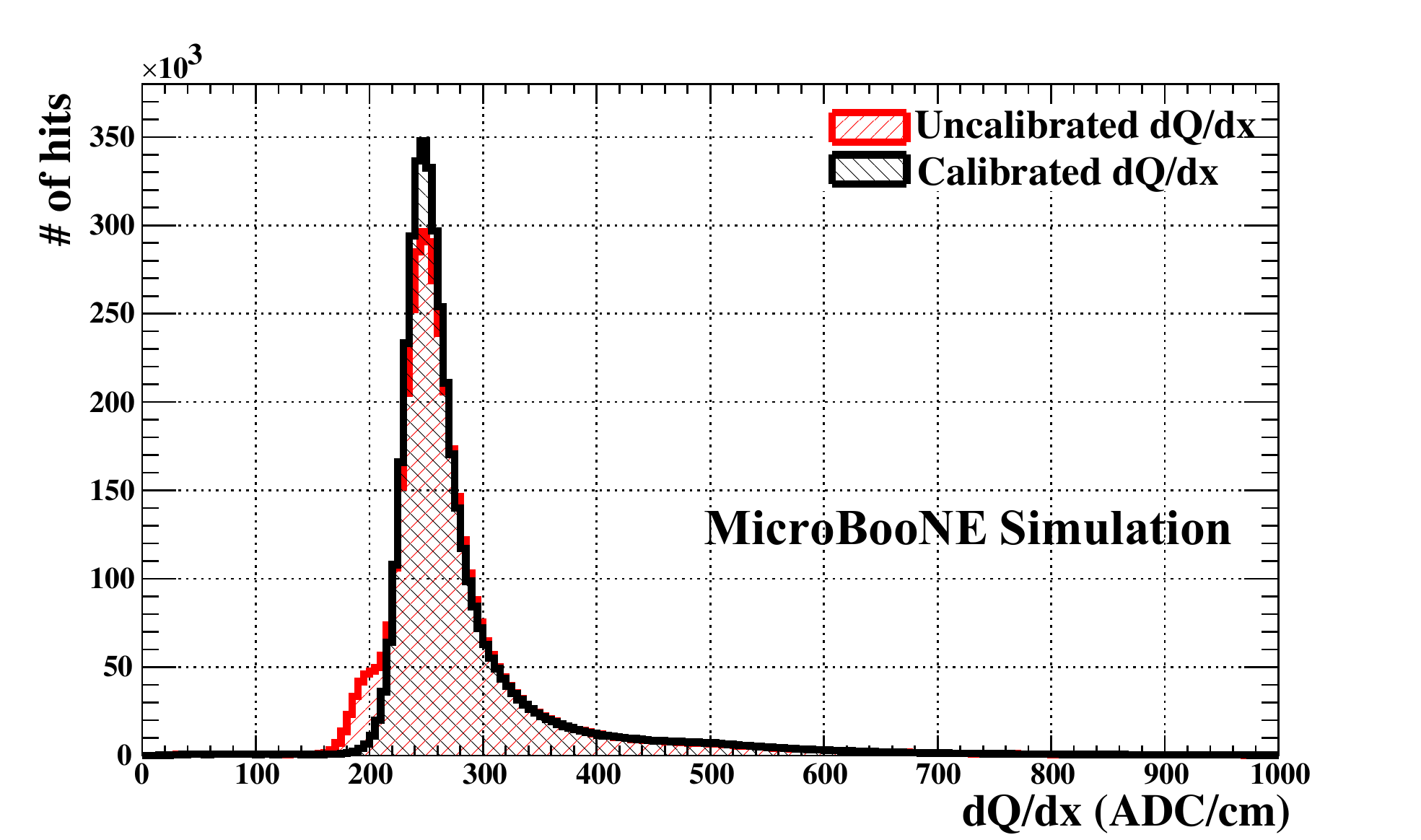}
\includegraphics[width=0.49\textwidth]{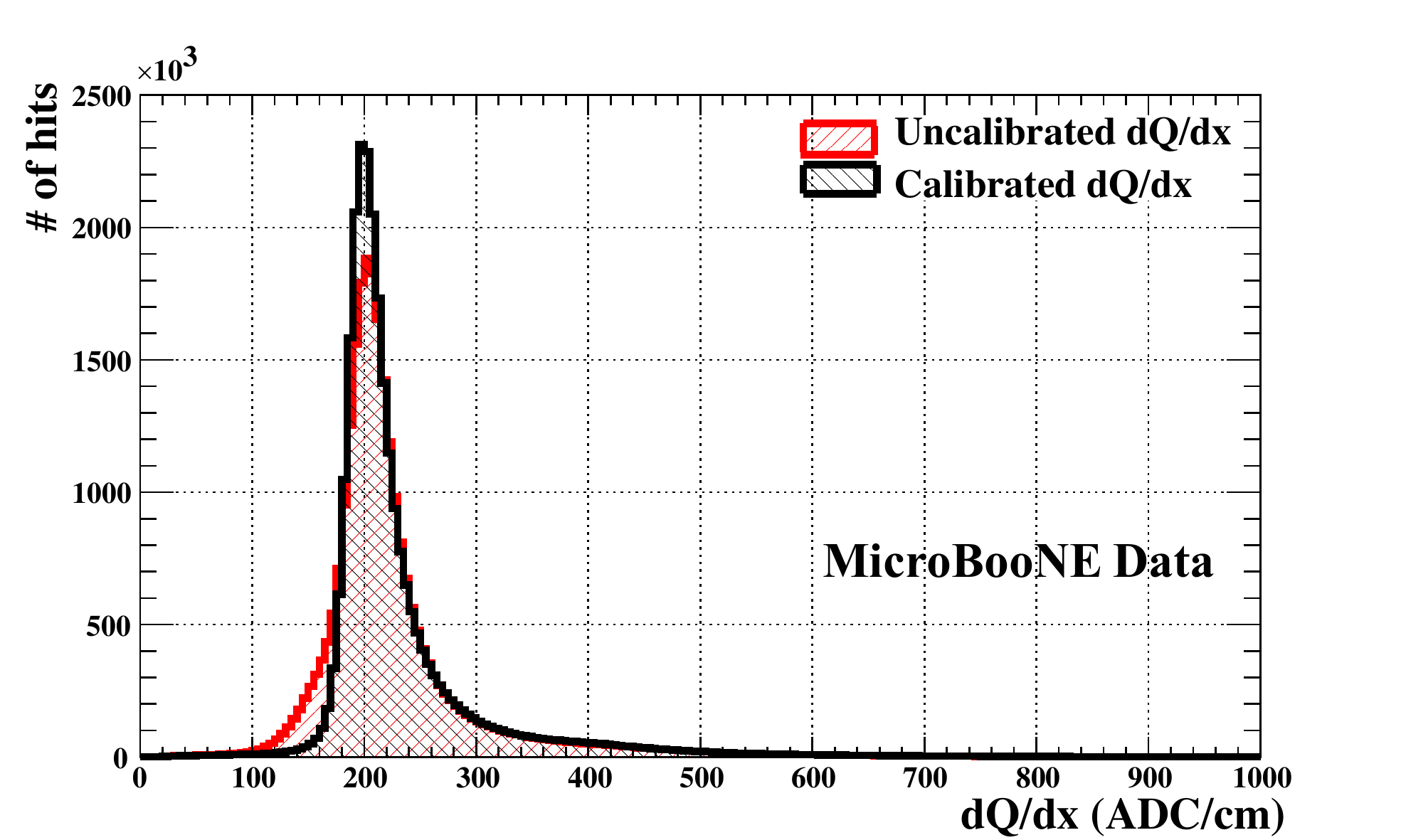}
\caption{Left: Calibrated and uncalibrated $dQ/dx$ in simulation using crossing CR in the collection plane. Right: Calibrated and uncalibrated $dQ/dx$ in data using crossing CR in the collection plane. In these plots the Y-axis label \# of hits refers to the  number of 3D hits of tracks.}
\label{fig:dq_dx_com_mc_data_cross_mu}
\end{figure}

The drift direction corrections factors and time corrections are derived for the time period from February 2016 to October 2016 and the time period from September 2017 to March 2018. $C(t)$ is calculated for each day. Due to intermittent periods of detector downtime, some days are skipped in this measurement. See Figure~\ref{fig:time_cor_data} for the variation of the time-dependent correction factor over time. Figure~\ref{fig:dq_dx_com_mc_data_cross_mu} shows the impact of the $dQ/dx$ calibration on the $dQ/dx$ distribution in both MC and data. 
In both data and MC, we see that the width of the central peak of
the $dQ/dx$ distribution has narrowed by 2\%, and the significant
excess of low $dQ/dx$ values in the data before calibration has
been corrected.  This represents a significant improvement in the
measurement of $dE/dx$, particularly in regions requiring the
largest yz corrections, which is important for the measurement of
total deposited energy and particle identification based on
energy loss per unit length as a function of residual range.

\subsection{Angular dependence study}
The calibration procedure described in this paper does not make any angular-dependent corrections because of the limited angular coverage of anode-cathode crossing CR. The angular distributions of the anode-cathode crossing CR tracks are shown in figure~\ref{fig:ang_cut_pl2}. In order to more closely study the angular dependence of the $dQ/dx$ calibration, we divide the crossing CR into four angular subsamples:
\begin{itemize}
\item $0^{\circ}<|\theta_{XZ}|<50^{\circ}$,
\item $50^{\circ}<|\theta_{XZ}|<75^{\circ}$,
\item $105^{\circ}<|\theta_{XZ}|<130^{\circ}$ and
\item $130^{\circ}<|\theta_{XZ}|<180^{\circ}$.
\end{itemize}
and derive the yz and x correction factors for each subsample following the same procedure described above. All of the other selection criteria remain the same.

\begin{figure}
\centering
\subfloat[Collection plane, MC]{
\includegraphics[width=0.49\textwidth]{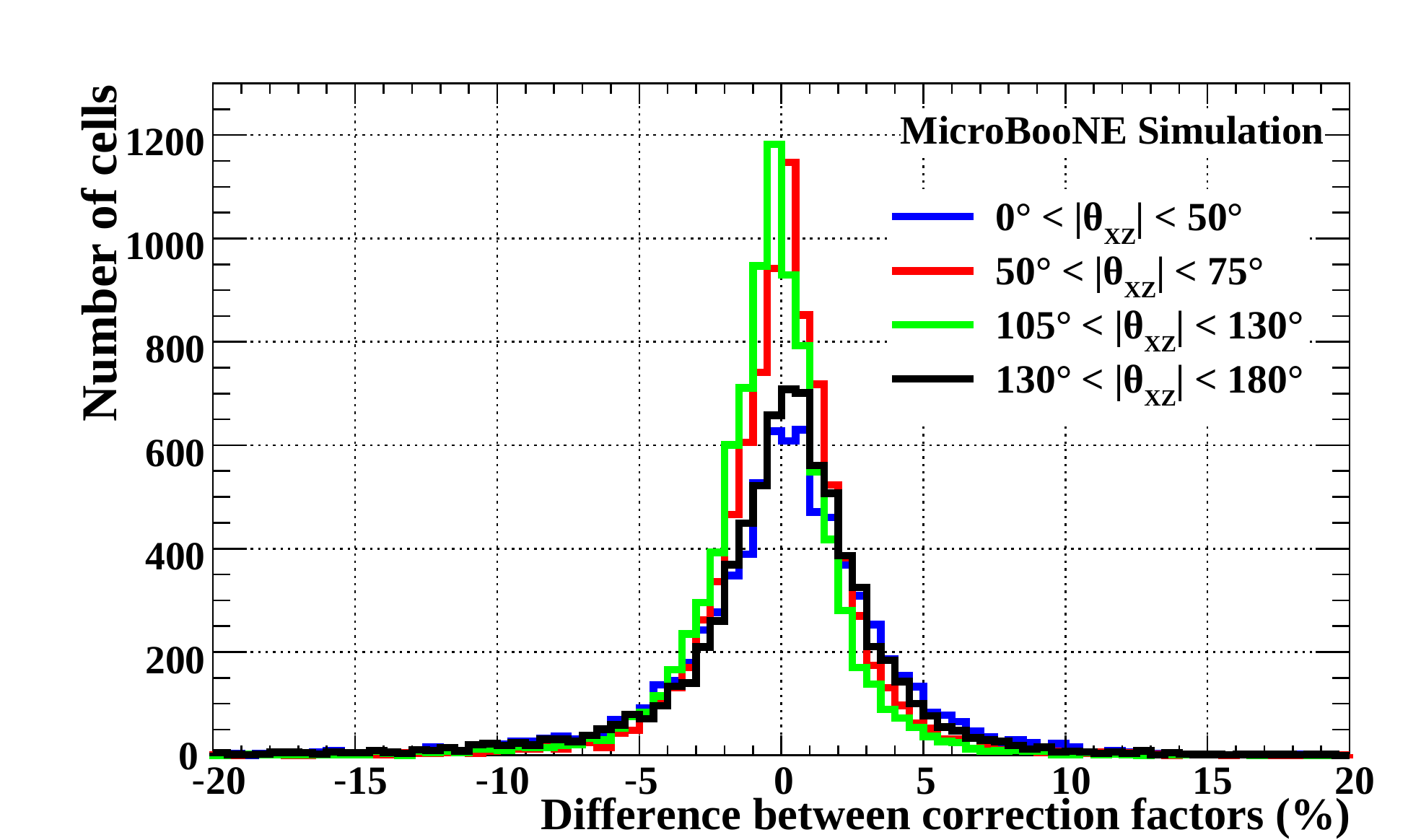}
\label{fig:p2mcyz}
}
\subfloat[Collection plane, data]{
  \includegraphics[width=0.49\textwidth]{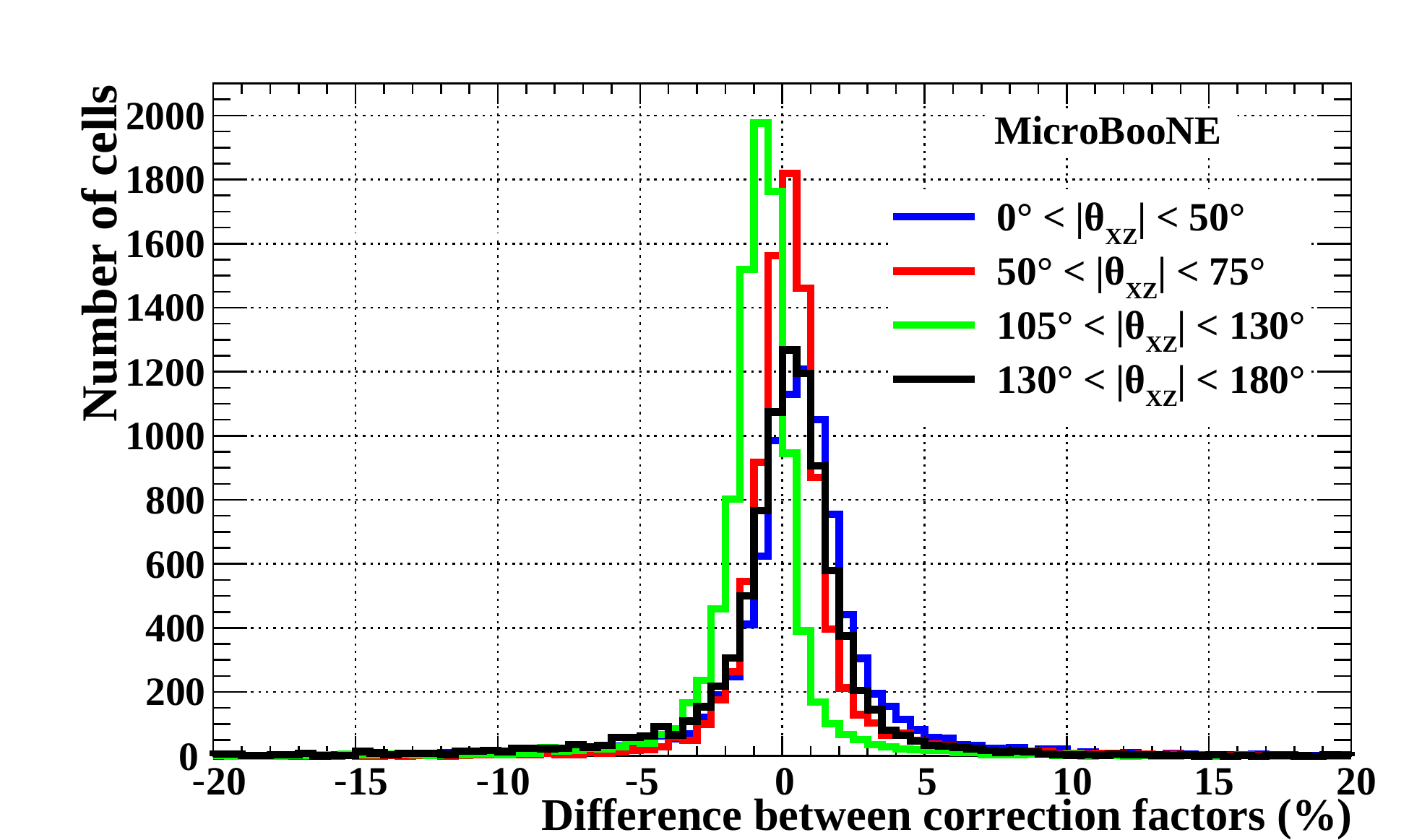}
  \label{fig:p2datayz}
}
\caption{Relative differences of the yz correction factors for each angular subsample with respect to the complete sample, as defined in text.}
\label{fig:yzsys}
\end{figure}

For each yz cell in each subsample, we compute the relative difference of the yz correction factor for the subsample $C_\text{subsample}(y_i,z_i)$ from the correction factor found using all crossing CR $C(y_i,z_i)$.
Figure~\ref{fig:yzsys} shows the distribution of the relative differences of the yz correction factors relative to the complete sample, $\left[C_\text{subsample}(y_i, z_i) \allowbreak - C(y_i,z_i)\right] / C(y_i,z_i)$,
for each angular subsample.
The shift in the mean of the distribution away from zero is indicative of the amount of bias observed in a given angular bin. 
The peaks of the four distributions all agree within 2\%.

\begin{figure}
\centering
\subfloat[Collection plane, MC]{
 \includegraphics[width=0.49\textwidth]{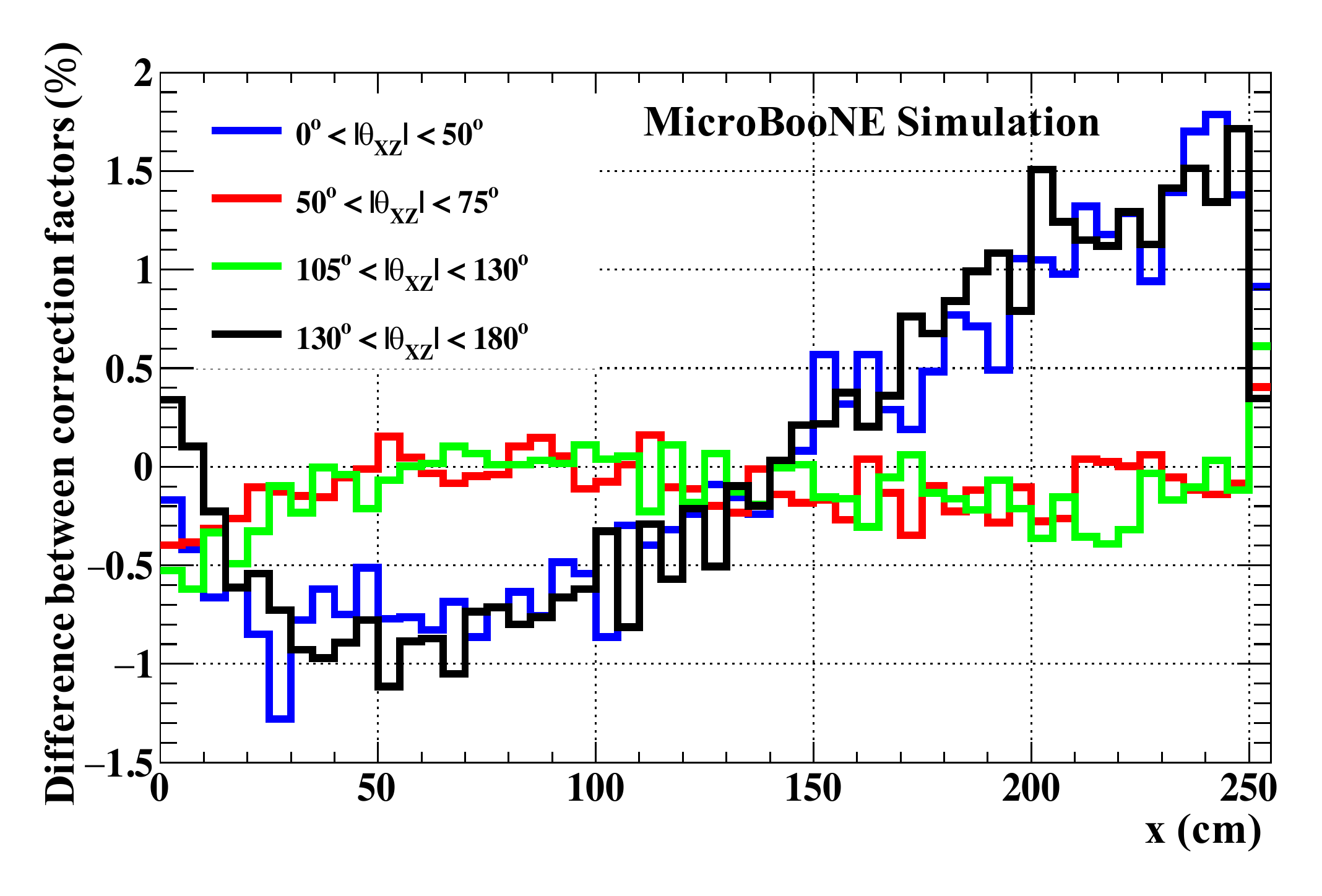}
  \label{fig:p2mcx} 
}
\subfloat[Collection plane, data]{
 \includegraphics[width=0.49\textwidth]{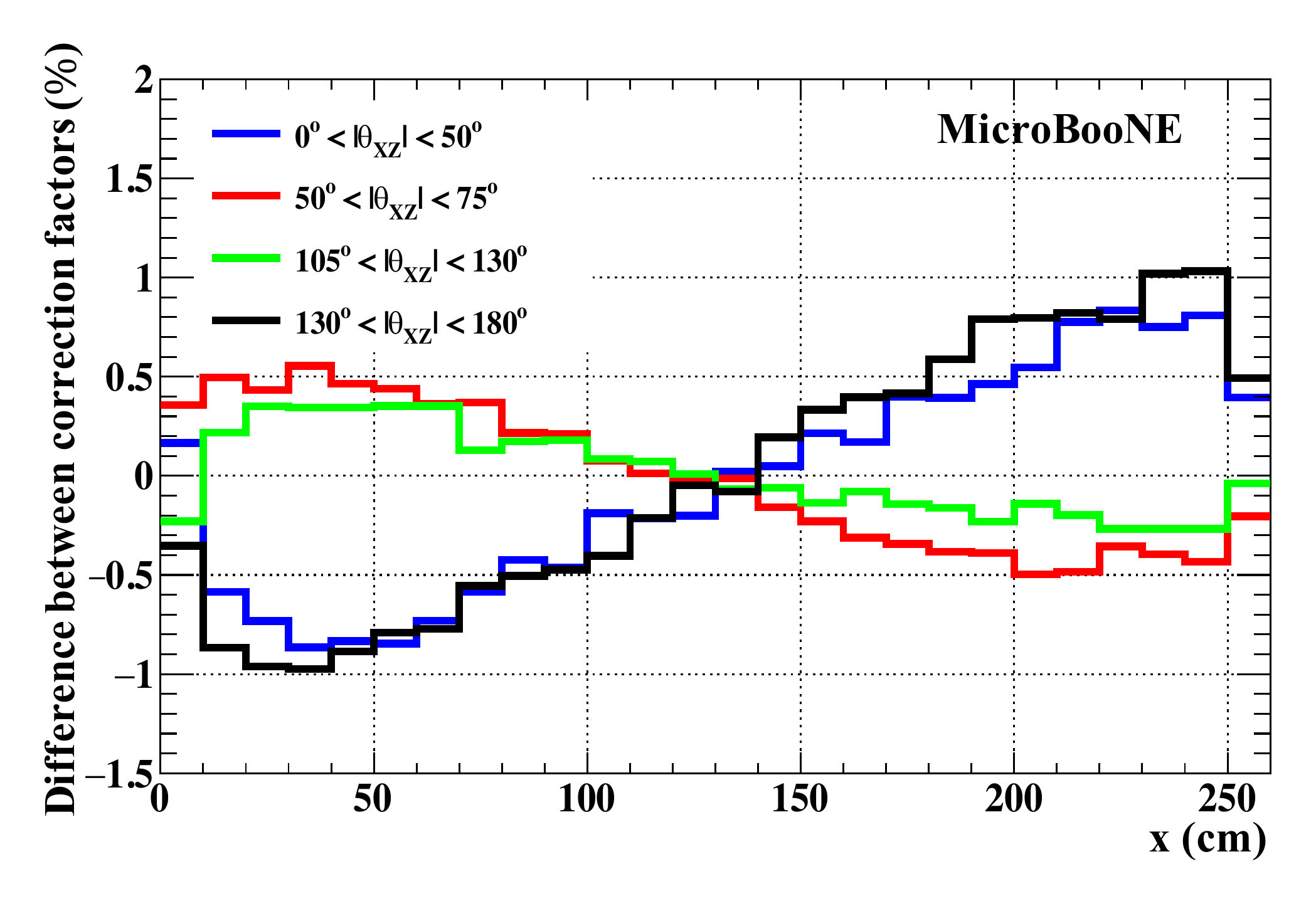}
  \label{fig:p2datax} 
}
\caption{Relative differences in x correction factors between different angular subsamples and the combined sample. The differences between simulation (left) and data (right) could be due to a mis-modeling of space charge effects or wire response in our simulation.}
\label{fig:xsys}
\end{figure}

Figure~\ref{fig:xsys} shows the relative differences in the x correction factors between the different angular subsamples and the combined sample. The maximum variation is $1.5\%$. 

Based on the above studies, we assign a $2\%$ systematic error on the yz correction and a $1.5\%$ systematic error on the x correction for the collection plane. This study shows very little angular dependence for the sample selected for the $dQ/dx$ calibration. However, because of the limited angular coverage of this sample, more studies are underway to further refine the angular dependence of the detector response.

\section{dE/dx Calibration}
\label{sec:dedx}
\subsection{Introduction}
After the $dQ/dx$ correction is complete we can determine the absolute energy loss per unit length scale. To move from the calibrated $dQ/dx$ to $dE/dx$, we use the modified box model~\cite{recomb} for recombination:

\begin{equation}
\label{eqn:de_dx}
\left(\frac{dE}{dx}\right)_\textsubscript{calibrated}=\frac{\exp\left(\frac{(\frac{dQ}{dx})\textsubscript{calibrated}}{C\textsubscript{cal}}\frac{\beta^{\prime} W\textsubscript{ion}}{\rho\mathscr{E}}\right)-\alpha}{\frac{\beta^{\prime}}{\rho\mathscr{E}}},
\end{equation}
with\newline
$C\textsubscript{cal}$ is a calibration constant used to convert ADC values to number of electrons,
 \begin{conditions*}
W\textsubscript{ion} & 23.6 x 10\textsuperscript{-6} MeV/electron (work function of                                argon), \\
\mathscr{E} & 0.273 kV/cm (MicroBooNE drift electric field), \\
\rho & 1.38 g/cm\textsuperscript{3} (liquid argon density at a pressure 124.106 kPa), \\ \beta\textsuperscript{$\prime$} & 0.212 (kV/cm)(g/cm\textsuperscript2)/MeV, and \\
\alpha & 0.93. \\
\end{conditions*}

The last two parameters were measured by the ArgoNeuT experiment~\cite{recomb} at an electric field of 0.481 kV/cm.
The modified box model is applied at MicroBooNE's electric field of 0.273 kV/cm. 

According to the above equation, precise determination of the calibration constant $C\textsubscript{cal}$ which translates \say{ADC/cm} to \say{(number of electrons)/cm}, is important in determining the absolute energy scale.  The charge $Q$ is measured by the integral of the pulse of the deconvoluted anode response signal where the simulated electronics and field responses are removed.  $C\textsubscript{cal}$ is normalized so that the unit (\say{ADC}) corresponds to 200 electrons. In the case where the detector response is perfectly modeled ({\it e.g.} in the simulation), the calibration constant $C\textsubscript{cal}$ should be exactly 1/200. Here the goal of the $dE/dx$ calibration is to determine the calibration constant $C\textsubscript{cal}$ using stopping muons as the standard candle, because they have a well-understood energy loss profile. A sample of stopping muons can either be identified from cosmic data or neutrino interactions. The method we use for $dE/dx$ calibration is discussed in section~\ref{sec:ana_met_dedx}.

\subsection{Data sample}
The $dE/dx$ calibration of the detector is performed on both MC and data samples. For MC we use a CORSIKA CR simulation overlaid with neutrino interactions to obtain stopping muons induced by the neutrino interactions. In data we use datasets collected from February 2016 to October 2016 and from September 2017 to March 2018, with the final states of the inclusive charged current $\nu_{\mu}$ interactions~\cite{numu_cc_filt},
to obtain candidate stopping muons from charged current interactions. In both MC and data we use the Pandora~\cite{pandora} pattern recognition program to reconstruct the neutrino and cosmic interactions.

\subsection{Event selection}
\label{sec:sel_cut_abs}
In the $dE/dx$ calibration of the detector, we employ stopping muons induced by neutrino interactions. The motivation is to get a sample of muons from neutrino interactions with known muon energy loss per unit length so that a  portion of the track can be selected that corresponds closely to a minimum ionizing particle (MIP), for which the $dE/dx$ is very well understood theoretically to better than 1\%.
This sample contains the particles with the energy regime and the angular profile well representing the signatures of neutrino interactions.
Further, most events have forward-going muons, providing us with plenty hits in the collection plane.
The well known starting time of the stopping muons, determined by the arrival of the neutrinos produced in BNB, allows us to apply the dQ/dx calibration corrections.

\begin{figure}[!ht]
\centering
\includegraphics[width=0.6\textwidth]{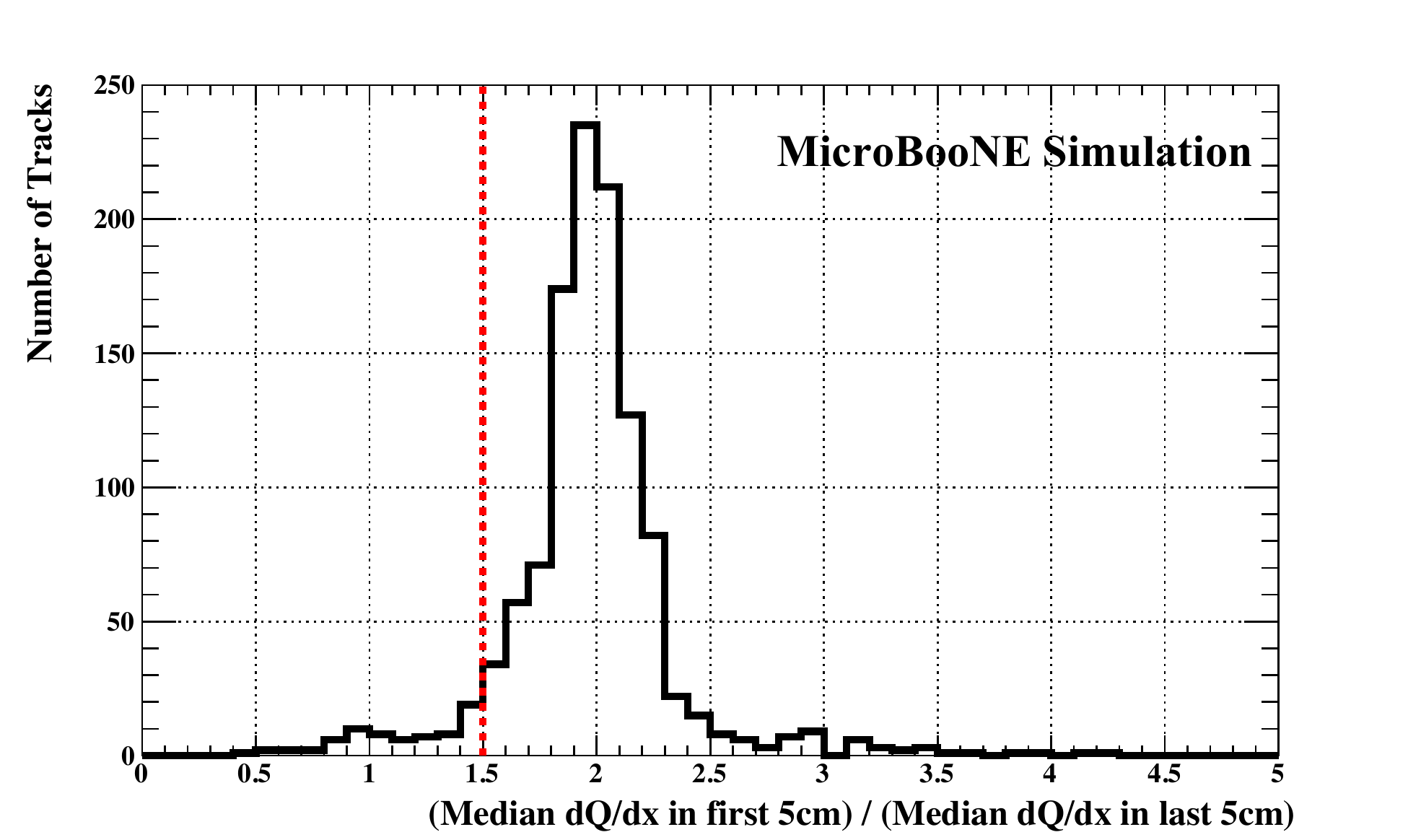}
\caption{Ratio of the $dQ/dx$ in the last to first 5 cm segment of stopping muons in MicroBooNE simulation. 
We select a cut value of 1.5 for the ratio considered to find well reconstructed stopping muon tracks in data.}
\label{fig:dq_dx_ratio} 
\end{figure}

We select the candidate muon tracks with the reconstructed start and end points 10~cm away from the TPC border in each dimension.
The tracks with the reconstructed angles, $75^{\circ} < \theta_{XZ} < 105^{\circ}$ and $80^{\circ} < \theta_{YZ} < 100^{\circ}$, are discarded to exclude tracks pointing into the wire planes and tracks nearly parallel to the collection plane wires, respectively.
To further ensure the quality of the reconstruction and the feature of minimum ionization, the tracks are required to have a minimum length of 150~cm.
In the MC sample, the candidate muon track has to match the true stopping muon coming from a neutrino interaction, and the distance between the start (end) point of the reconstructed track and that of the true muon has to be shorter than 5~cm.
In the data sample, the ratio of the median $dQ/dx$ value in the last 5 cm segment to that of the first 5 cm segment of the candidate track is required to be greater than 1.5, ensuring the selected tracks to have the profile of a stopping muon.
The selection criterion based on the $dQ/dx$ profile is determined from a MC simulation sample, as shown in figure~\ref{fig:dq_dx_ratio}. The non-muon contamination in the selected data sample is estimated to be less than 1\% using MC.

\subsection{Analysis method}
\label{sec:ana_met_dedx}
The following explains the procedure to perform the $dE/dx$ calibration both in MC and data:

\begin{enumerate}[label=(\roman*)]
\item Start with a set of tracks that satisfy the selection in section~\ref{sec:sel_cut_abs}.


\item Segment the last 200 cm residual ranges (residual range is defined as the path of the particle before its stopping point) of tracks into 5 cm bins, which leads to a total of 40 residual range bins. If the track length is less than 200 cm, the entire track is used, otherwise, the last 200 cm segment with respect to the stopping point is used. 

\item Loop over all the 3D hits of each selected track and fill the residual range
bins with $dE/dx$ values derived using equation~\ref{eqn:de_dx} by setting the calibration constant $C\textsubscript{cal}$ to an arbitrary value. Moreover, we consider only the 3D hits which are separated by 0.3 to 0.4 cm; as the minimum spacing between 3D hits is set by the wire spacing of 0.3 cm, this selects forward going muons.



\item After looping over all the tracks, fit each $dE/dx$ distribution to a Landau-convoluted Gaussian function \cite{lgfit} to extract the most probable $dE/dx$ value (MPV) representing that particular residual range bin.

\item Plot the MPV $dE/dx$ values against the kinetic energy of the particle; for each residual range bin, we take the middle bin value as the representative residual range value of that bin and transform that to kinetic energy. In this transformation we use a cubic spline fit to the tabulated values of  CSDA (continuous slowing down approximation) residual range {\it vs.} kinetic energy for stopping muons in liquid argon~\cite{csda_res_range}.

\item Compare the curve generated in the previous step with the prediction made by the Landau-Vavilov function~\cite{landau_vavilov} in the kinetic energy range of the muons from 250 MeV to 450 MeV, which is in the region expected for MIPs, and a $\chi$\textsuperscript{2} value is calculated using equation~\ref{eqn:chi_min}. The Landau-Vavilov function describes the energy loss probability distribution for a particle in a given medium. The most probable energy loss of a particle is dependent on the thickness of the energy absorber. To get the predicted MPV $dE/dx$, we set the absorber thickness to be 0.35 cm, the average value of $dx$.

\begin{equation}
\label{eqn:chi_min}
\chi^2 = \sum{\Bigg(\frac{(MPV(dE/dx)\textsubscript{Predicted}-MPV(dE/dx)\textsubscript{Measured})^2}{\sigma^2}\Bigg)}~,
\end{equation}
where we sum all the data points in the kinetic energy region of 250 MeV to 450 MeV.  

For MC, $\sigma^2=\delta_{\text{fit}}^{2}$, where $\delta\textsubscript{fit}$ is the uncertainty associated with the MPV $dE/dx$ extracted by fitting a Landau-convoluted Gaussian function to the energy distribution. 

For data, $\sigma^2=\delta_{\text{fit}}^{2} + \delta_{\text{recombination}}^2$, where $\delta\textsubscript{recombination}$ is the systematic error associated with recombination model uncertainties (here we take the uncertainty to be 1.5$\%$ of the measured $dE/dx$. This uncertainty is coming from the 1$\sigma$ level uncertainties of the two free parameters $\beta^{\prime}$ and $\alpha$ in the modified box model for recombination in  equation~\ref{eqn:de_dx}).

Note the MC sample was simulated and reconstructed with the same recombination model. Therefore, there is no systematic uncertainty associated with the recombination model for MC.

\item Iterate through the steps iii to vi described above several times to generate tabulated set of data between a given calibration constant ($C\textsubscript{cal}$) and $\chi$\textsuperscript{2} value (here we change the parameter $C\textsubscript{cal}$ in each iteration and calculate the $\chi$\textsuperscript{2} value. The total number of iterations in this procedure is the total number of data points in the plots shown in figure~\ref{fig:chi_min_plots_p2}).

\item In the final step, plot the $\chi$\textsuperscript{2} values generated against calibration constants and fit that distribution with a second order polynomial to get the calibration constant which corresponds to the lowest $\chi$\textsuperscript{2} value.

\item With the newly derived calibration constant ($C\textsubscript{cal}$), calculate the $dE/dx$ values using equation~\ref{eqn:de_dx} with calibrated $dQ/dx$ values as the input. 
\end{enumerate}

After the $dE/dx$ calibration, we compare the newly derived $dE/dx$ values with uncorrected $dE/dx$ values to see the effects of applying the algorithm. (See section~\ref{sec:results_dedx}).





\subsection{Results}
\label{sec:results_dedx}
\begin{figure}[!ht]
\centering
\includegraphics[width=0.49\textwidth]{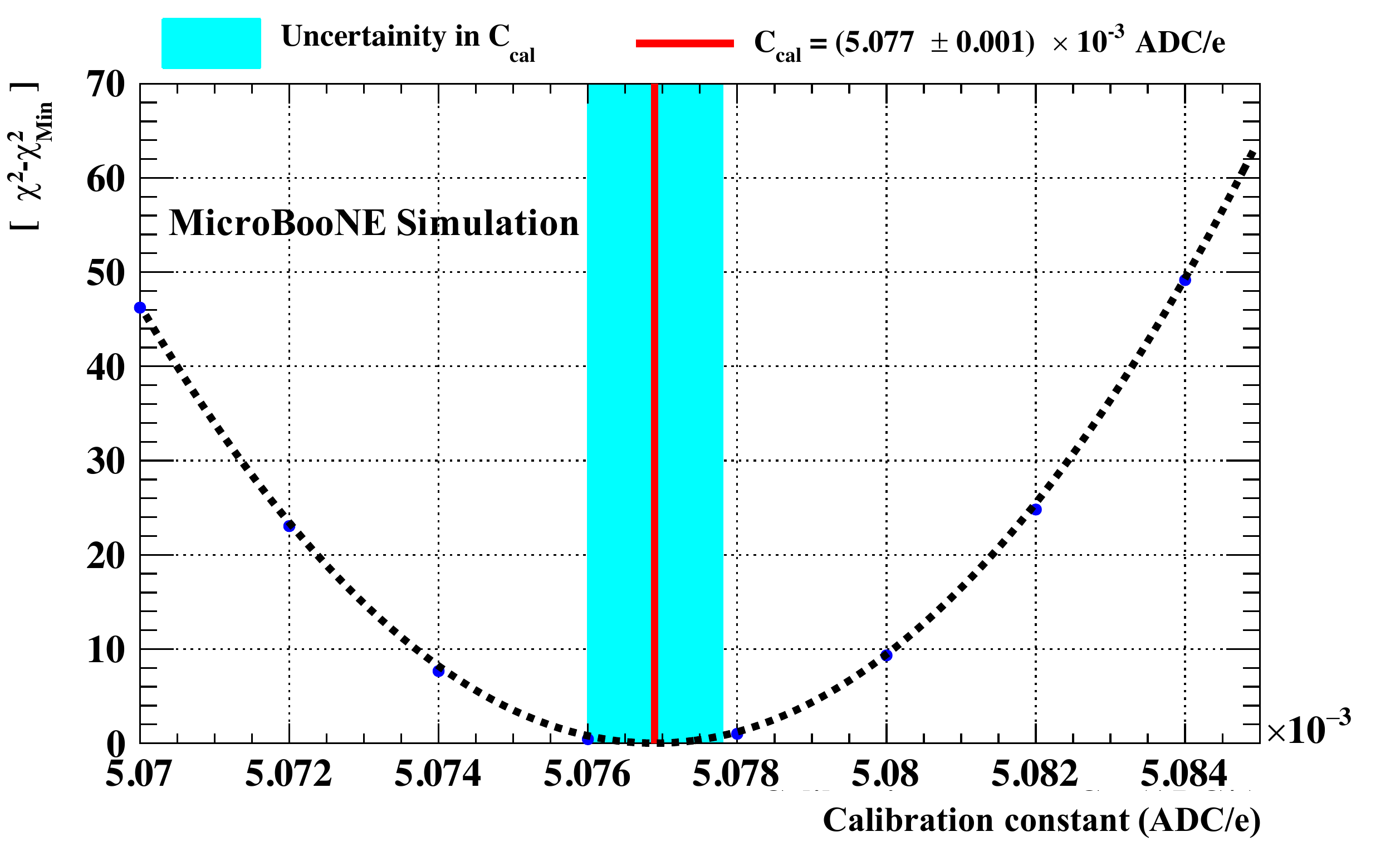} 
\includegraphics[width=0.49\textwidth]{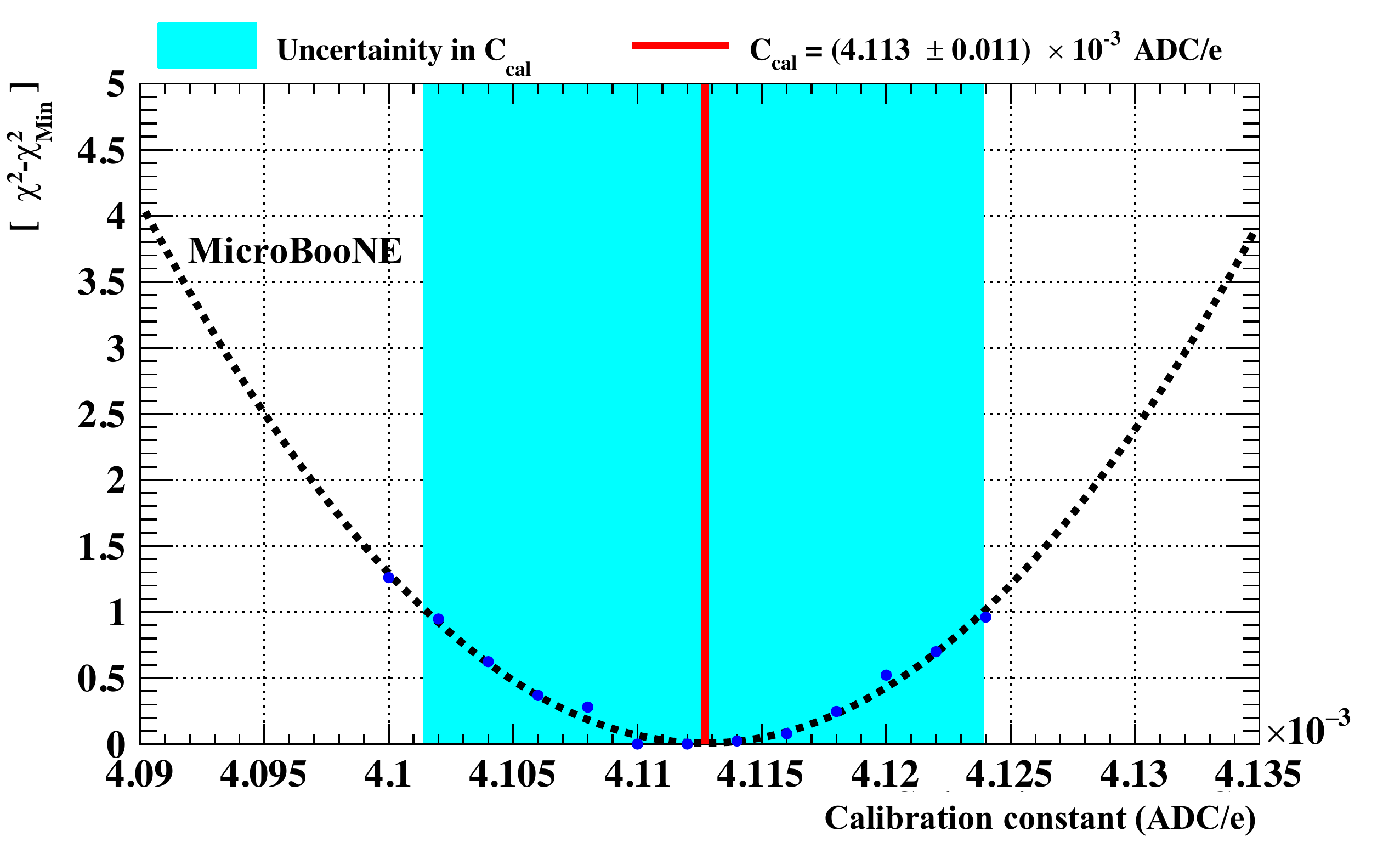} 
\caption{Distributions of $\chi^{2} - \chi_{\text{Min}}^{2}$ {\it vs.} calibration constant $C\textsubscript{cal}$ for the collection plane. The color bands show the uncertainty associated with calibration constant $C\textsubscript{cal}$. (Left) MC. (Right) Data. Results are applicable to all periods of data-taking, as temporal variations are taken into account by the relative $dQ/dx$ corrections.}
\label{fig:chi_min_plots_p2}
\end{figure}

The absolute calibration of the detector is performed on both MC and data samples. Figure~\ref{fig:chi_min_plots_p2} shows $\chi^{2} - \chi_{\text{Min}}^{2}$ {\it vs.} the calibration constant $C\textsubscript{cal}$ for the collection plane. The best fit value is the one that gives the minimal $\chi^{2}$. The uncertainty on the extracted value of $C\textsubscript{cal}$ is given by $\Delta\chi^{2} = \chi^{2} - \chi_{\text{Min}}^{2} = 1$. 

The calibration constants derived for the collection plane are shown in table~\ref{tab:cal_cons}. The MC calibration constant uncertainty is statistical only. The data calibration constant uncertainty includes both the statistical uncertainty and the recombination uncertainty. The difference between data and MC calibration constants is due to the fact that the MC simulation of detector response is not in a perfect agreement with data.

\begin{table}[ht]
\centering
\caption{Calibration constants and $\chi_{\text{Min}}^{2}/d.o.f.$ for the collection plane in MC and data.} 
\begin{tabular}[t]{lcc} 
\toprule
 &MC &Data \\
\midrule
Fitted value of $C\textsubscript{cal}$  &(5.077 $\pm$ 0.001) $\times$$10^{-3}$~ADC/e &(4.113 $\pm$ 0.011) $\times$ $10^{-3}$~ADC/e\\
$\chi_{\text{Min}}^{2}/d.o.f$ &15.0/18 $\sim$ 0.84 &5.12/18 $\sim$ 0.28\\
\bottomrule
\end{tabular}
\label{tab:cal_cons}
\end{table}

\begin{figure}[!ht]
\centering
\includegraphics[width=0.49\textwidth]{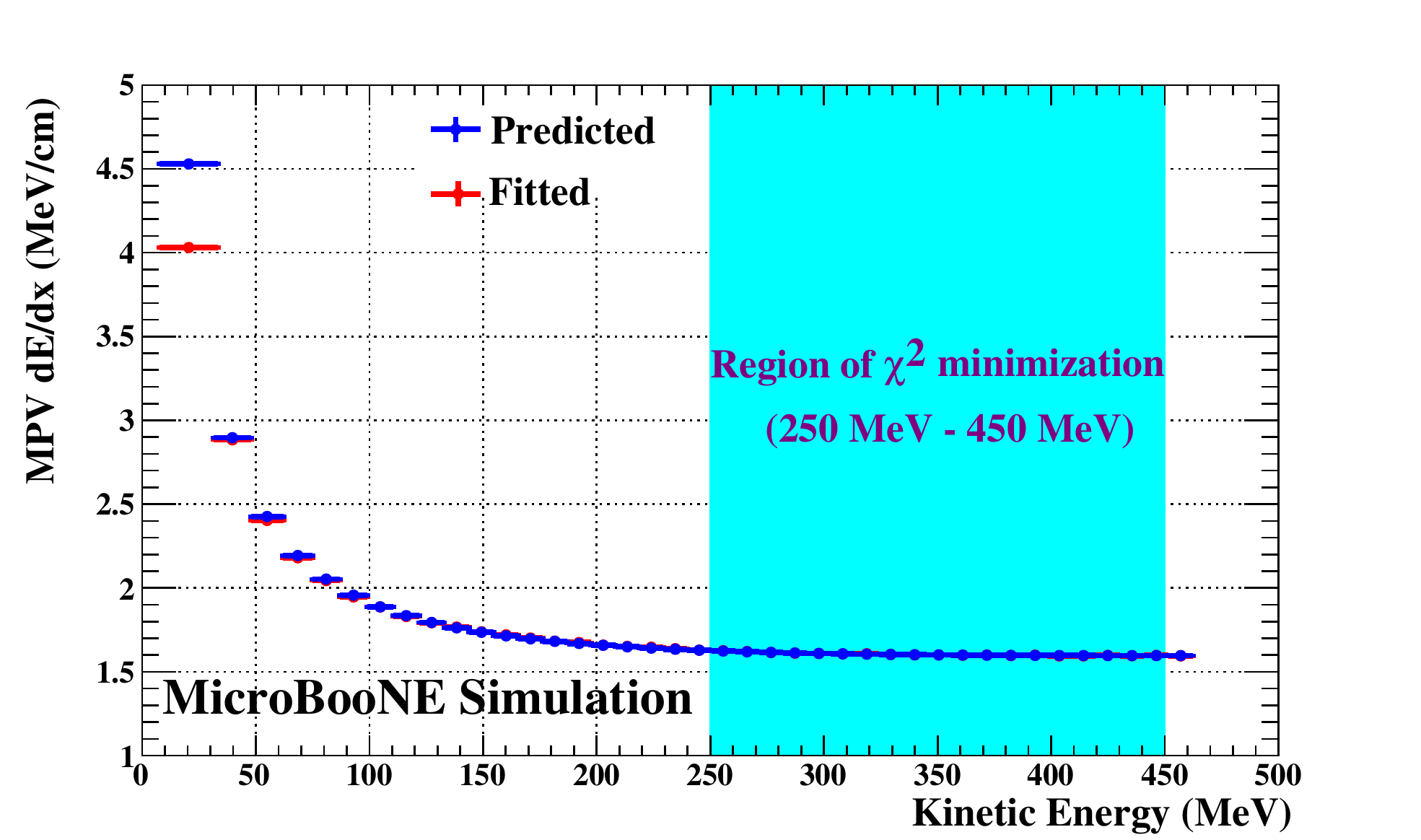} 
\includegraphics[width=0.49\textwidth]{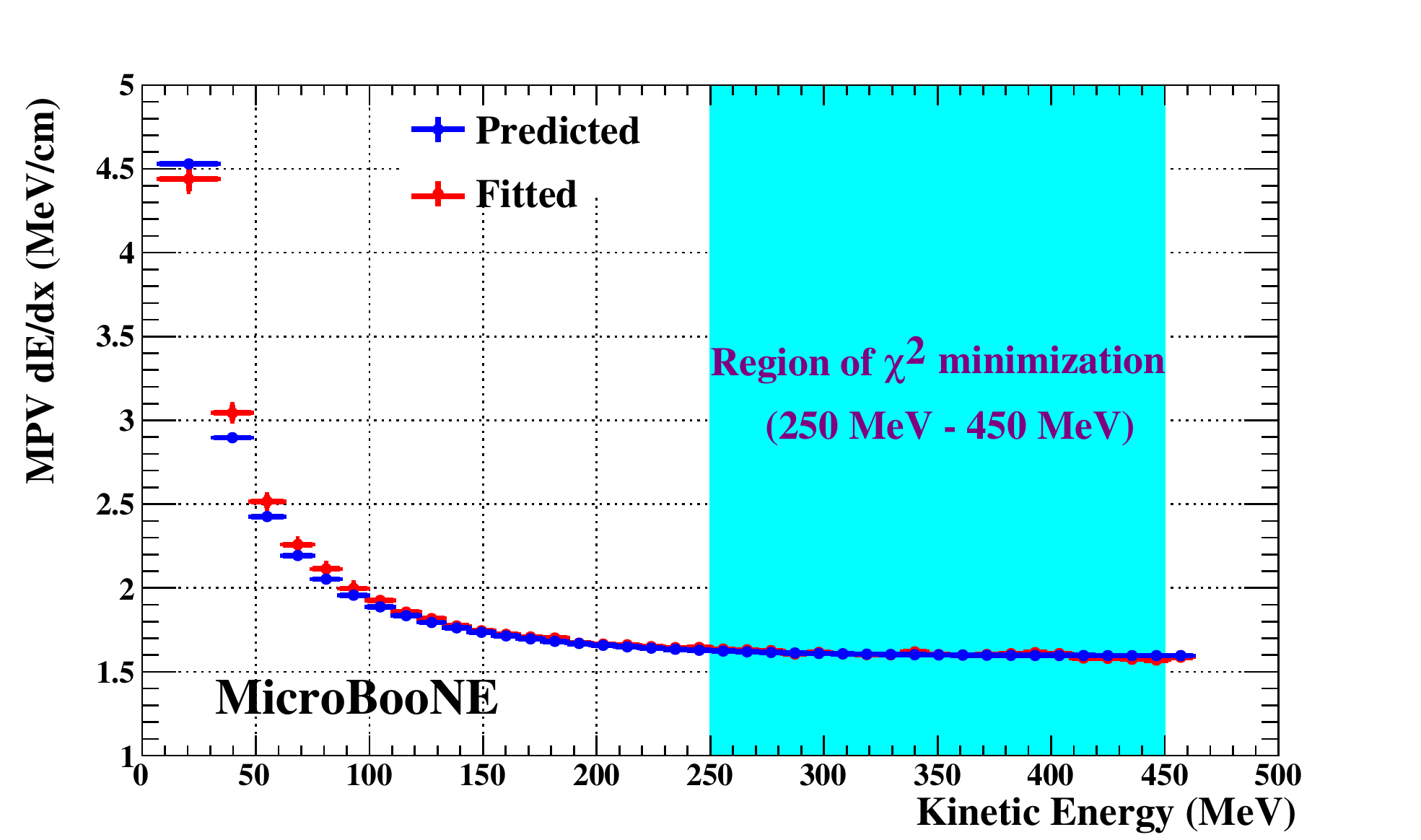} 
\caption{(Left) Comparison between the prediction and the fitted MPV $dE/dx$ for stopping muons in MC using the collection plane. (Right) Comparison between the prediction and the fitted MPV $dE/dx$ for stopping muons in MicroBooNE data from 2016 using the collection plane.}
\label{fig:cali_plots} 
\end{figure}

Figure~\ref{fig:cali_plots} shows the comparison between the predicted most probable energy loss~\cite{landau_vavilov} with the fitted most probable energy loss using the calibration constants shown in table~\ref{tab:cal_cons} for stopping muons both in MC and data. Here the disagreement between the fitted distribution and theory at lower kinetic energies in data is mainly due to the recombination model uncertainty. Results can be applied to all time ranges, since temporal variations are taken into account by the relative $dQ/dx$ corrections. The first fitted point in the distributions corresponds to the end of the tracks. Most of the time the end of the track can happen in between two wires. Therefore calorimetric information for the end points of tracks is not properly reconstructed. The significant deviation between prediction and our fitted distribution for MC in the first bin is due to this reason.

\begin{figure}
\centering
\includegraphics[width=0.49\textwidth]{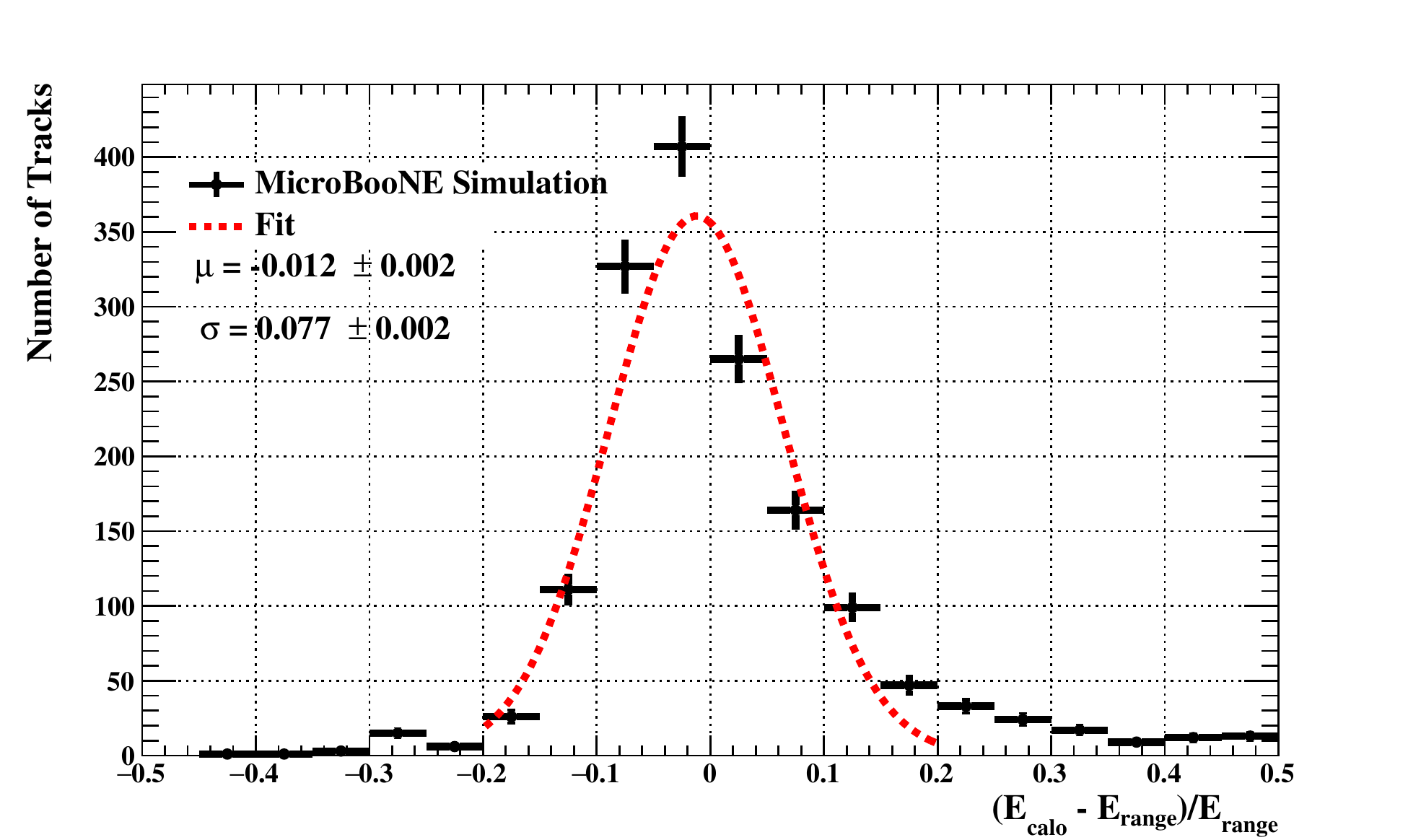} 
\includegraphics[width=0.49\textwidth]{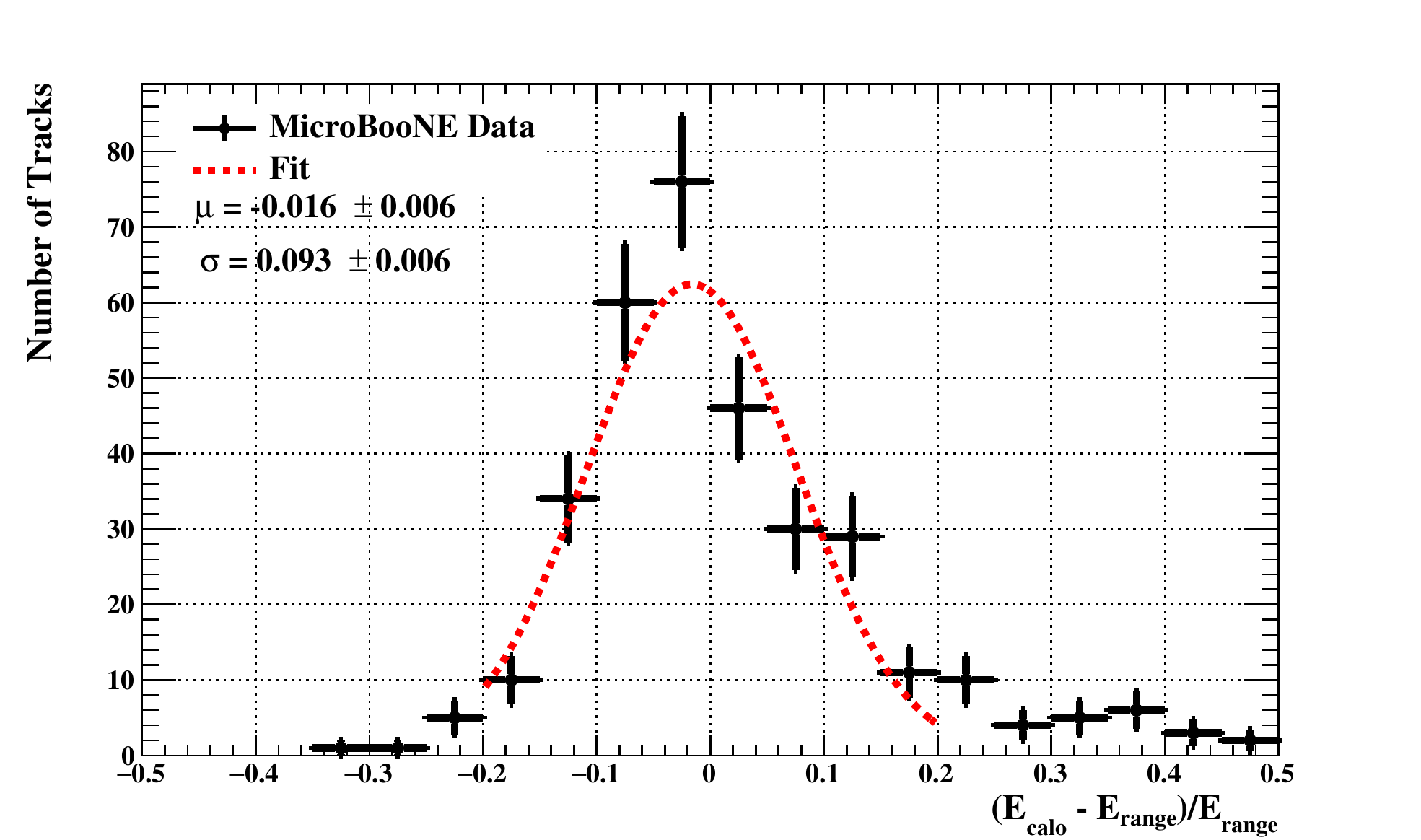} 
\caption{Relative difference in range {\it vs.} calorimetric kinetic energy for selected stopping muons in MC (left) and data (right). The distributions are fit to a Gaussian which returns a mean $\mu = -0.012 \pm 0.002 $ and width $\sigma = 0.077 \pm 0.002$ for MC and a mean $\mu = -0.016 \pm 0.006 $ and width $\sigma = 0.093 \pm 0.006$ for data.}
\label{fig:closure} 
\end{figure}

\paragraph{}The absolute energy-scale calibration can be validated in a data-driven way by comparing the range-based energy to that obtained via calorimetry for selected stopping muon candidates in data and MC. For each selected stopping muon track, the kinetic energy calculated by range and by calorimetry is computed, and the relative difference between these methods is shown in figure~\ref{fig:closure}. Here, in calculating the kinetic energies, the entire track is used. The agreement between the two, approximately 2\% for data (1\% in MC), gives confidence in the proper absolute energy-scale calibration to within this level of agreement, and provides a closure test for this calibration procedure. The tails on the positive side are caused by contributions of other particles to the muon track near the neutrino interaction vertex. A MC study shows that after the calibration procedure, the bias in $(E_{calo}-E_{range})/E_{range}$ reduces from -3.2\% to -1.2\% and the resolution improves from 8.1\% to 7.7\%.

\section{Energy correction with protons}

\subsection{Introduction}

After the nonuniformities in detector response have been corrected for and the absolute energy scale has been determined as described in the previous sections, we perform studies using a proton sample to investigate the effect of electron-ion recombination.

The calibration of $dQ/dx$ and $dE/dx$ presented in the previous sections sets the correct scale for energy loss per unit length for MIPs, for which $dE/dx$ remains nearly a constant. As particles approach the end of their range, they become increasingly ionizing. Electron-ion recombination introduces a non-linear relationship between $dQ/dx$ and $dE/dx$, which has been extensively studied~\cite{Thomas:1987zz, Birks:1964zz}. In this study, 
we use stopping protons to measure the non-linear relationship between $dQ/dx$ and $dE/dx$. 
The variation of $dE/dx$ with the proton energy allows us to measure the recombination effect for a wide range of stopping powers up to about 20 MeV/cm. 

The detector simulation used by MicroBooNE employs the modified box model by default to simulate the recombination effect, with the parameters obtained by the ArgoNeuT experiment~\cite{recomb}. The charge deposition $dQ/dx$ at each Geant4 step is calculated using the energy loss $dE/dx$ following:

\begin{equation}
\label{eqn:dq_dx_modbox}
\frac{dQ}{dx}(e/cm) =\frac{\ln(\frac{dE}{dx}\frac{\beta^{\prime}}{\rho\mathscr{E}}+\alpha)}{\frac{\beta^{\prime}}{\rho\mathscr{E}} W_\text{ion}}~,
\end{equation}
with the parameters $W\textsubscript{ion}$, $\mathscr{E}$, $\rho$, $\beta\textsuperscript{$\prime$}$ and $\alpha$, already described in the inverse of this equation, eq.~\ref{eqn:de_dx}.


In this analysis, we compare the proton $dQ/dx$ {\it vs.} $dE/dx$ measurements with two recombination models: the modified box model as described in eq.~\ref{eqn:dq_dx_modbox} and Birks' law, following eq.~\ref{eqn:Birks_de_dx} with $A_B=0.8$ and $k=0.0486\,\text{(kV/cm)(g/cm}^2\text{)/MeV}$, as measured by the ICARUS collaboration~\cite{recombICARUS}: 
\begin{equation}
\label{eqn:Birks_de_dx}
\frac{dQ}{dx}(e/cm)=\frac{A_B}{W_\text{ion}}\left(\frac{\frac{dE}{dx}}{1+\frac{k}{\rho\mathscr{E}}\frac{dE}{dx}}\right).
\end{equation}
The two models are parameterized by ArgoNeuT and ICARUS at different electric fields from the operating electric field at MicroBooNE. Therefore, it is interesting to compare both models with the MicroBooNE measurement to study the dependence of recombination on electric field. 

To perform these studies, we use the calibrated $dQ/dx$ information and calculate $dE/dx$ using an empirical fit to Bethe-Bloch energy loss using the range information,  $(dE/dx)_{\text{range}}$. The $(dE/dx)_{\text{range}}$ values for protons are calculated using the residual range information per hit, as described in Ref.~\cite{recomb} according to eq.~\ref{eqn:de_dx_theory}
\begin{equation}
\label{eqn:de_dx_theory}
\left(\frac{dE}{dx}\right)_{\text{range}}=A\cdot R^b
\end{equation}
where $R$ is the residual range and $A$ and $b$ are empirical constants (the constant $A$ in eq.~\ref{eqn:de_dx_theory} should not be confused with $A_B$ from Birks' law in eq.~\ref{eqn:Birks_de_dx}). We use the values obtained by ArgoNeuT, $A = 17~\text{MeV}/\text{cm}^{1-b}$ and $b = -0.42$ in the calculation. The $dE/dx$ values calculated in the way are independent of the charge measurement, and therefore can be compared to the calibrated $dQ/dx$ values to derive recombination model parameters.

\subsection{Event selection}
\label{sec:sel_protons}

In this study we use protons emerging from neutrino interactions with argon. We choose a sample of events with one muon and two protons. This sample is referred to as $1\mu2p$. Such events are produced almost exclusively by the neutrino interactions, rather than the cosmic ray interactions, as verified using data taken when the BNB beam was off. Having three tracks sharing the same vertex improves the resolution of vertex and track reconstruction. This sample provides a highly pure proton sample with well reconstructed track trajectories, which is ideal for the recombination measurement. 


The event selection is based on a set of cuts designed to select charged-current $\nu_{\mu}$ interactions producing $1\mu2p$ final state in the TPC fiducial volume (FV). We take advantage of the very specific number of charged particles and the absence of $e^\pm/\pi^0$ in the final state, which yields a sample of well reconstructed tracks.

For the $1\mu2p$ events, the signature is three tracks originating from the same vertex, one consistent with a muon and the other two consistent with protons. The selection criteria are described as follows:

\begin{itemize}
\item {\bf Beam window}: 32 PMTs are used in MicroBooNE to record scintillation light information from particles traveling inside the TPC. In order to select neutrino interactions, we require there should be a scintillation light signal detected in the PMTs coincident with the neutrino beam spill.  

\item {\bf Vertex}: The reconstructed neutrino vertex is required to be in the FV. The FV is defined as a rectangular box shaped as follows:  the boundary from the anode plane and  the  cathode  plane is 10 cm,  the boundary from the upstream and downstream ends is 10 cm, and the boundary from the top and bottom of the TPC is 20 cm.
It is the same FV as used by the MicroBooNE physics analysis reported in \cite{numu_cc_Marco}.
The more stringent requirement in the vertical direction ensures a higher CR background removal.


\item {\bf Three tracks cut}: Only three tracks are connected to the vertex within a 5~cm radius. 
If more than one neutrino candidate is present in an event, the one where the sum of the distance between the start position of each track to each other is smallest is selected. This suppresses cosmic ray background mimicking neutrino interactions.

\item {\bf Proton containment}: The leading particle, defined as the longest track, is taken as the muon candidate; the remaining two tracks are taken as proton candidates. These two proton candidates are required to be contained in the TPC FV to aid in particle identification (PID).

\item {\bf Minimum number of hits}: For each track we require a minimum number of 10 hits in total, summing all three planes. In order to have the most reliable calorimetry information, we also require at least 5 hits in the collection plane. This requirement is applied to all tracks in a $1\mu2p$ event.

\item {\bf PID cut}: We use the $\chi^2$ test with respect to a proton hypothesis in the $dE/dx$ {\it vs.} residual range curves, from Bethe-Bloch predictions, to discriminate MIP with respect to non-MIP particles. The predicted proton $dE/dx$ {\it vs.} residual range curve is calculated using the Geant4 simulation, where the residual range is sampled with a step size of 0.08 cm.

For each selected track (either in MC or data), we reconstruct $dE/dx$ per hit in the collection plane as described in section~\ref{eqn:de_dx} and calculate the $\chi^2$ between the reconstructed and the Geant4-determined mean $dE/dx$ at each given residual range bin. These values are then summed for all hits on the selected track. 

We exclude the first and last hits on the track from this calculation because of the uncertainty in the $dx$ calculation. This is due to the fact that the $dx$ calculation for the first and last hits may be wrong since the exact start and end positions of the track between the wires are unknown. 
It also mitigates the complications caused by energy deposition overlap at the interaction vertex from other particles ({\it e.g.} short protons) produced by the neutrino interaction.

The value obtained is then normalized by the number of degrees of freedom (d.o.f.), which corresponds to the number of hits in the collection plane. This calculation can be determined for several particle assumptions, but in this analysis we exclusively use the proton hypothesis. 

\begin{equation}
\label{eqn:PIDeq}
PID = \chi^2_{\textrm{proton}}/d.o.f. = \sum_{\text{Track hits}} \left(\frac{(dE/dx\textsubscript{measured}- dE/dx\textsubscript{theory})}{ \sigma_{dE/dx_{}}}\right)^2/d.o.f.,
\end{equation}

where $\sigma_{dE/dx}$ is the estimated resolution in measuring a fixed value of $dE/dx$, the subscript {\it measured} and {\it theory} corresponds to the $dE/dx$ measured at the wires using the charge deposition information and the predicted by Geant4 respectively. Values used for resolution in $dE/dx$ are taken from studies performed by the ArgoNeuT Collaboration. This does not take into account the $3$ mm wire spacing in MicroBooNE (compared to the $4$ mm wire spacing in ArgoNeuT), which leads to an overestimated resolution in $dE/dx$ when compared to the resolution we obtain from the MicroBooNE MC studies. We prefer to use the values from ArgoNeuT as a conservative approach since the recombination model is not tuned based on MicroBooNE data at this stage yet. 
In order to select a sample of highly ionizing particles, an initial guess of recombination correction is needed, for which we employ the modified box model with the parameters measured by ArgoNeuT. This model is sufficiently close to the final measurement to ensure a good separation of protons and muons.

Taking into account both the particle discrimination (muon {\it vs.} proton) and the data/MC shape agreement, we apply a PID requirement that identifies protons as the tracks with $\chi^2\textsubscript{proton}/d.o.f.<88$, and muons as tracks with $\chi^2\textsubscript{proton}/d.o.f>88$. The values are optimized by maximizing the efficiency and purity for the proton identification.


Table~\ref{tab:PID_purity} shows the purity and efficiency of the muon and proton selections, estimated from the MC sample. Purity is defined as the quantity of particles from an specific species which has been correctly identified, divided by the total number of particles of that species in the sample. For comparison, the same values are shown before applying the equalization procedure described in section~\ref{sec:dqdx}. Note that in the MC simulation, we reconstruct $dE/dx$ using the recombination parameters obtained by ArgoNeuT, as they are the values used in the simulation. 
The equalization calibration improves both the purity and efficiency of the muon and proton selections. 
\begin{table}[!htp]
\begin{center}

\caption{Purity and efficiency for selecting muons and protons in the MC, after applying the PID selection. For comparison, the purity and efficiency before equalization calibration is also shown.} 
\begin{tabular}[t]{lccccc} 
\toprule
particle type  & purity & efficiency & purity before& efficiency before\\
  &  &  & $dQ/dx$ calibration & $dQ/dx$  calibration\\
\midrule
selected muons & $94\%$ & $99\%$ & $93.4\%$ & $98.7\%$\\
selected protons & $93\%$ & $85\%$ & $92.7\%$ & $83.1\%$\\

\bottomrule
\end{tabular}
\label{tab:PID_purity}
\end{center}
\end{table}

\item {\bf Shower veto}: We remove events with a shower reconstructed within 15~cm of the vertex, both to reduce $\pi^0$ contamination and to reduce background caused by tracks misreconstructed as showers.
\end{itemize}
In this analysis, a total of 226 proton candidates are selected from the neutrino data.

\subsection{Analysis method}
\label{sec:ana_met_recombination}

\begin{figure}[!htp]
  \centering
  \subfloat[Data]{
    \includegraphics[width=0.49\textwidth]{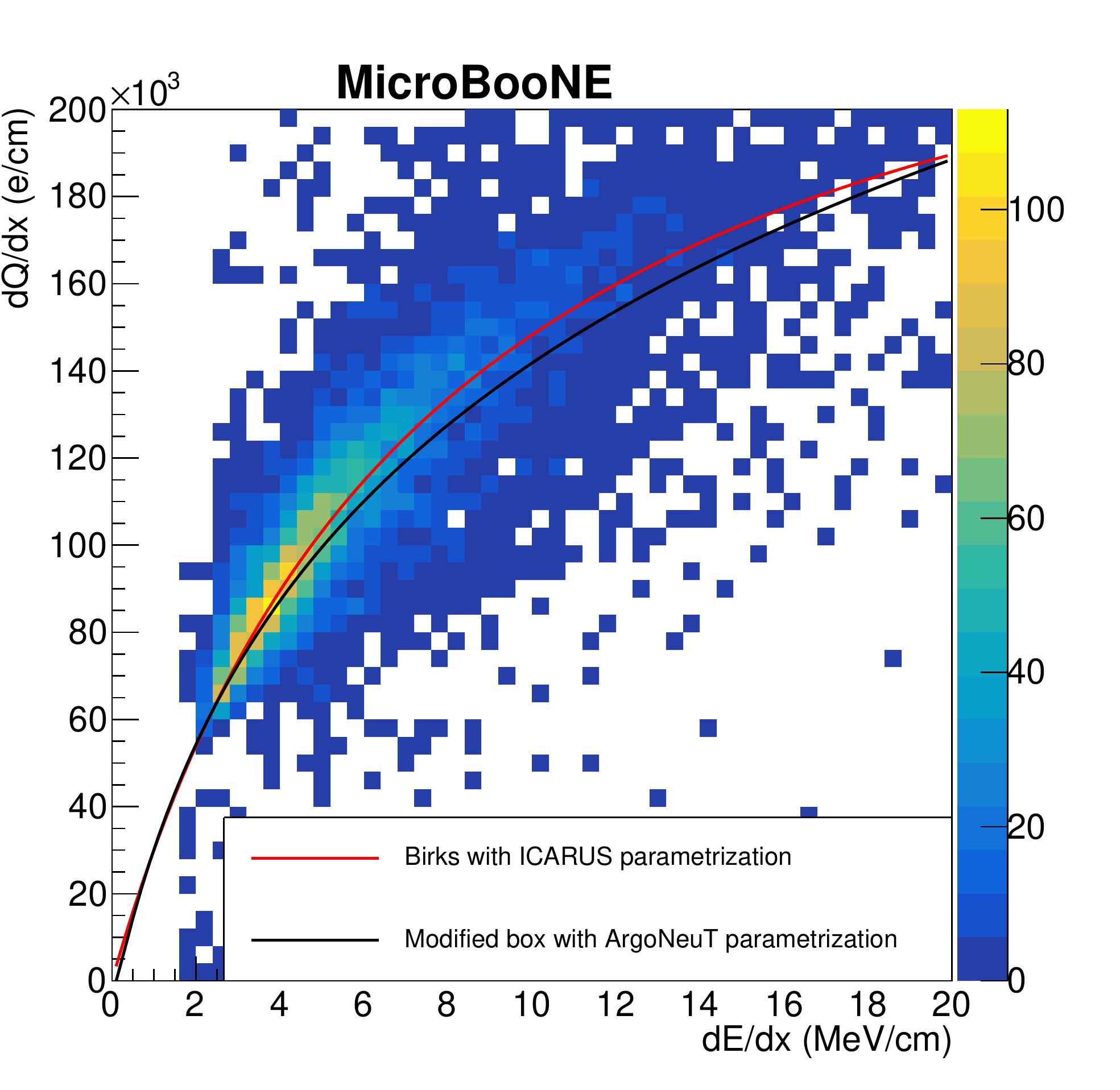}
    \label{fig:protons2D_defaultBirksBox_data}
  }
  \subfloat[MC simulation]{
    \includegraphics[width=0.49\textwidth]{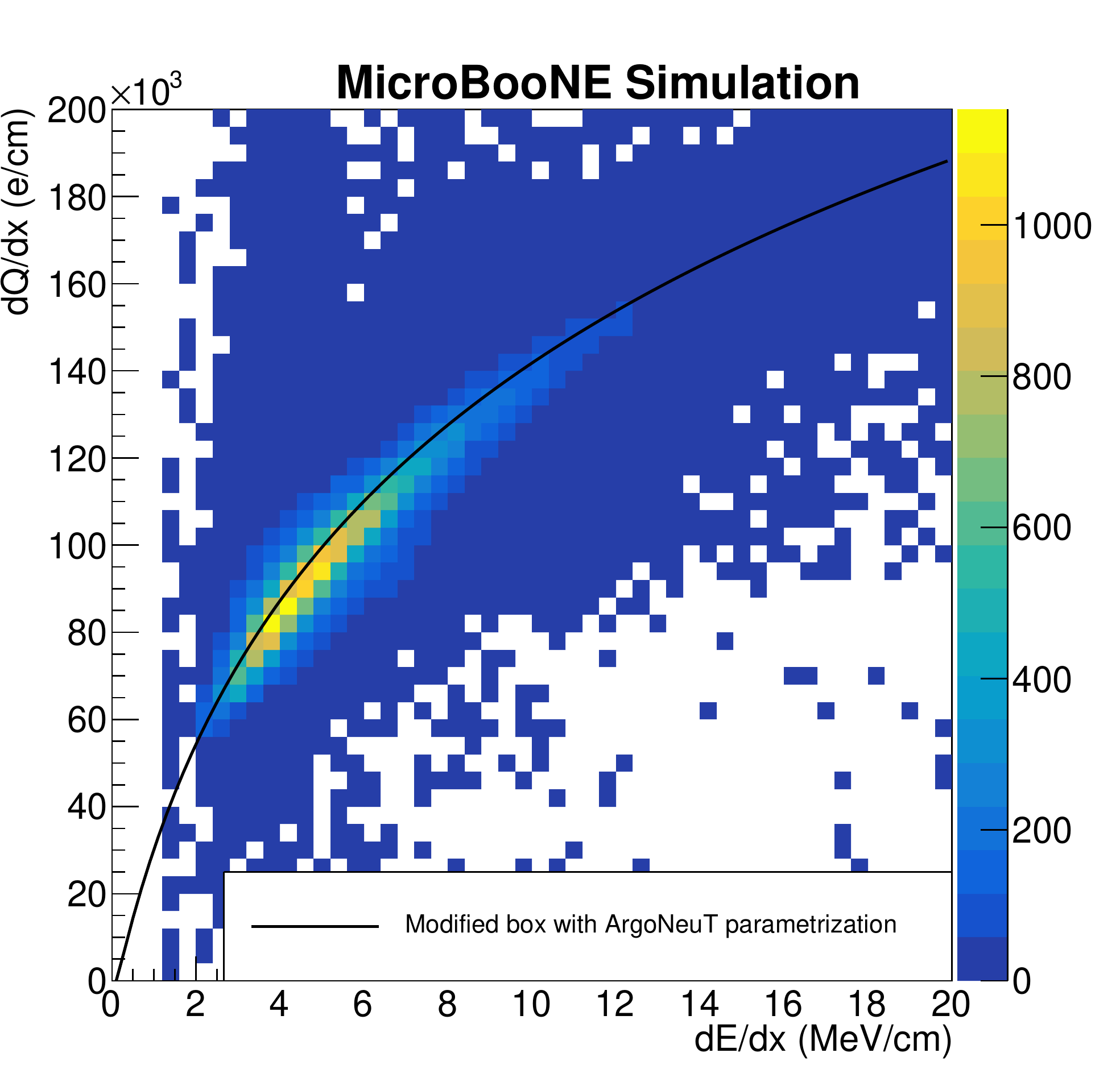}
    \label{fig:protons2D_defaultBirksBox_mc}
  }
  \caption{\protect\subref{fig:protons2D_defaultBirksBox_data} Comparison of the measured $dQ/dx$ {\it vs.} $(dE/dx)_{\text{range}}$ distribution with recombination models for selected proton tracks in data; \protect\subref{fig:protons2D_defaultBirksBox_mc} same comparison for selected protons in the MC simulation. The red curve corresponds to Birks' law and the black curve corresponds to the modified box model, using the parameters measured by the ICARUS and ArgoNeuT collaborations. Both model predictions are calculated at the nominal electric field of 0.273 kV/cm for MicroBooNE.}
  \label{fig:protons2D_defaultBirksBox}
\end{figure}

With the proton sample described we study two different recombination models:
the modified box model, eq.~\ref{eqn:dq_dx_modbox}, and Birks' law, eq.~\ref{eqn:Birks_de_dx}. The comparison of these two models with the MicroBooNE proton dataset (described in section~\ref{sec:sel_protons}) is shown in figure~\ref{fig:protons2D_defaultBirksBox}. In this figure, the measured $dQ/dx$, after calibration, is shown with respect to the theoretically predicted energy deposit $(dE/dx)_{\text{range}}$, computed from the proton residual range following eq.~\ref{eqn:de_dx_theory}, for both data (left) and MC (right). Both model predictions are calculated at the nominal electric field of 0.273 kV/cm for MicroBooNE.


We observe a good agreement of data with Birks' law and the modified box model at low $dE/dx$ in figure~\subref*{fig:protons2D_defaultBirksBox_data}, as expected. Low $dE/dx$ corresponds to the MIP region, which has been calibrated with muons previously. At higher $dE/dx$, data shows a lower recombination effect than the one predicted by the modified box model. On the other hand, Birks' law agrees much better with data at higher ionization. This is not surprising since Birks' law was tuned with ICARUS data taken at drift fields of 0.2, 0.35 and 0.5 kV/cm while the modified box model was tuned with ArgoNeuT data taken at a single drift field of 0.481 kV/cm. As expected, the MC distribution agrees well with the modified box model using the ArgoNeuT parameters, shown as the black curve in figure~\subref*{fig:protons2D_defaultBirksBox_mc}, since it is the same model used for the simulation.

Because the parameters $\beta^{\prime}$ in the modified box model and $k$ in Birks' law are coupled to the electric field $\mathscr{E}$ in eqs~\ref{eqn:dq_dx_modbox} and~\ref{eqn:Birks_de_dx}, it is the quantity $\beta^{\prime}/\mathscr{E}$ or $k/\mathscr{E}$ that we are able to measure. The accumulation of positive ions produced by CR in the liquid argon leads to a nonuniform electric field. 
Even though the nonuniformities are largely reduced by the equalization procedure described in section~\ref{sec:dqdx}, there could still be residual nonuniformities in the electric field distribution.
The discrepancies shown in figure~\subref*{fig:protons2D_defaultBirksBox_data} could be due to the fact that the recombination model parameters derived at other experiments are not applicable at the MicroBooNE electric field, or that the average electric field at MicroBooNE is different from the nominal electric field of 0.273 kV/cm.
For the recombination measurements reported in this paper, we adopt an effective approach by fixing the electric field at 0.273 kV/cm and tune the other model parameters to improve the agreement with the MicroBooNE proton data.

We obtain the {\it effective recombination} parameters through a $\chi^{2}$ fit. We construct a $\chi^2$ function based on  the mean $dQ/dx$ value per each $dE/dx$ bin in data and MC. We then minimize the $\chi^{2}$ function to obtain the best fit parameters ($\alpha$ and $\beta^{\prime}$ for the modified box model, and $A_{B}$ and $k$ for Birks' law). For data, $dQ/dx$ in $(e/cm)$ is calculated by dividing the calibrated $dQ/dx$ in $(ADC/cm)$, as described in section~\ref{sec:dqdx}, by the calibration constant $C_\text{cal}$, as described in section~\ref{sec:results_dedx}; {\it i.e.}
\begin{equation}
  \frac{dQ}{dx}(e/cm)^\textsc{data} = \frac{dQ}{dx}(ADC/cm)^\textsc{data}/C_\text{cal},
\end{equation}
and $dE/dx$ is calculated using the residual range following eq.~\ref{eqn:de_dx_theory}. For MC, $dE/dx$ is also calculated using residual range and $dQ/dx$ is converted from $(dE/dx)_{\text{range}}$ using the recombination model (the modified box model eq.~\ref{eqn:dq_dx_modbox} or Birks' law eq.~\ref{eqn:Birks_de_dx}). The $\chi^2$ is defined as:


\begin{equation}
\label{eqn:chi2_fitRecombination}
\chi^2 = \sum_{dE/dx\ \text{bins}} \left(\frac{\frac{dQ}{dx}(e/cm)^\textsc{data}-\frac{dQ}{dx}(\alpha,\beta^{\prime},(dE/dx)\textsubscript{range})^\textsc{MC}}{ \sqrt{\left(\sigma_{(dQ/dx)^\textsc{data}}^2+\sigma_{(dQ/dx)^\textsc{MC}}^2\right)}}\right)^2,
\end{equation}
where $\sigma_{(dQ/dx)^\textsc{data}}$ and $\sigma_{(dQ/dx)^\textsc{MC}}$ are the statistical errors from data and MC respectively.

\subsection{Results}
\label{sec:results_recombination}

\begin{figure}[!htp]
  \centering
  \subfloat[Comparison with the modified box model]{
    \includegraphics[width=0.49\textwidth]{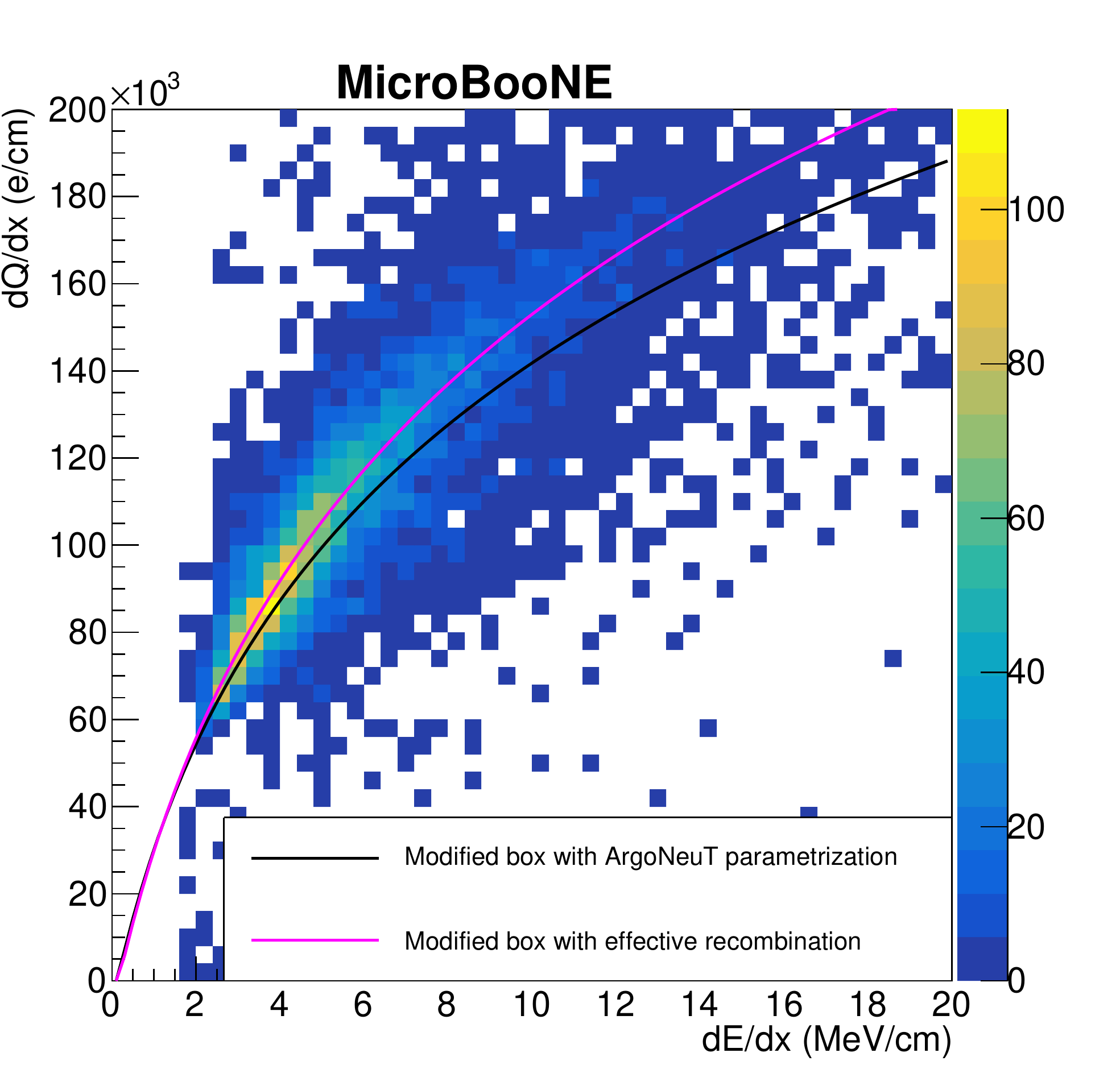}
    \label{fig:protons2D_resultBirksBox_box}
  }
  \subfloat[Comparison with Birks' law]{
    \includegraphics[width=0.49\textwidth]{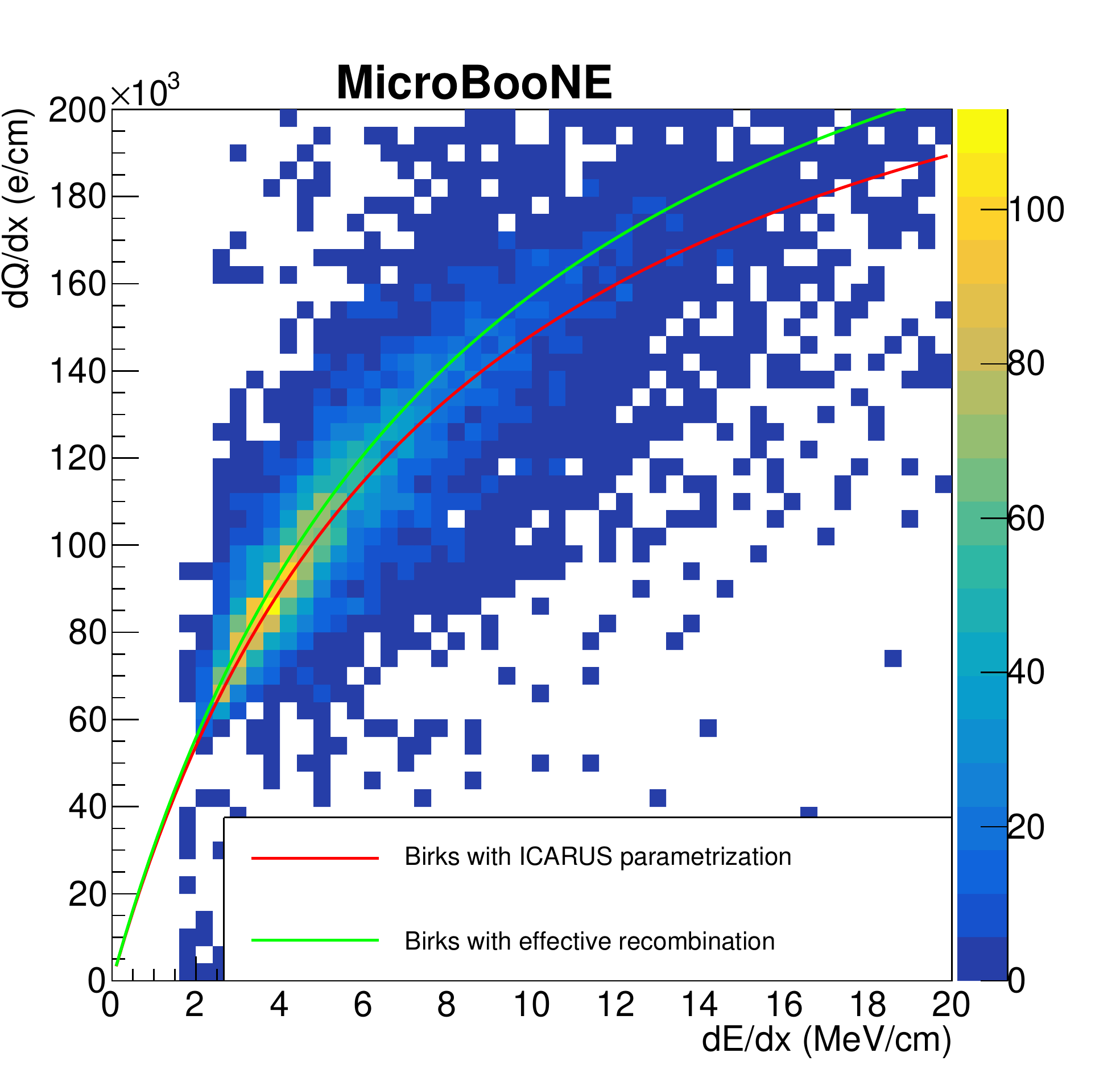}
    \label{fig:protons2D_resultBirksBox_birks}
  }
  \caption{Comparison of the measured $dQ/dx$ {\it vs.} $(dE/dx)_{\text{range}}$ distribution with recombination models for selected proton tracks in data. \protect\subref{fig:protons2D_resultBirksBox_box} the black curve represents the modified box model with the original parameters and the magenta curve uses the new parameters from table~\ref{tab:proton_recombination}; \protect\subref{fig:protons2D_resultBirksBox_birks} the red curve is Birks' law with the original parameters and the green curve uses the new parameters from table~\ref{tab:proton_recombination}.}
  \label{fig:protons2D_resultBirksBox}
\end{figure}

We perform the $\chi^{2}$ fit using both the modified box model and Birks' law. The best fit results are summarized in table~\ref{tab:proton_recombination}, including the statistical uncertainties from the fit. To obtain these results, we assume an electric field of $0.273~\text{kV/cm}$, which is the nominal electric field in MicroBooNE. Under this assumption, the parameters that are not coupled to the electric field, $\alpha$ and $A_{B}$, are consistent with the measurements from ArgoNeuT and ICARUS, while the parameters that are coupled to the electric field, $\beta^{\prime}$ and $k$, are in a disagreement with the measurements from ArgoNeuT and ICARUS. The effect of the obtained result can be clearly observed in figure~\ref{fig:protons2D_resultBirksBox}, which is the same as figure~\subref*{fig:protons2D_defaultBirksBox_data} now showing also the model curves using {\it effective recombination} parameters. We observe a better agreement of the data to the two models with the new parameters.

\begin{table}[!htp]
\begin{center}
\caption{{\it Effective recombination} parameters obtained from the fit of MicroBooNE proton data to the models. Statistical errors on the fit parameters are reported. The parameters obtained by the ArgoNeuT and ICARUS collaborations are also shown for comparison.} 
\begin{tabular}[t]{lcc}  
\toprule
 & values from ref. ~\cite{recomb}~\cite{Birks:1964zz} & new value \\
\midrule
modified box model $\alpha$ & (0.93 $\pm$ 0.02) &(0.92 $\pm$ 0.02) \\
\hline
modified box model $\beta^{\prime}$ & (0.212 $\pm$ 0.002) &(0.184 $\pm$ 0.002) \\
$\text{(kV/cm)(g/cm\textsuperscript2)/MeV}$ & &\\
\hline
Birks' law $A_{B}$ & (0.800 $\pm$ 0.003) &(0.816 $\pm$ 0.012) \\
\hline
Birks' law $k$ & (0.0486 $\pm$ 0.0006) &(0.045 $\pm$ 0.001) \\
$\text{(kV/cm)(g/cm}^2\text{)/MeV}$ & & \\
\bottomrule
\end{tabular}
\label{tab:proton_recombination}
\end{center}
\end{table}


\begin{figure}[!htp]
  \centering
  \subfloat[With ArgoNeuT parameters]{
    \includegraphics[width=0.49\textwidth]{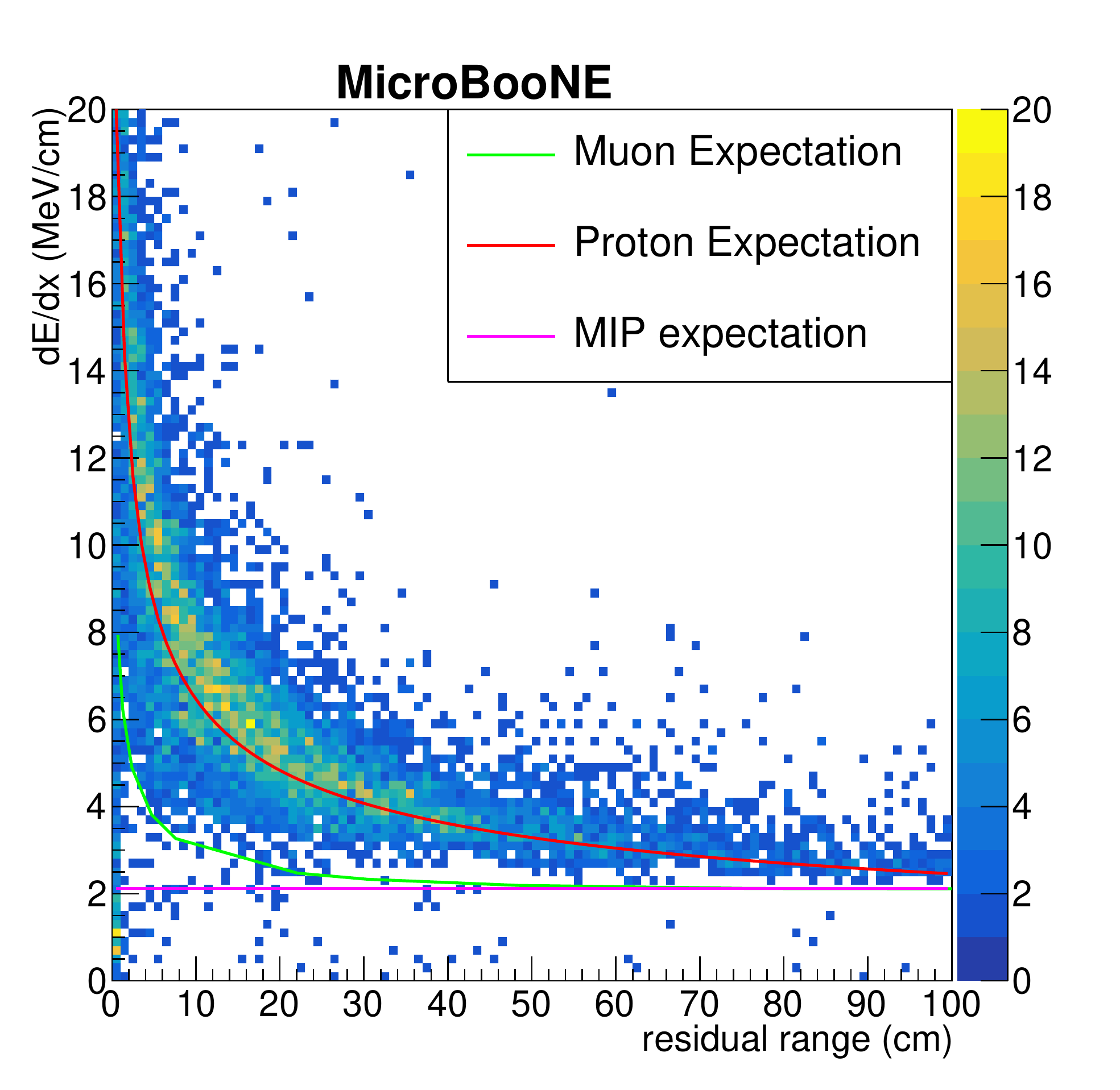}
    \label{fig:allprotons_dEdxvsrr_OnBeam_argo}
  }
  \subfloat[With MicroBooNE {\it effective recombination} parameters]{
    \includegraphics[width=0.49\textwidth]{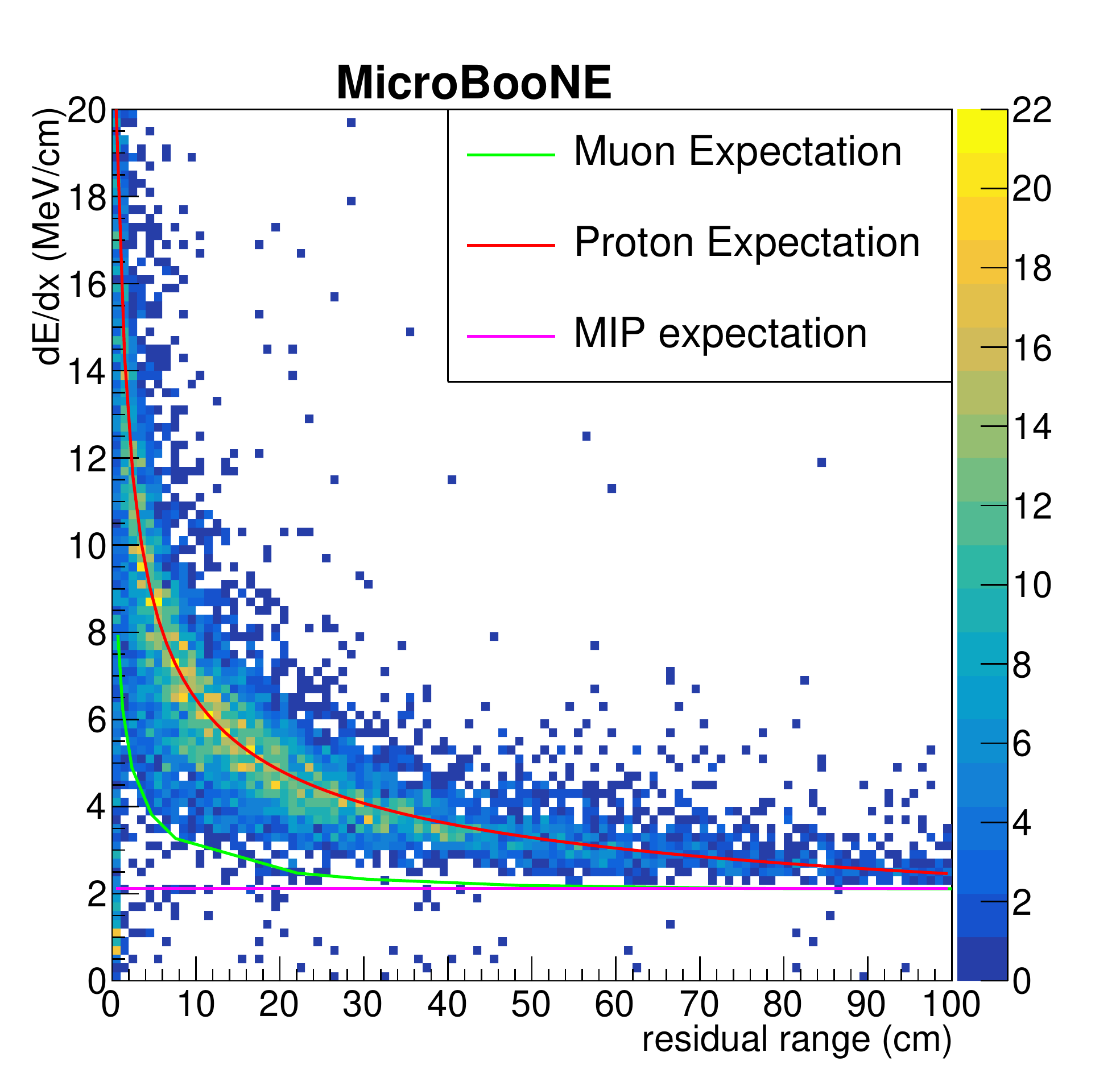}
    \label{fig:allprotons_dEdxvsrr_OnBeam_ub}
  }
  \caption{$dE/dx$ {\it vs} residual range for contained tracks within the selected $\nu_{\mu}$ CC inclusive sample, with a proton PID requirement, for neutrino data. \protect\subref{fig:allprotons_dEdxvsrr_OnBeam_argo} shows the $dE/dx$ reconstructed using the modified box model with the ArgoNeuT parameters. \protect\subref{fig:allprotons_dEdxvsrr_OnBeam_ub} shows the $dE/dx$ reconstructed using the modified box model with the MicroBooNE {\it effective recombination} parameters, which leads to an improved agreement with the theoretical prediction using the Bethe-Bloch equation.}
  \label{fig:allprotons_dEdxvsrr_OnBeam}
\end{figure}

The {\it effective recombination} parameters are validated using the MicroBooNE $\nu_{\mu}$ CC inclusive data sample. The calculated $dE/dx$ values using the {\it effective recombination} parameters are compared with the Bethe-Bloch prediction with respect to residual range in figure~\subref*{fig:allprotons_dEdxvsrr_OnBeam_ub}. As a comparison, we show in figure~\subref*{fig:allprotons_dEdxvsrr_OnBeam_argo} the reconstructed $dE/dx$ using the modified box model with the ArgoNeuT parametrization. Improvement between the calculated $dE/dx$ distribution and the Bethe-Bloch prediction for protons can be seen when using the {\it effective recombination} parameters (right), with respect to the same model using the parameters obtained by ArgoNeuT (left).

\begin{figure}[!htp]
  \centering
  \subfloat[With uncalibrated $dQ/dx$ and ArgoNeuT parameters]{
    \includegraphics[width=0.49\textwidth]{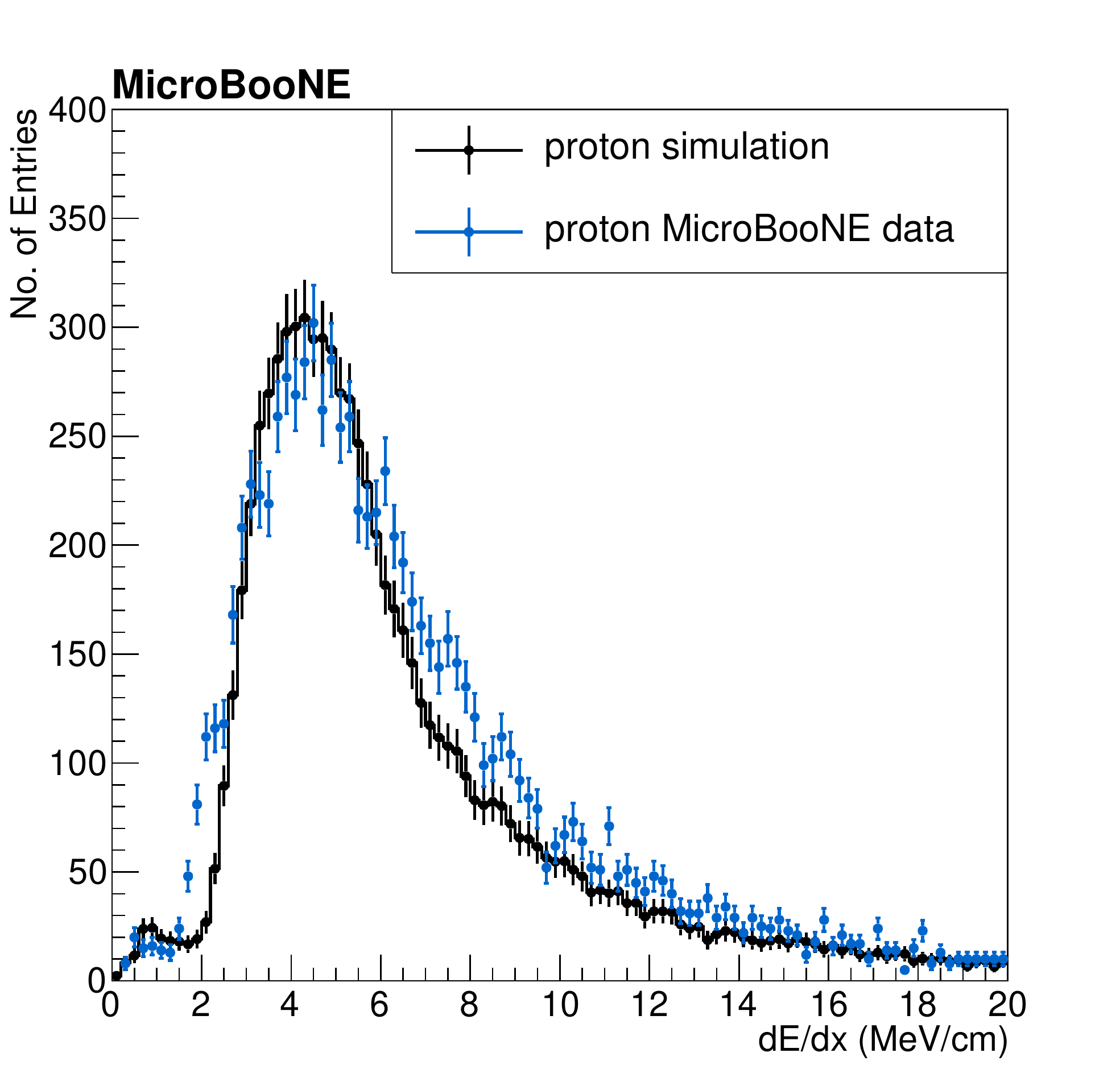}
    \label{fig:comparison_protondEdx_a}
  }
  \subfloat[With calibrated $dQ/dx$ and ArgoNeuT parameters]{
    \includegraphics[width=0.49\textwidth]{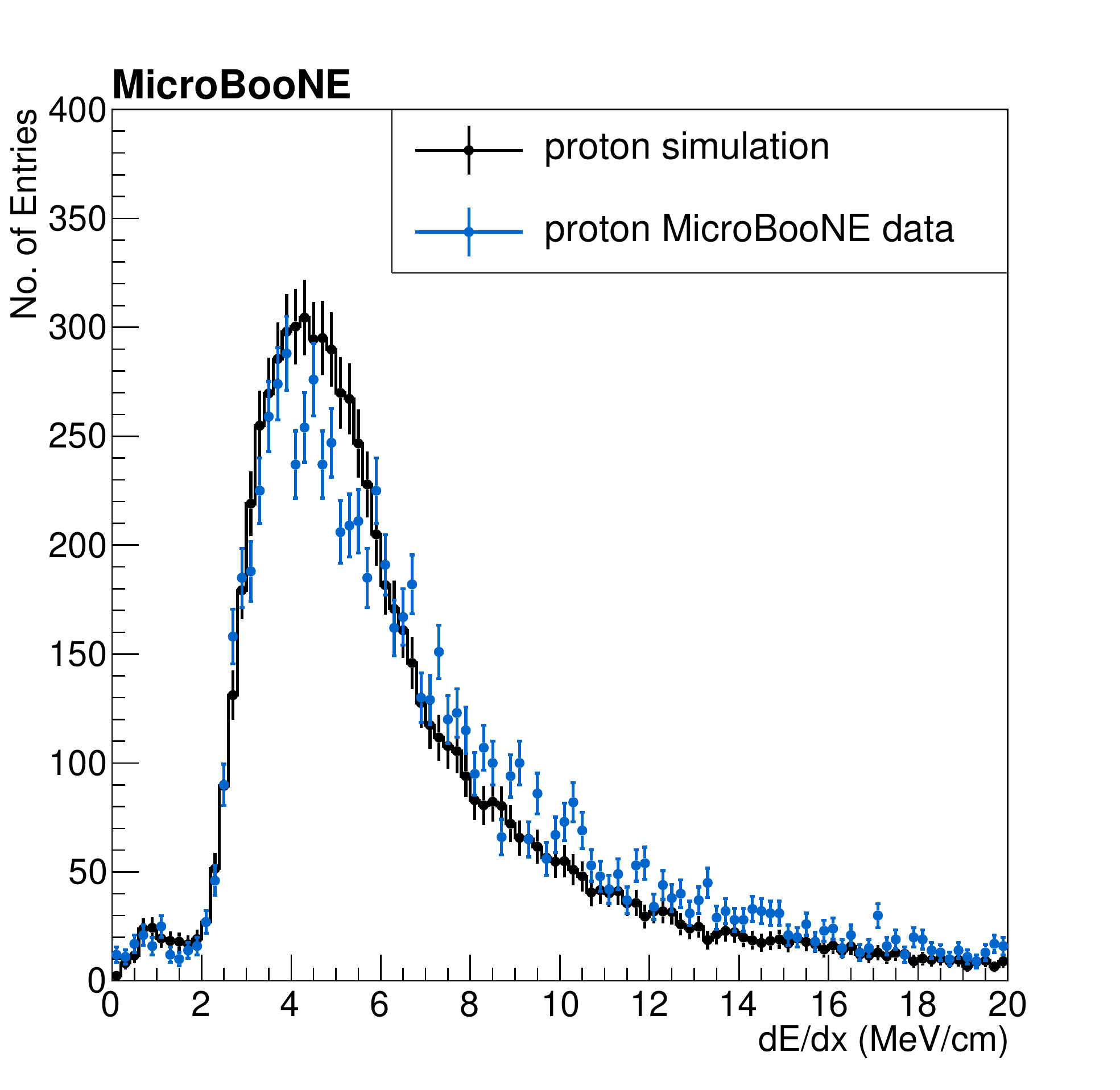}
    \label{fig:comparison_protondEdx_b}
  }
  \\
  \subfloat[With calibrated $dQ/dx$ and MicroBooNE parameters]{
    \includegraphics[width=0.49\textwidth]{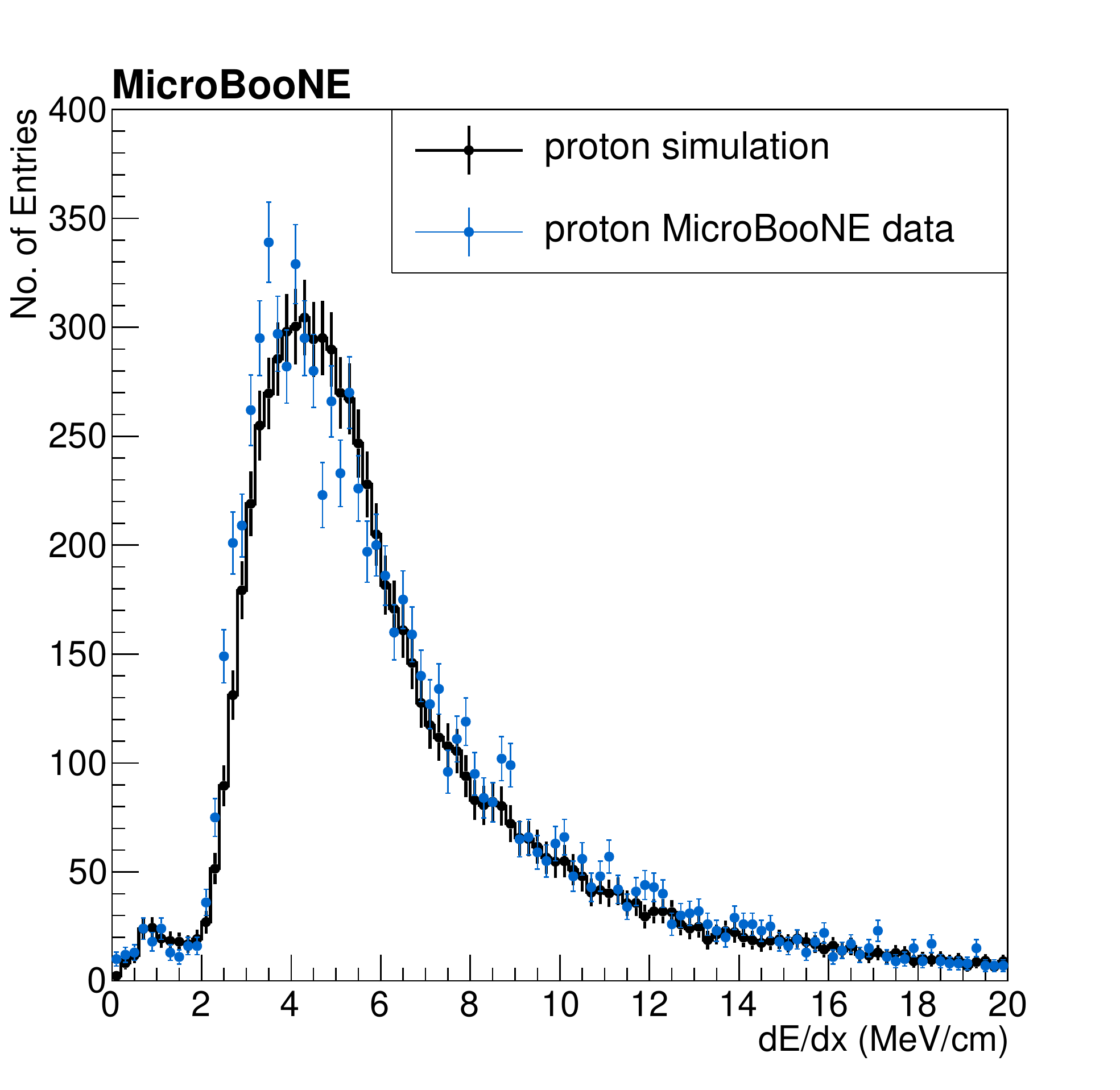}
    \label{fig:comparison_protondEdx_c}
  }
  \subfloat[$dE/dx$ distributions of muons and protons]{
    \includegraphics[width=0.49\textwidth]{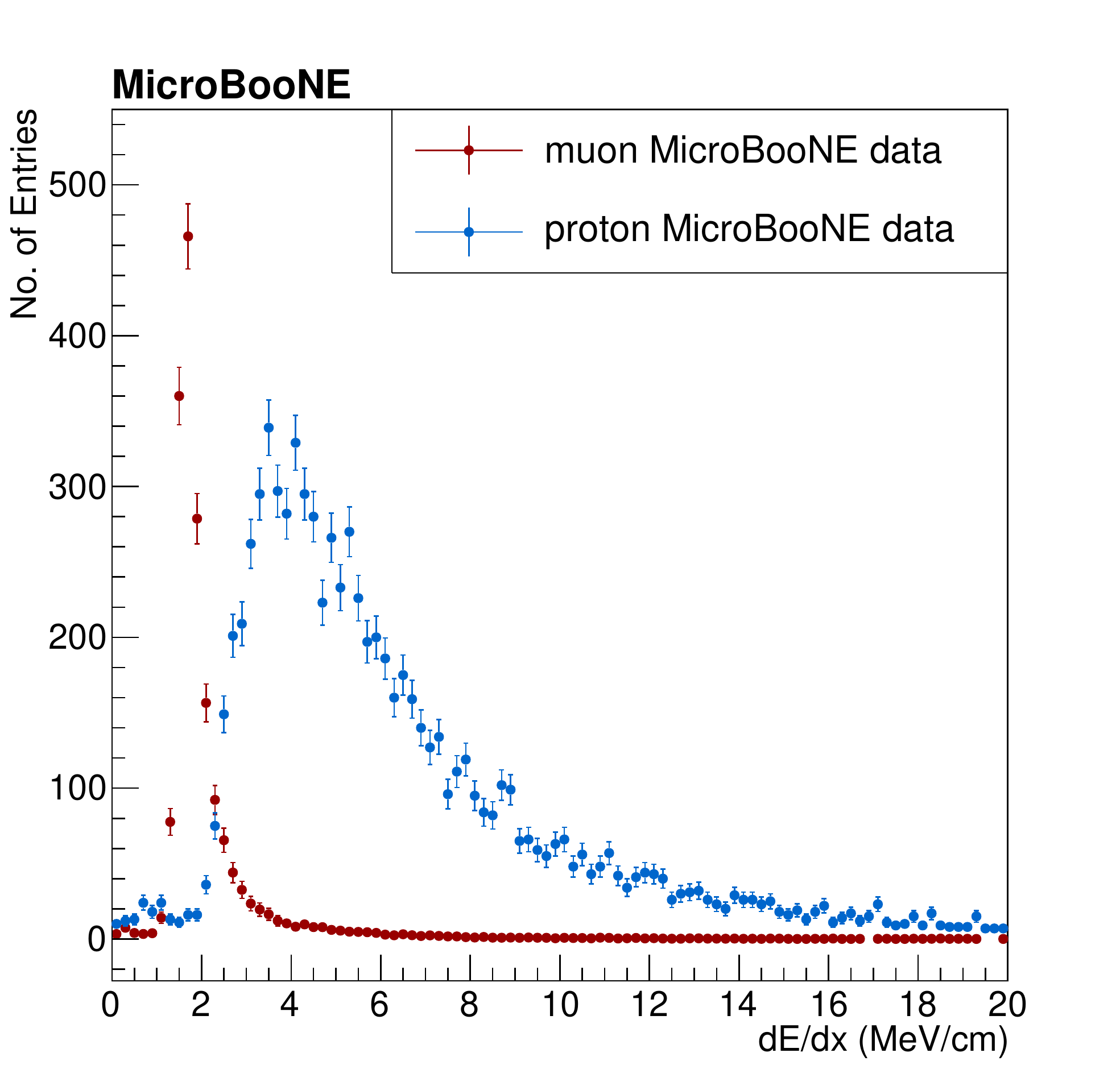}
    \label{fig:comparison_protondEdx_d}
  }
  \caption{$dE/dx$ per hit for tracks selected as a proton candidate from MicroBooNE data compared to simulation. \protect\subref{fig:comparison_protondEdx_a} data use uncalibrated $dQ/dx$ and recombination parameters from ArgoNeuT; \protect\subref{fig:comparison_protondEdx_b} data use calibrated $dQ/dx$ and recombination parameters from ArgoNeuT; \protect\subref{fig:comparison_protondEdx_c} data use calibrated $dQ/dx$ and parameters in table~\ref{tab:proton_recombination}; \protect\subref{fig:comparison_protondEdx_d} $dE/dx$ per hit for tracks selected as muons compared to the ones selected as protons in data.}
  \label{fig:comparison_protondEdx}
\end{figure}

The impact of the full calibration procedure on proton $dE/dx$ reconstruction is shown in figure~\ref{fig:comparison_protondEdx}. For figures~\subref*{fig:comparison_protondEdx_a},~\subref*{fig:comparison_protondEdx_b} and \subref*{fig:comparison_protondEdx_c}, the distributions show the $dE/dx$ per hit of the proton candidates in data compared to those from the simulation. Simulated $dE/dx$ is calculated using calibrated $dQ/dx$ and the modified box model with ArgoNeuT parameters. Data $dE/dx$ in figures~\subref*{fig:comparison_protondEdx_a} is calculated using uncalibrated $dQ/dx$ and the modified box model with ArgoNeuT parameters. Data $dE/dx$ in figures~\subref*{fig:comparison_protondEdx_b} is calculated using calibrated $dQ/dx$ and the modified box model with ArgoNeuT parameters. The equalization calibration clearly improves the $dE/dx$ reconstruction, but there is a discrepancy at high $dE/dx$ values. Data $dE/dx$ in figures~\subref*{fig:comparison_protondEdx_c} plot is calculated using calibrated $dQ/dx$ and the modified box model with the MicroBooNE {\it effective recombination} parameters as reported in table~\ref{tab:proton_recombination}. The agreement is improved for high $dE/dx$ values.  The agreement for low $dE/dx$ values gets slightly worse. This will be improved by retuning the $dE/dx$ scale using stopping muons, as described in section~\ref{sec:dedx}, after applying the {\it effective recombination} parameters in the future. 

We also show the $dE/dx$ per hit distributions for selected protons and muons from data in figures~\subref*{fig:comparison_protondEdx_d}. In this plot, the equalization calibration and the {\it effective recombination} correction are applied. We can clearly see from the figure that protons and muons can be well identified and have different widths in $dE/dx$ as expected. 

\section{Conclusions}
\label{sec:conclusion}
In this paper, we describe for the first time a method that calibrates the calorimetric response of the ionization signal in a LArTPC using muons. We use crossing muons to remove spatial and temporal variations in the detector response. We use stopping muons to determine the absolute energy loss per unit length scale for minimum ionizing particles.  We use stopping protons to further refine the relation between the measured charge and energy loss for highly-ionizing particles. The most probable value of calibrated energy loss agrees with expectation. The calibrated energy loss per unit length is used in particle identification and energy reconstruction. This method can be followed by other LArTPC experiments to calibrate their detectors. 

\section*{Acknowledgements}
This document was prepared by the MicroBooNE collaboration using the
resources of the Fermi National Accelerator Laboratory (Fermilab), a
U.S. Department of Energy, Office of Science, HEP User Facility.
Fermilab is managed by Fermi Research Alliance, LLC (FRA), acting
under Contract No. DE-AC02-07CH11359.  MicroBooNE is supported by the
following: the U.S. Department of Energy, Office of Science, Offices
of High Energy Physics and Nuclear Physics; the U.S. National Science
Foundation; the Swiss National Science Foundation; the Science and
Technology Facilities Council (STFC), part of the United Kingdom Research and Innovation; and The Royal Society (United Kingdom).  Additional support for the laser calibration system and cosmic ray tagger was provided by the Albert
Einstein Center for Fundamental Physics, Bern, Switzerland.

\end{document}